\documentclass[preprint,amsmath,amssymb,aps,prd,nofootinbib]{revtex4}
\usepackage{epsfig,natbib,graphicx,color,appendix,fge}
\begin{document}
%\begin{flushright}
%{ CAU-THEP-19-08}
%\end{flushright}
\def\CP{{\it CP}~}
\def\cp{{\it CP}}
\title{\mbox{}\\[15pt]
 Neutrino Mass Origin and Flavored-QCD axion\\ in an Extra-Dimension}

\author{Y. H. Ahn \footnote{Email: yhahn@htu.edu.cn}
}
\affiliation{Institute of Particle and Nuclear Physics, Henan Normal University, Xinxiang, Henan 453007, China}

%\email[]{Your e-mail address}
%\homepage[]{Your web page}
%\thanks{}
%\altaffiliation{}

%Collaboration name if desired (requires use of superscriptaddress
%option in \documentclass). \noaffiliation is required (may also be
%used with the \author command).
%\collaboration can be followed by \email, \homepage, \thanks as well.
%\collaboration{}
%\noaffiliation

%\date{\today}
\begin{abstract}
We propose a unified flavor model with the Standard Model fields on two 3-branes within an extra-dimensional setup, incorporating $\Gamma_N\times U(1)_X$ symmetry with a modulus and scalar field responsible for symmetry breaking. When compactified to four dimensions, Yukawa couplings, initially expressed as modular forms with mass dimensions, are normalized to conform to canonical four-dimensional theory, with the Yukawa coefficients being complex numbers of unit absolute value. We show that this model naturally explains the mass and mixing hierarchies of quarks and leptons, solves the strong CP problem, provides a natural solution to the hierarchy problem, and  can inherently satisfy no axionic domain-wall problem. The $U(1)_X$ mixed gravitational anomaly-free condition necessitates that electrically neutral mirror bulk fermions couple to the normal neutrino field on the 3-brane, consistent with the boundary condition. Consequently, we demonstrate a mechanism for generating light neutrino masses, similar to the Weinberg operator, by transmitting the information of $U(1)_X$ breakdown between the two 3-branes. The scale of $U(1)_X$ breaking is estimated from neutrino data to be around $10^{15}$ GeV, leading to a QCD axion mass of approximately $2.5\times10^{-9}$ eV. Through numerical analysis, we demonstrate that the model yields results consistent with current experimental data on quarks and leptons, and it also provides predictions for neutrinos.

\end{abstract}

\maketitle 
%%%%%%%%%%%%%%%%%%%%%%%%%%%%%%%%%%%%%%%%%%%%%%%%%%%%%%%%%%%%%%%%%%%%%%%%%%
\section{Introduction}
Although the Standard Model (SM) is theoretically consistent and has been validated by low-energy experimental results, it leaves several unanswered theoretical and cosmological challenges. 
Various attempts have been made to extend the SM to address the open questions and account for experimental results that the SM cannot explain. For instance, the canonical seesaw mechanism\,\cite{Minkowski:1977sc} explains the small masses of neutrinos by introducing new heavy neutral fermions alongside existing SM particles. The Peccei-Quinn (PQ) mechanism\,\cite{Peccei-Quinn} aims to solve the strong CP problem in quantum chromodynamics (QCD) by incorporating an anomalous $U(1)_X$ symmetry. Feruglio's work\,\cite{Feruglio:2017spp} is a string-derived mechanism that naturally restricts the possible variations in the flavor structure of quarks and leptons, which are unconstrained by the SM gauge invariance (see references\,\cite{modular, modular2}). Building on this, the authors in Ref.\cite{Ahn:2023iqa} advanced the mechanism more generally by making it modular anomaly-free.

To address the aforementioned open questions, we propose a unified SM with modular invariance, based on a simple effective 5-dimensional (5D) geometry within a supersymmetric framework derived from superstring theory. This model incorporates a $G_{\rm SM}\times \Gamma_N\times U(1)_X$ symmetry, providing a suitable ultraviolet (UV) completion of the-brane-world setup. This configuration should manifest at the fundamental scale $\Lambda_5$ (5D Planck scale). In addition to the SM gauge group with its associated field content, a newly introduced gauge group may undergo spontaneous symmetry breaking at the UV scale, leaving behind a global subgroup. Consequently, the low-energy theory features a global symmetry group\,\footnote{The non-Abelian discrete symmetry group $\Gamma_N$ (with $N\geq2$) is a subgroup of the fundamental mathematical structure of the modular group $SL(2,\mathbb{Z})$. Additionally, the anomalous global $U(1)_X$ symmetry originates from a gauged $U(1)_X$ symmetry. Consequently, both symmetries are not affected by quantum gravity effects\,\cite{Ahn:2023iqa}.} $G_F=U(1)_X\times\Gamma_N$, which includes at least one scalar field and one modulus responsible for the spontaneous symmetry breaking. The non-Abelian discrete symmetry $\Gamma_N$ (with $N=2,3,4,5$) plays a role of modular invariance\,\cite{Feruglio:2017spp, Ahn:2023iqa}, and may originate from superstring theory in compactified extra dimensions, where it acts as a finite subgroup of the modular group\,\cite{deAdelhartToorop:2011re}. The modular invariance of the superpotential under the modular group $\Gamma_N$ is summarized in Appendix-\ref{A4_i} and Appendix-\ref{A4_2}. The model is simply illustrated in Fig.\ref{Fig1} and has following features:
\begin{description}
\item[ (i)] Thanks to the orbifold comapctification, we set all the SM elementary fermions form a chiral set because chirality can enter the theory. All ordinary matter and Higgs fields charged under $G_{\rm SM}\times G_F$ are localized at either brane. Then all the SM $SU(2)$ singlets such as right-handed quarks ($q^c$) and right-handed charged leptons ($\ell^c$) are localized at $y=0$ brane, while the $SU(2)$ doublets such as left-handed quarks ($Q$), left-handed leptons ($L$), and two electroweak (EW) Higgs $H_{u(d)}$ are localized at $y=L$ brane. Additionally, the newly introduced SM gauge singlet scalar field ${\cal S}$, responsible for $U(1)_X$ symmetry breaking, is localized at the $y=0$ brane.
%%%%%%%%%%%%%%%
%   Fig A-1   %
%%%%%%%%%%%%%%%
\begin{figure}[t]
%%\vspace*{-5.0cm}
%\hspace*{-0.5cm}
\centering
\includegraphics[width=10.0cm]{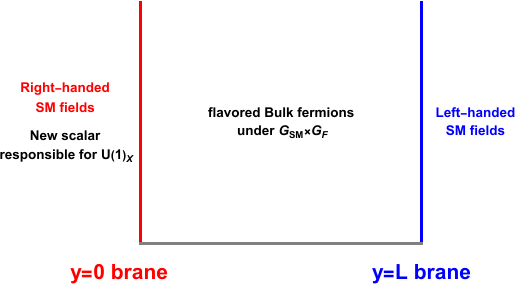}
%%\vspace*{-5.5cm}
\caption{\label{Fig1} A simple extra dimension scenario introducing the $G_{\rm SM}\times G_F$.}
\end{figure}
\item[ (ii)] In a slice of AdS$_5$, there are bulk fermions propagating in a 5-dimensional space. These fermions are singlets under $SU(2)$ with hypercharge $Y_f$ and masses $M^f_i$. They interact with the ordinary matter fields confined to the branes $y=0$ and $y=L$. The bulk fermions can play a crucial role in exchanging information, such as the breaking of flavor symmetry $G_F=U(1)_X\times\Gamma_N$ and the quantum number of SM fields, between the two branes. Consequently, the 4-dimensional theory can naturally explain the observed SM fermion mixing and masses. To conserve charge under $G_{\rm SM}\times G_F$, two types of $SU(2)$ singlet bulk fermions are introduced: bulk fermions and their mirror fermions. Bulk fermions with a common hypercharge can be distinguished by a flavor symmetry $G_F$, hence termed flavored-bulk fermions. $i$-th $f$-type bulk fermion $\Psi_{f_i}(x,y)$ connects the corresponding $f$-type SM fermion confined at both branes, where $f=u$ (up-type quark), $d$ (down-type quark), $\ell$ (charged-lepton), $\nu$ (neutrino), and $i=1,2,3,...$ (the number of generation stemming from the non-Abelian discrete symmetry $\Gamma_N$). Since all bulk fermions reside in a slice of AdS$_5$, we assume their masses are quasi-degenerate:
\begin{eqnarray}
M^u_i\approx M^d_i\approx M^\ell_i\approx M^\nu_i= M_f\,.
\label{f_M1}
\end{eqnarray}
This quasi-degeneracy guarantees that the low-energy effective Yukawa Lagrangian produces the experimentally observed quark and lepton masses (see Eqs.(\ref{4d_a}) and (\ref{4d_a1})).
\item[ (iii)]  As there are no right-handed neutrinos in the SM, we stipulate the absence of a corresponding right-handed neutrino at the $y=0$ brane. To consistently couple gravity to matter, the $U(1)_X$ mixed gravitational anomaly must be canceled. To satisfy this anomaly-free condition, the electrically neutral mirror bulk fermion $\Psi^c_\nu$ should couple to the normal neutrino field on the 3-brane, see Eq.(\ref{lep1}) and above Eq.(\ref{gr_ano}). Consequently, the electrically neutral bulk fermion $\Psi_\nu$ can couple to itself with the scalar field ${\cal S}$ at the $y=0$ brane, while its mirror bulk fermion can couple to the $SU(2)$ lepton doublet at the $y=L$ brane. This enables the generation of light neutrino masses by communicating the breakdown of $G_F$ between the two 3-branes (see Eqs.(\ref{4d_b}, \ref{4d_a2})). The scale at which $U(1)_X$ symmetry breaks is pivotal in this process and can be inferred from experimental constraints on light neutrino masses, see Eq.(\ref{MR5}).
\item[ (iv)] In a slice of AdS$_5$ the scalar mass $\mu_{\cal S}$ represents a value near the 5D cutoff scale, as expected for a scalar field. This scalar is responsible for the spontaneous breaking of flavor symmetry $U(1)_X$ at a very high energy scale. On the brane, the global $U(1)_X$ symmetry, originating from the gauged $U(1)_X$, is anomalous. When the $U(1)_X$ is spontaneously broken, a Nambu-Goldstone (NG) mode $A$ emerges. This mode interacts with ordinary quarks and leptons through Yukawa interactions in the flavored-axion framework. Consequently, this anomaly could lead to the axionic domain-wall problem in QCD instanton backgrounds within the flavored-axion framework\,\cite{Ahn:2014gva}. Notably, this issue can be resolved in the extra-dimension framework without introducing two anomalous axial $U(1)$ symmetries. The condition $N_{\rm DW}=1$ can be satisfied by the presence of an additional scalar field ${\cal S}$, charged under $U(1)_X$, in the $y=0$ brane operators (see Eqs.(\ref{nc1}, \ref{up1}, \ref{down1})). 
\item[ (v)] In higher-dimensional theories, upon compactification to four dimensions, Yukawa couplings, which have mass dimensions and appear as modular forms, undergo normalization to align with canonical four-dimensional theory. This normalization process yields canonically normalized Yukawa modular forms (see Eq.(\ref{mf1})). Additionally, all Yukawa coefficients in the superpotential can be complex numbers with unit absolute value (see Eq.(\ref{4d_a1})).
\end{description}

The rest of this paper is organized as follows. In the next section, we set up an extra-dimension model based on $G_F \times G_{\rm SM}$ symmetry by introducing flavored bulk fermions, the SM fields on the branes, and one scalar field and a modulus responsible for the spontaneous symmetry breakdown of $G_F$. Consecutively, in Sec.\ref{sec3}, we discuss the low-energy 4D effective Lagrangian, presenting the effective quark and lepton Yukawa Lagrangian, including a mechanism for generating light neutrino masses and the interactions of the 4D KK modes of bulk neutrinos. Here, we address the challenges of tiny neutrino masses, the strong CP problem, and the hierarchies of SM fermion mass and mixing.
 In Sec.\ref{sec4}, we visually demonstrate the interconnections between quarks, leptons, and the flavored-QCD axion. Here, we also present numerical values of physical parameters that satisfy the current experimental data on flavor mixing and mass for quarks and leptons. The study predicts the Dirac CP phases of quarks and leptons, as well as the mass of the flavored-QCD axion and its coupling to photons and electrons. The final section provides a summary of our work.
%%%%%%%%%%%%%%%%%%%%%%%%%%%%%%%%%%%%%%%%%%%%%%%%%%%%%%%%%%%%%%%%%%%%%%%%%%%
\section{Minimal model set-up}
\label{sec2}
First, consider the general setup of an extra dimension\,\cite{Randall:1999ee, Horava:1995qa}, based on a 5D theory with the extra dimension compactified in an orbifold, $S^1/Z_2$. This setup involves a circle $S^1$ with the extra identification of $y$ with $-y$ ($y$ is the physical distance along the extra-dimension)\,\cite{Gabadadze:2003ii}. The orbifold fixed points at $y=0$ and $y=L$ are the locations of two 3-branes, which form the boundaries of the 5D spacetime.

Assume that this 5D theory has a cosmological constant in the bulk $\Lambda$, and on the two boundaries $\Lambda_0$ and $\Lambda_L$. The 5D gravity action is given by\,\cite{Randall:1999ee}
\begin{eqnarray}
S\supset-\int d^4xdy\sqrt{g}\Big[-\frac{1}{2}\Lambda^3_5{\cal R}+\Lambda+\frac{\delta(y)}{\sqrt{-g_{55}}}\big(\Lambda_0+{\cal L}_0\big)+\frac{\delta(y-L)}{\sqrt{-g_{55}}}\big(\Lambda_L+{\cal L}_L\big)\Big]
\label{ac_g_w}
\end{eqnarray}
where $\Lambda_5$ is the 5D reduced Planck mass, ${\cal R}$ is the 5D Ricci scalar, $g\equiv \det(g_{MN})$ is the determinant of the 5D metric $g_{MN}$ with the 5D coordinates $M=(\mu, 5)$ with $\mu=0,1,2,3$, and ${\cal L}_0({\cal L}_L)$ is the brane localized Lagrangian, where all ordinary matter fields localized at either brane are charged under a flavored global symmetry, $G_F=\Gamma_N\times U(1)_X$.
The 5D Einstein's equations, which respects four-dimensional Poincare invariance in the $x^\mu$ direction, are solved by
\begin{eqnarray}
ds^2=g_{MN}dx^Mdx^N=e^{-2\sigma(y)}\eta_{\mu\nu}dx^{\mu}dx^{\nu}-dy^2
\label{met1}
\end{eqnarray}
compactified on an interval $y\in[0,L]$, where
\begin{eqnarray}
\sigma(y)=ky\qquad\text{with}~k=\sqrt{-\frac{\Lambda}{6\Lambda^3_5}}>0\,,
\end{eqnarray}
and $1/k$ is the 5D anti-deSitter (AdS) curvature radius. The 4D Minkowski flat metric is $\eta_{\mu\nu}={\rm diag}(+,-,-,-)$. Consistently, the metric solution of Eq.(\ref{met1}) requires $\Lambda_0=-\Lambda_L=-\Lambda$. (In Eq.(\ref{met1}) we can always take $\sigma(0)=0$ by rescaling the $x^\mu$). The 4D reduced Planck mass $M_P\simeq2.43\times10^{18}$ GeV can be extracted in terms of the 5D Planck mass $\Lambda_5$ as
\begin{eqnarray}
M^2_P=\Lambda^3_5\int^{L}_{-L}dy\,e^{-2\sigma(y)}=\frac{\Lambda^3_5}{k}\big(1-e^{-2kL}\big)\,.
\label{re_M}
\end{eqnarray}
From the form of the metric solution Eq.(\ref{met1}), the spacetime between the branes located at $y=0$ and $y=L$ brane is simply a slice of AdS$_5$ geometry. The slice of AdS$_5$ provides a low energy effective field theory below the Planck scale $M_P$. The fundamental gravity scale of 5D theory, $\Lambda_5$, is assumed to be higher than the EW scale since there is no evidence of quantum gravity well up to energies around few hundred GeV, while $\Lambda_5$ and $k$ are lower than the Planck scale $M_P$. Then the UV cutoff, which is the scale of flavor dynamics, would be given by the quantum gravity scale $\Lambda_5$, above which the theory must be UV-completed. The compactification length $L$ is associated with the VEV of a massless 4D scalar field\,\cite{Randall:1999ee}, which has zero potential and thus does not determine $L$. While various models have been proposed to stabilize $L$\,\cite{Goldberger:1999uk, Steinhardt:1999eh}, we will not consider such stabilization in this paper. 

In the model where {\it flavored bulk fermions} charged under $G_{\rm SM}\times G_F$ are propagating in the extra dimension, we consider the case where the warp size $\sigma(L)$ is much smaller than one. Interestingly, $\sigma(L)$ could be estimated as $\sigma(L)\ll1$ based on flavor physics and the normalization of modular forms, as detailed later (see section-\ref{4DYL}).
The separation between the two 3-branes, $L[m]$, can be determined by the UV cutoff scale $\Lambda_5$:
\begin{eqnarray}
L[m]\simeq5.82\times10^{20}\,\frac{e^{\sigma(L)}\sigma(L)}{\sinh\sigma(L)}\Big(\frac{{\rm GeV}}{\Lambda_5}\Big)^3\approx5.8\times10^{20}\,\Big(\frac{{\rm GeV}}{\Lambda_5}\Big)^3\,.
\label{re_M1}
\end{eqnarray}
In the second approximation, $\sigma(L)\ll1$ has been used. Given a specific value of $\Lambda_5$ with the condition $\sigma(L)\ll1$, the possible compactification lengths $L$ as examples are
\begin{eqnarray}
%&\Lambda_5=10^{16}{\rm GeV}\Rightarrow\sigma(L)=[0.01, 10^{-8}]\Leftrightarrow L[{\rm m}]\sim6\times10^{-28}\Leftrightarrow k[{\rm GeV}]=[3\times10^{9}, 3\times10^{3}]\,\nonumber\\
\Lambda_5=10^{15}{\rm GeV}\Rightarrow L[{\rm m}]\sim6\times10^{-25}\,,\quad \Lambda_5=10^{14}{\rm GeV}\Rightarrow L[{\rm m}]\sim6\times10^{-22}\,.
\label{re_M2}
\end{eqnarray}
The $U(1)_X$ breaking scale can be strongly constrained by experimental data of light neutrinos (see Eqs.(\ref{4d_a2}) and (\ref{MR5})). Consequently, $\Lambda_5$ can also be inferred via $\langle{\cal S}\rangle/\Lambda_5\lesssim1$ governed by flavor dynamics, necessitating a small warp factor $\sigma(L)\ll1$ (see Eq.(\ref{4d_a1})).

%%%%%%%%%%%%%%%%%%%%%%%%%%%%%%%%%%%%%%%%%%%%%%%%%%%%%%%%%%%%%%%%%%%%%%%%%%%%%%%%%%%%%%%%%%
\subsection{Flavored bulk fermions}
The flavored bulk fermions play a crucial role in exchanging information, such as the breaking of flavor symmetry $G_F=U(1)_X\times\Gamma_N$ and the quantum number of SM fields, between the two 3-branes. For instance, the bulk fermion $\Psi_f(x,y)$ with $f=u,d,\ell,\nu$ has one-to-one corresponding hypercharge $Y_f$ to the SM $SU(2)$ singlet fermion confined at $y=0$ brane, while its mirror bulk fermion $\Psi^c_{f}(x,y)$ has the opposite hypercharge $-Y_f$. For simplicity, we assign quantum numbers of $G_{\rm SM}\times G_F$ with $\Gamma_3\simeq A_4$\,\cite{deAdelhartToorop:2011re}, especially, by ensuring $\tau$-independent modular form according to Ref.\cite{Ahn:2023iqa}, as shown in Table-\ref{reps_1}. Details of the $A_4$ group are provided in Appendix \ref{A4_i}.
  \begin{table}[h]
\begin{widetext}
\begin{center}
\caption{\label{reps_1} Representations of the flavored bulk quarks and leptons living in a slice of AdS$_5$ under $G_{\rm SM}\times \Gamma_3\times U(1)_X$ with $\Gamma_3\simeq A_4$ and modular weight $k_I$.}
\begin{ruledtabular}
\begin{tabular}{ccccc}
Field &$\Psi_{u_1}$\,,~$\Psi_{u_2}$\,,~$\Psi_{u_3}$&$\Psi_{d_1}$\,,~$\Psi_{d_2}$\,,~$\Psi_{d_3}$&$\Psi_{e}$\,,~$\Psi_{\mu}$\,,~$\Psi_{\tau}$ & $\Psi_\nu$   \\
$G_{\rm SM}$&$(3,1)_{2/3}$&$(3,1)_{-1/3}$  & $(1,1)_{-1}$ & $(1,1)_{0}$ \\
\hline
$A_4$&${\bf 1}$ \, ${\bf 1}''$ \, ${\bf 1}'$&${\bf 1}$ \, ${\bf 1}''$ \, ${\bf 1}'$ & ${\bf 1}$ \, ${\bf 1}'$ \, ${\bf 1}''$ & ${\bf 3}$ \\
$k_I$&$\frac{h}{2}$&$\frac{h}{2}$&$\frac{h}{2}$ & $\frac{h}{2}$\\
$U(1)_X$&$X_{Q_1}\,,X_{Q_2}\,,X_{Q_3}$&$X_{Q_1}\,,X_{Q_2}\,, X_{Q_3}$&$X_L$ & $X_L$\\
\end{tabular}
\end{ruledtabular}
\end{center}
\end{widetext}
\end{table} 
Therefore, the exchange of flavored bulk fermions between the two 3-branes can induce non-local interactions between right- and left-handed SM fermions (see Eqs.(\ref{up1}, \ref{down1}, \ref{lep1})), where all bulk fermions (their mirrors) are left-handed particles (antiparticles).
For an orbifold compactification (leading to the physical region in the extra-dimension $[0,L]$), chirality enters the theory and the corresponding gauge theory is anomalous\,\cite{Adler:1969gk}.

We consider $U(1)$ gauge field, $A_M$, and flavored bulk fermions, $\Psi_{f_i}$, living in a slice of AdS$_5$ given by the metric Eq.(\ref{met1}). The 5D action for flavored bulk fermions $\Psi_{f_i}(x,y)$ ($i=1,2,3,...$ and $f=u, d, \ell,\nu$) reads
\begin{eqnarray}
&&S\supset\int d^4x dy\sqrt{g}\Big[\bar{\Psi}_{f_i}\Big\{\frac{i}{2}e^{M}_{~A}\Gamma^A\overleftrightarrow{D}_M-M^f_{i}(y)\Big\}\Psi_{f_i}
-\frac{1}{4}g^{MN}g^{KL}{\cal F}_{MK}{\cal F}_{NL}\Big]
\label{fba1}
\end{eqnarray}
where the 5D metric $g_{MN}$ is decomposed into vierbeins $e^A_{~M}$: $g_{MN}=\eta_{AB}\,e^A_{~M}e^B_{~N}$, $\Gamma^A=(\gamma^\mu, i\gamma_5)$ and $\Gamma_A=(\gamma_\mu, -i\gamma_5)$ satisfy the Dirac-Clifford algebra $\{\Gamma^A, \Gamma^B\}=2\eta^{AB}$ where $\eta^{AB}$ is the 5D flat metric $={\rm diag}(\eta_{\mu\nu}, -1)$. 
The $i$-th $f$-type flavored bulk fermion $\Psi_{f_i}(x,y)$ couples to the $U(1)_{Y}$ gauge field in 5D, denoted by $A_M$. There gauge fields consist of 4D gauge field component $A_\mu$ and 4D scalar component $A_5$. The $U(1)$ gauge covariant derivative is given by $D_M=\partial_M+iY_fA_M$, where $Y_f$ is the hypercharge of the fermion. The $U(1)$ gauge field strength in 5D is defined as ${\cal F}_{MN}=\partial_M A_N-\partial_N A_M$, with the gauge coupling absorbed into the gauge boson $A_M$.
In order for the action Eq.(\ref{fba1}) to conserve $y$-parity, as required by the $Z_2$ orbifold symmetry, the following transformations are imposed:
\begin{eqnarray}
\Psi_{f_i}(x, y)\rightarrow\gamma_5\,\Psi_{f_i}(x, -y)\,;\quad A_\mu(x, y)\rightarrow A_{\mu}(x, -y)\,;\quad
A_5(x, y)\rightarrow-A_{5}(x, -y)\,.
\label{ref1}
\end{eqnarray}
Additionally, for the action in Eq.(\ref{fba1}) to be well-defined under the transformation $y\rightarrow-y$, the mass function should satisfy $M^f_i(y)\rightarrow-M^f_i(-y)$.
It can be seen that the action in Eq.(\ref{fba1}) is invariant under the gauge transformations: 
\begin{eqnarray}
&\Psi_{f_i}(x,y)\rightarrow e^{i\xi(x,y)}\Psi_{f_i}(x,y)\,,\quad A_M(x,y)\rightarrow A_M(x,y)-\partial_M\xi(x,y)\,
\label{gauT}
\end{eqnarray}
with $\xi(x,y)=\xi(x,-y)$. Consequently, it is inevitable that the current is conserved, $\partial_M J^M_{Y}=\partial_\mu J^\mu_{Y}+\partial_5J^5_{Y}=0$, at classical level. 
However, the 5D current is anomalous at quantum level for a 5D fermion coupled to an external gauge potential $A_M(x, y)$ on an $S^1/Z_2$ orbifold with the restricted interval $[0, L]$\,\cite{ArkaniHamed:2001is, Callan:1984sa}. The same framework can be extended to non-Abelian $SU(3)_C$ gauge fields.
To ensure the consistency of the orbifold gauge theory, these anomalies must be canceled. In the 5-dimensional bulk, with the assignment of $SU(3)_C$ and $U(1)_{Y(X)}$ charges to the flavored bulk fermions (as shown in Table-\ref{reps_1}), the gauge anomalies $[SU(3)_C]^2\times U(1)_{Y}$, $[U(1)_Y]^3$, and $U(1)_{Y}\times[{\rm gravity}]^2$ are automatically canceled due to their mirror charges: ${\rm tr}(Y_{f_i})=0$.

In terms of $\Psi_{f_i}(x,y)$ the KK wavefunctions for bulk fermions forming a complete, orthogonal set
\begin{eqnarray}
\Psi_{f_iL(R)}(x, y)=\frac{e^{\frac{3}{2}\sigma(y)}}{\sqrt{L}}\sum_n\psi^{n}_{f_iL(R)}(x)\,f^n_{iL(R)}(y)
\label{KKw1}
\end{eqnarray} 
are chosen to obey the 4D equation of motion (EOM) 
\begin{eqnarray}
S=\sum_n\int d^4x\bar{\psi}^{n}_{f_i}(i\gamma_\mu D^\mu-m^{f_i}_n)\psi^{n}_{f_i}
\label{4Dkk}
\end{eqnarray} 
where $m_n^{f_i}$ is the 4D mass of the $n$-th KK mode, with the normalization condition
\begin{eqnarray}
\frac{1}{L}\int^L_0dy\,f^m_{iL(R)}f^n_{iL(R)}=\delta_{mn}\,.
\label{KKnor}
\end{eqnarray} 
In this context, the gauge $A_5=0$ is chosen to ensure that the KK wave functions are independent of the gauge fields.

In order to discuss the physical effects of the flavored bulk fermion we vary the action of Eq.(\ref{fba1}) and obtain the EOM and boundary condition: Requiring $\delta S=0$ for any $\delta\bar{\Psi}_{f_i}$, the EOM is 
\begin{eqnarray}
ie^{\sigma}\gamma^\mu D_\mu\Psi_{f_i}-\gamma_5\partial_y\Psi_{f_i}+\frac{1}{2}(\partial_y\sigma)\gamma_5\Psi_{f_i}-M^f_i\Psi_{f_i}=0\,,
\label{bulk_1}
\end{eqnarray} 
and the boundary condition
\begin{eqnarray}
\delta \bar{\Psi}_{f_i}\gamma_5\Psi_{f_i}\big|^{y=L}_{y=0}=0\,.
\label{bulk_2}
\end{eqnarray} 
Plugging Eq.(\ref{KKw1}) into Eqs.(\ref{bulk_1}, \ref{bulk_2}), in terms of left- and right-handed spinors $f^n_{iL,R}$ the EOM of Eq.(\ref{bulk_1}) becomes
\begin{eqnarray}
&&\Big(\partial_y-\frac{1}{2}\sigma'+M^f_i\Big)f^n_{iR}(y)=e^{\sigma}m^{f_i}_n\,f^n_{iL}(y)\,,\nonumber\\
&&\Big(\partial_y-\frac{1}{2}\sigma'-M^f_i\Big)f^n_{iL}(y)=-e^{\sigma}m^{f_i}_n\,f^n_{iR}(y)\,,
\label{bulk_11}
\end{eqnarray} 
where $\sigma'=\partial_y\sigma$, and the boundary condition of Eq.(\ref{bulk_2}) becomes
\begin{eqnarray}
\delta f^n_{iL}(y)\,f^n_{iR}(y)-\delta f^n_{iR}(y)\,f^n_{iL}(y)\big|^{y=L}_{y=0}=0\,.
\label{bulk_22}
\end{eqnarray} 
The nonzero KK modes can be obtained by solving the first-order equations of motion in Eq.(\ref{bulk_11}) for the Dirac spinors $f^n_{iL,R}$, subject to the Dirichlet boundary conditions, such as those given in Eq.(\ref{dich}).
At energies much lower than the mass scale $1/L$ associated with the first KK mode ($E\ll1/L$), only the zero mode is significant, and higher modes are suppressed. Conversely, at higher energies ($E\gtrsim1/L$), all KK modes contribute significantly. Given our interest in energies much lower than $1/L$, the 4D covariant derivative term in Eq.(\ref{bulk_1}) can be neglected. This simplifies the solutions to
\begin{eqnarray}
f^0_{iR}(y)= f^0_{iR}(0)\,e^{\frac{1}{2}\sigma(y)-M_fy}\,,\qquad f^0_{iL}(y)= f^0_{iL}(L)\,e^{\frac{1}{2}\{\sigma(y)-\sigma(L)\}+M_f(y-L)}\,,
\label{bc_3}
\end{eqnarray} 
where $\sigma(0)=0$ and $M^f_{i}(y)=M_f=$ constant, see Eq.(\ref{f_M1}), are used.
Choosing the boundary conditions as 
\begin{eqnarray}
\delta f^n_{iL}(L)=\delta f^n_{iR}(0)=0\,,
\label{bc_1}
\end{eqnarray} 
ensures that all left-handed and right-handed KK modes vanish at the $y=L$ and $y=0$ branes, respectively. Consequently, only the right-handed (left-handed) modes can couple to the SM fermions located on the $y=L$ ($y=0$) brane. A crucial point to note is that since there are no right-handed neutrinos at the $y=0$ brane, we impose an additional condition for electrically neutral bulk fermions
\begin{eqnarray}
\delta f^n_{\nu L}(0)=0\,.
\label{bc_2}
\end{eqnarray} 

%%%%%%%%%%%%%%%%%%%%%%%%%%%%%%%%%%%%%%%%%%%%%%%%%%%%%%%%%%%%%%%%%%%%%%%%%%%
\subsection{Modular invariant supersymmetric potential}
\label{exam01}
According to Ref.\cite{Ahn:2023iqa}, a modulus-independent scalar potential can be constructed by introducing minimal supermultiplets. These supermultiplets are naturally introduced by incorporating a new symmetry, $U(1)_X$, into the theory. The supermultiplets include SM singlet fields: $\chi_0$, which has a modular weight of $h$, and $\chi$ and $\tilde{\chi}$, which both have a modular weight of zero. The field $\chi_0$ can also act as an inflaton\,\cite{Ahn:2017dpf}. The fields ${\cal S}=\{\chi$ and $\tilde{\chi}\}$ are charged by $+1$ and $-1$, respectively, and are ensured by the extended $U(1)_X$ symmetry due to the holomorphy of the superpotential. Under the modular transformation Eq.(\ref{mt1}), along with the K{\"a}hler transformation Eq.(\ref{tr1}), the $A_4$-singlet $\chi_0$ field with modular weight $h$ ensures that the modular functions in Eq.(\ref{ms1}) are independent of $\tau$. 
Additionally, the theory includes the usual two Higgs doublets, $H_{u}$ and $H_d$, which have a modular weight of zero and are responsible for EW symmetry breaking. Under $k_I\times A_4\times U(1)_X$, we assign the two Higgs doublets $H_{u(d)}$ to be $(0,{\bf 1},0)$ and three SM gauge singlets $\chi, \tilde{\chi}, \chi_0$ to be $(0,{\bf 1},+1), (0,{\bf 1}, -1), (h, {\bf 1}, 0)$ respectively. 

Then, the brane-localized supersymmetric scalar potential invariant under $G_{\rm SM}\times U(1)_{X}\times A_4$ is given by
\begin{eqnarray}
W_v&=&\delta(y)\big\{\hat{\chi}_0(c_{\chi}\hat{\chi}\hat{\tilde{\chi}}-\mu^2_\chi)\big\}
\label{d_pot}
\end{eqnarray}
where $\mu_{\chi}$ corresponds to the scale of the spontaneous $U(1)_X$ symmetry breaking, and dimensionless coupling constant $c_\chi$ is a complex number assumed to be $|c_{\chi}|=1$. However, this coupling constant is modified to Eq.(\ref{AFN1}) by considering all higher-order terms induced by $\chi\tilde{\chi}$ combinations. 
%%%%%%%%%%%%%%%%%%%%%%%%%%%%%%%%%%%%%%%%%%%%%%%%%%%%%%%%%%%%%%%%%%%%%%%%%%%%%%%%%%%%%%%%%%
\subsection{Brane-localized Yukawa superpotential}
All ordinary matter and Higgs fields are localized on either brane. Thanks to the orbifold compactification, we set all elementary fermions form a chiral set. Then all SM $SU(2)$ singlets such as right-handed quarks ($q^c$) and right-handed charged-leptons ($\ell^c$) are localized at $y=0$ brane, while $SU(2)$ doublets such as left-handed quarks ($Q_i$), left-handed leptons ($L$), and two EW Higgs $H_{u(d)}$ are localized at $y=L$ brane. Beyond the SM gauge group, a newly introduced gauge group might experience spontaneous symmetry breaking at the UV scale, resulting in the emergence of a global subgroup. Consequently, at low energies, the theory exhibits a global symmetry group $G_F=U(1)_X\times\Gamma_N$, incorporating at least one scalar field ${\cal S}$ and a modulus $\tau$
responsible for their spontaneous symmetry breakdown. Then the newly introduced SM gauge singlet field ${\cal S}$ is localized at the $y=0$ brane. 
Under $G_{\rm SM}\times G_F$, alongside the flavored bulk fermions in Table \ref{reps_1} and Yukawa superpotentials (\ref{up1}, \ref{down1}, \ref{lep1}), their quantum numbers are summarized in Table \ref{reps_2}. The additional quantum number, denoted as $g_\alpha=\pm1$, arises due to the emergence of the field ${\cal S}$ in the superpotential on the $y=0$ brane, as described in superpotentials (\ref{up1}, \ref{down1}, \ref{lep1}).
  \begin{table}[h]
\begin{widetext}
\begin{center}
\caption{\label{reps_2} Representations of the quark and lepton fields under $G_{\rm SM}\times A_4\times U(1)_X$ with modular weight $k_I$. In $({\cal Q}_1, {\cal Q}_2)_Y$ of $G_{\rm SM}$, ${\cal Q}_1$ and ${\cal Q}_2$ are the representations under $SU(3)_C$ and $SU(2)_L$, and the script $Y$ denotes the $U(1)$ hypercharge. All fields are left-handed particles/antiparticles.}
\begin{ruledtabular}
\begin{tabular}{cccccc}
Field &$G_{\rm SM}$&$A_4$&$k_I$ & $U(1)_X$ & \text{For instance: see Eqs.(\ref{para_sp}) and (\ref{para_sp_nu})} \\
\hline
$Q_1$ & $(3,2)_{1/6}$ & ${\bf 1}$  & $\frac{h}{2}-6$ & $f_b+g_b-f_d-g_d$ & $|f_b|=5$ \\
$Q_2$ & $(3,2)_{1/6}$ & ${\bf 1}''$ & $\frac{h}{2}-6$ & $f_b+g_b-f_s-g_s$ & $|f_s|=11$ \\
$Q_3$ & $(3,2)_{1/6}$ & ${\bf 1}'$ & $\frac{h}{2}-6$ & $0$ & $|f_d|=14$ \\
$D^c$ & $(3,1)_{1/3}$ & ${\bf 3}$  & $\frac{h}{2}-6$ & $-f_b-g_b$ &  \\
$u^c$ & $(3,1)_{-2/3}$ & ${\bf 1}$  & $\frac{h}{2}-6$ & $f_d+g_d-f_b-g_b-f_u-g_u$ & $|f_u|=22$ \\
$c^c$ & $(3,1)_{-2/3}$ & ${\bf 1}'$  & $\frac{h}{2}-6$ & $f_s+g_s-f_b-g_b-f_c-g_c$ & $|f_{c}|=9$ \\
$t^c$ & $(3,1)_{-2/3}$ & ${\bf 1}''$  & $\frac{h}{2}-6$ & $-g_t$ & $|g_{d,s,b,u,c,t}|=1$ \\
\hline
$L$ & $(1,2)_{-1/2}$ & ${\bf 3}$  & $\frac{h}{2}-2$ & $-\frac{1}{2}$ & $|g_{e,\mu,\tau}|=1$ \\
$e^c$ & $(1,1)_{1}$ & ${\bf 1}$  & $\frac{h}{2}-6$ & $\frac{1}{2}-f_e-g_e$ & $|f_{e}|=21$ \\
$\mu^c$ & $(1,1)_{1}$ & ${\bf 1}''$  & $\frac{h}{2}-6$ & $\frac{1}{2}-f_\mu-g_\mu$ & $|f_{\mu}|=12$ \\
$\tau^c$ & $(1,1)_{1}$ & ${\bf 1}'$  & $\frac{h}{2}-6$ & $\frac{1}{2}-f_\tau-g_\tau$ & $|f_{\tau}|=7$ \\
\end{tabular}
\end{ruledtabular}
\end{center}
\end{widetext}
\end{table} 
The $U(1)_X$ symmetry, generated by the charge $X$ normalized to one, is anomalous.
Then the color anomaly coefficient of $U(1)_{X}\times[SU(3)_C]^2$ defined as $N_C\delta^{ab}=2\sum_{\psi}X_{\psi}{\rm Tr}(T^aT^b)$ in the QCD instanton backgrounds, where the $T^a$ are the generators of the representation of $SU(3)_C$ to which Dirac fermion  belongs with $X$-charge, reads
\begin{eqnarray}
N_C=-f_b-f_s-f_d-f_c-f_u-g_b-g_s-g_d-g_t-g_c-g_u\,,
\label{nc1}
\end{eqnarray}
where the quantum numbers $f_{d,s,b,u,c}$ are correlated with quark mass and mixing, while $g_{d,s,b,u,c,t}$ are extra quantum numbers that are not related to quark mass and mixing, see Eqs.(\ref{4d_a1}, \ref{Ch2}, \ref{Ch1}).
Note that $U(n)$ generators ($n\geq2$) are normalized according to ${\rm Tr}[T^aT^b]=\delta^{ab}/2$. The $U(1)_X$ is broken down to its discrete subgroup $Z_{N_{\rm DW}}$ in the backgrounds of the QCD instanton, and the quantity $N_C$ (non-zero integer) is given by the axionic domain-wall number $N_{\rm DW}=|N_C|$. At the QCD phase transition, each axionic string becomes the edge to $N_{\rm DW}$ domain walls, and the process of axion radiation stops. To avoid the domain-wall problem, it is necessary to ensure either that $N_{\rm DW}=1$ or that the PQ phase transition occurred during (or before) inflation if $N_{\rm DW}>1$\,\cite{PDG}. Interestingly, the condition $N_{\rm DW}=1$ can be inherently satisfied for $|f_b+f_s+f_d+f_c+f_u|=1,3,5,7$ due to the presence of an additional scalar field ${\cal S}$ in the $y=0$ brane operators (see Eqs.(\ref{nc1}, \ref{up1}, \ref{down1})), whose quantum number $g_\alpha=\pm1$. This can circumvent the domain-wall problem.

Furthermore, nonperturbative quantum gravitational anomaly effects \cite{Kamionkowski:1992mf} violate the conservation of the corresponding current, $\partial_\mu J^{\mu}_X\propto R\tilde{R}$, where $R$ is the Riemann tensor and $\tilde{R}$ is its dual, thereby making the axion solution to the strong CP problem problematic.
To eliminate the breaking effects of the axionic shift symmetry by gravity and to consistently couple gravity to matter, the mixed gravitational anomaly $U(1)_X\times[{\rm gravity}]^2$ (related to the color anomaly $U(1)_X\times[SU(3)_C]^2$) must be canceled\,\cite{Ahn:2016hbn, Ahn:2018cau, Ahn:2021ndu}. An important aspect of achieving this anomaly-free condition is that the neutral mirror bulk fermion, $\Psi^c_\nu$, should couple to the normal neutrino field on the brane, which is consistent with the boundary condition of Eq.(\ref{bc_1}). This leads to the relations $U(1)_X\times[{\rm gravity}]^2|_{\rm Quark}=3N_C$ and $U(1)_X\times[{\rm gravity}]^2|_{\rm Lepton}=6X_L+X_{e^c}+X_{\mu^c}+X_{\tau^c}+3X_{\Psi^c_\nu}$, that is,
\begin{eqnarray}
3N_C=f_e+g_e+f_\mu+g_\mu+f_\tau+g_\tau\,,
\label{gr_ano}
\end{eqnarray}
where $N_C$ is equivalent to the value given in Eq.(\ref{nc1}).
This, combined with the condition $N_{\rm DW}=1$, imposes a constraint on the $U(1)_X$ quantum numbers of the quark and lepton fields.

The brane-localized Yukawa superpotentials for up- and down-type quark fields and lepton fields, invariant under $G_{\rm SM}\times A_4\times U(1)_X$ with modular forms, are given with ${\cal \hat{S}}=\{\hat{\chi}$ or $\hat{\tilde{\chi}}\}$ by
\begin{eqnarray}
W^u_q&=&\delta(y-L)\big\{\hat{Y}^{(6)}_{\bf 1}\hat{\Psi}^c_{u_3}\hat{Q}_3\hat{H}_u+\hat{Y}^{(6)}_{\bf 1}\hat{\Psi}^c_{u_2}\hat{Q}_2\hat{H}_u+\hat{Y}^{(6)}_{\bf 1}\hat{\Psi}^c_{u_1}\hat{Q}_1\hat{H}_u\big\}\nonumber\\
&+&\delta(y)\Big\{\alpha_t\hat{Y}^{(6)}_{\bf 1}\hat{t}^c\hat{\Psi}_{u_3}\hat{\cal S}+\alpha_c\Big(\frac{\hat{\cal S}}{\Lambda_5}\Big)^{|f_c|}\hat{Y}^{(6)}_{\bf 1}\hat{c}^c\hat{\Psi}_{u_2}\hat{\cal S}+\alpha_u\Big(\frac{\hat{\cal S}}{\Lambda_5}\Big)^{|f_u|}\hat{Y}^{(6)}_{\bf 1}\hat{u}^c\hat{\Psi}_{u_1}\hat{\cal S}\Big\}\,,
\label{up1}
\end{eqnarray}
\begin{eqnarray}
W^d_q&=&\delta(y-L)\big\{\hat{Y}^{(6)}_{\bf 1}\hat{\Psi}^c_{d_3}\hat{Q}_3\hat{H}_d+\hat{Y}^{(6)}_{\bf 1}\hat{\Psi}^c_{d_2}\hat{Q}_2\hat{H}_d+\hat{Y}^{(6)}_{\bf 1}\hat{\Psi}^c_{d_1}\hat{Q}_1\hat{H}_d\big\}\nonumber\\
&+&\delta(y)\Big\{\alpha_b\Big(\frac{\hat{\cal S}}{\Lambda_5}\Big)^{|f_b|}(\hat{Y}^{(6)}_{\bf 3}\hat{D}^c)_{{\bf 1}''}\hat{\Psi}_{d_3}\hat{\cal S}+\alpha_s\Big(\frac{\hat{\cal S}}{\Lambda_5}\Big)^{|f_s|}(\hat{Y}^{(6)}_{\bf 3}\hat{D}^c)_{{\bf 1}'}\hat{\Psi}_{d_2}\hat{\cal S}\nonumber\\
&&\qquad+\alpha_d\Big(\frac{\hat{\cal S}}{\Lambda_5}\Big)^{|f_d|}(\hat{Y}^{(6)}_{\bf 3}\hat{D}^c)_{{\bf 1}}\hat{\Psi}_{d_1}\hat{\cal S}\Big\}\,,
\label{down1}
\end{eqnarray}
\begin{eqnarray}
W_{\ell\nu}&=&\delta(y-L)\big\{(\hat{Y}^{(2)}_{\bf 3}\hat{L})_{{\bf 1}''}\hat{\Psi}^c_{\tau}\hat{H}_d+(\hat{Y}^{(2)}_{\bf 3}\hat{L})_{{\bf 1}'}\hat{\Psi}^c_{\mu}\hat{H}_d+(\hat{Y}^{(2)}_{\bf 3}\hat{L})_{{\bf 1}}\hat{\Psi}^c_{e}\hat{H}_d+(\hat{Y}^{(2)}_{\bf 3}\hat{\Psi}^c_\nu\hat{L})_{\bf 1}\hat{H}_u\big\}\nonumber\\
&+&\delta(y)\Big\{\alpha_\tau\Big(\frac{\hat{\cal S}}{\Lambda_5}\Big)^{|f_\tau|}\hat{Y}^{(6)}_{\bf 1}\hat{\tau}^c\hat{\Psi}_{\tau}\hat{\cal S}+\alpha_\mu\Big(\frac{\hat{\cal S}}{\Lambda_5}\Big)^{|f_\mu|}\hat{Y}^{(6)}_{\bf 1}\hat{\mu}^c\hat{\Psi}_{\mu}\hat{\cal S}+\alpha_e\Big(\frac{\hat{\cal S}}{\Lambda_5}\Big)^{|f_e|}\hat{Y}^{(6)}_{\bf 1}\hat{e}^c\hat{\Psi}_{e}\hat{\cal S}\nonumber\\
&&\qquad+\frac{1}{2}\hat{y}_\nu\hat{\Psi}_\nu\hat{\Psi}_\nu\hat{\chi}\Big\}\,,
\label{lep1}
\end{eqnarray}
where\,\footnote{Here, the hat fields represent all superfields, where ordinary superfields have mass dimension $1$, while bulk superfields have mass dimension $3/2$.} hat modular form $\hat{Y}$ has a mass dimension $-1/2$, while hat Yukawa coupling $\hat{y}_\nu$ has a mass dimension $-1$. Clearly, it shows that SM fermions localized at the two branes could form ordinary interactions between left- and right-handed fermions via the exchange of their flavored bulk fermions. Any additive finite correction terms that could potentially be generated by higher weight modular forms are prohibited because the modular weight of the $\chi$ ($\tilde{\chi}$) fields located on the $y=0$ brane is zero. However, higher-order corrections arising from the combination $\chi\tilde{\chi}$ are allowed, but they do not modify the leading-order flavor structure. On the $y=0$ brane, the Yukawa coefficients $\tilde{\alpha}_i$ are all complex numbers with unit absolute values ($|\alpha_i|=1$). Note that there exists an infinite series of higher-dimensional operators induced by the combination of $\chi\tilde{\chi}$ in the supersymmetric limit. These operators are constructed by multiplying the leading-order operators by $\sum^{\infty}_{n=1}\Big(\hat{\chi}\tilde{\hat{\chi}}/\Lambda^{2}_5\Big)^n$. These higher-dimensional operators can be absorbed into the finite leading-order terms, effectively modifying the coefficients at the leading order, as will be shown later. Since it is challenging to reproduce the experimental data of fermion masses and mixing with Yukawa terms constructed from modular forms of weight 4 in the quark and charged-lepton sectors of this model, we consider Yukawa terms with modular forms of weight 6, which decompose as ${\bf1}\oplus{\bf 3}\oplus{\bf 3}$ under $A_4$ given explicitly by \cite{Feruglio:2017spp}
{\begin{eqnarray}
  &&Y^{(6)}_{\bf 1}=Y^3_1+Y^3_2+Y^3_3-3Y_1Y_2Y_3\,\nonumber\\
  &&Y^{(6)}_{{\bf 3},1}=(Y^3_1+2Y_1Y_2Y_3, Y^2_1Y_2+2Y^2_2Y_3, Y^2_1Y_3+2Y^2_3Y_2)\nonumber\\
  &&Y^{(6)}_{{\bf 3},2}=(Y^3_3+2Y_1Y_2Y_3, Y^2_3Y_1+2Y^2_1Y_2, Y^2_3Y_2+2Y^2_2Y_1)\,.
\label{modu6}
 \end{eqnarray}}
 
Then the action for quark and lepton fields localized on the branes reads
\begin{eqnarray}
S\supset\int d^4x dy\sqrt{g}\Big[\int d^2\vartheta\Big( W^u_q+W^d_q+W_{\ell\nu}\Big)+{\rm h.c.}\Big]
\label{Yu_su}
\end{eqnarray}
where $\vartheta$ is a Grassmann variable having mass dimension $-1/2$.

%%%%%%%%%%%%%%%%%%%%%%%%%%%%%%%%%%%%%%%%%%%%%%%%%%%%%%%%%%%%%%%%%%%%%%%%%%%%%%%%%
\section{low energy effective action}
\label{sec3}
After the scalar field ${\cal S}=\chi(\tilde{\chi})$ and modulus $\tau$ acquire VEVs, spontaneously breaking the flavored $U(1)_X$, the flavor group $G_F$ is also broken and becomes hidden. This process elucidates the flavor structure of mixing patterns and mass hierarchies of quarks and leptons, naturally providing a solution to the strong CP problem, and the generation of light neutrino masses by transmitting the information of $G_F$ breakdown between the two 3-branes.
%%%%%%%%%%%%%%%%%%%%%%%%%%%%%%%%%%%%%%%%%%%
\subsection{4D Kinetic terms and Scalar potential}
The kinetic terms and scalar potentials localized on the branes read
\begin{eqnarray}
&&S\supset\int d^4xdy\sqrt{g}\Big[\big\{-g^{\mu\nu}\big(\partial_\mu H^\dag_u\partial_\nu H_u+\partial_\mu H^\dag_d\partial_\nu H_d\big)-ie^\mu_{~\alpha}\bar{\psi}_L\gamma^\alpha\partial_\mu\psi_L+V_L(H_{u(d)})\big\}\delta(y-L)\nonumber\\
&&\qquad\qquad\qquad+\big\{
-g^{\mu\nu}\big(\partial_\mu\chi^\dag\partial_\nu\chi+\partial_\mu\tilde{\chi}^\dag\partial_\nu\tilde{\chi}\big)-ie^\mu_{~\alpha}\bar{\psi}_R\gamma^\alpha\partial_\mu\psi_R+V_0(\eta,\chi,\tilde{\chi})\big\}\delta(y)\Big],
\label{brL_a}
\end{eqnarray} 
where $\psi$ stands for all associated SM fermion fields and $g^{\mu\nu}=e^{2\sigma(y)}\eta^{\mu\nu}$ is the induced metric on the brane.
 In the above action we have omitted gauge interactions.  The brane-localized potentials in Eq.(\ref{brL_a}), denoted as $V_{L(0)}$, are given by $V_{L(0)}=\sum_i\Big|\frac{\partial W_{L(0)}}{\partial\varphi_i}\Big|^2+V^{\rm soft}_{L(0)}+V^{\rm D}_{L(0)} $, where $\varphi_i$ represents all the scalar fields on each brane. The terms $V^{\rm soft}_{L(0)}$ and $V^{\rm D}_{L(0)}$ correspond to soft-supersymmetric breaking terms and D-terms, respectively. 
By performing the rescaling of dimensionful parameters in the action (\ref{Yu_su}, \ref{brL_a}),
\begin{eqnarray}
&&{\cal S}'\rightarrow e^{-\sigma(y)}{\cal S}'\,,\qquad {\Psi}\rightarrow e^{-2\sigma(y)}{\Psi}\,,\qquad\psi_{L(R)}\rightarrow e^{-\frac{3}{2}\sigma(y)}\psi_{L(R)}\,,\nonumber\\
&&{\Lambda_5}\rightarrow e^{-\sigma(y)}{\Lambda_5}\,,\qquad {\mu_{i}}\rightarrow e^{-\sigma(y)}{\mu_i}\,,\qquad \hat{Y}\rightarrow e^{\frac{1}{2}\sigma(y)}\hat{Y}
\label{resc1}
\end{eqnarray} 
where ${\cal S}'=\chi(\tilde{\chi})$ and $H_{u(d)}$, and $\mu_i=\mu_{\chi(H)}$ are the scalar mass parameters, a canonical normalization of the fields on the branes is restored in Eq.(\ref{brL_a}). In a slice of AdS$_{5}$, the scalar mass $\mu_\chi$ represents a value near the 5D cutoff scale, $\langle {\cal S}\rangle/\Lambda_5\lesssim1$, as expected for a scalar field. The scalars $\chi$ and $\tilde{\chi}$, confined to the $y=0$ brane, are responsible for the spontaneous breaking of flavor symmetry $U(1)_X$ at a very high energy scale, much larger than EW scale. Additionally, we consider the Higgs potential  $V_L(H_{u(d)})$, where the Higgs mass parameter $\mu_H$ is generated radiatively via one-loop contributions at the EW scale\,\cite{Coleman:1973jx}.
Under the rescaling in Eq.(\ref{resc1}), the action for the two EW Higgs fields located on the $y=0$ brane is
\begin{eqnarray}
S_H&\supset&\int d^4x dy\sqrt{g}\Big\{g^{\mu\nu}\big(\partial_\mu H^\dag_u\partial_\nu H_u+\partial_\mu H^\dag_d\partial_\nu H_d\big)-\mu^2_H(|H_u|^2+|H_d|^2)+...\Big\}\delta(y)\nonumber\\
&=&\int d^4x\,\eta^{\mu\nu}\big(\partial_\mu H^\dag_u\partial_\nu H_u+\partial_\mu H^\dag_d\partial_\nu H_d\big)-\mu^2_H(|H_u|^2+|H_d|^2)+...
\end{eqnarray}
where the ellipsis represents the EW Higgs quartic coupling term arising from D-term contributions.
Clearly, if the Higgs mass parameter $\mu_H$ is radiatively generated at the EW scale, it remains unchanged by the rescaling in Eq.(\ref{resc1}). Consequently, this framework can provide a natural solution to the hierarchy problem.

 In the global SUSY limit, {\it i.e.} $M_P\rightarrow\infty$, the scalar potential derived from the $F$- and $D$-terms of all fields must vanish. The relevant $F$-term potential from Eq.(\ref{d_pot}) and the $D$-term potential for the anomalous $U(1)_X$ are given by
\begin{eqnarray}
 V_F^{\rm global}&=&\big|c_\chi\chi\tilde{\chi}-\mu^2_\chi\big|^2\,,\qquad  V_D^{\rm global}=\frac{|X|^2g^2_X}{2}\Big(-\frac{\xi^{\rm FI}_X}{|X|}+|\chi|^2-|\tilde{\chi}|^2\Big)^2\,\,.
  \label{super_1}
\end{eqnarray}
The scalar fields $\chi$ and $\tilde{\chi}$ have $X$-charges $+1$ and $-1$, respectively:
\begin{eqnarray}
 \chi\rightarrow e^{+i\xi}\chi\,,\qquad \tilde{\chi}\rightarrow e^{-i\xi}\tilde{\chi}\,,
  \label{super_2}
\end{eqnarray}
with a constant $\xi$. Thus, the potential $V_{\rm SUSY}$ exhibits $U(1)_X$ symmetry. Since SUSY is preserved after the spontaneous breaking of $U(1)_X$, the scalar potential in the limit $M_P\rightarrow\infty$ vanishes at its ground states, {\it i.e.}, $\langle V_F^{\rm global}\rangle=0$ and $\langle V_D^{\rm global}\rangle=0$, implying that a vanishing $F$-term also necessitates a vanishing $D$-term. From the minimization of the $F$-term scalar potential, we obtain
 \begin{eqnarray}
 \langle\chi\rangle=\langle\tilde{\chi}\rangle=\frac{v_{\chi}}{\sqrt{2}}\qquad\text{with}~\mu_\chi=v_\chi\sqrt{\frac{c_{\chi}}{2}}\,.
 \label{vev}
 \end{eqnarray}
This supersymmetric solution is consistent with the $D$-flatness condition for $\xi^{\rm FI}_X=0$ and $\langle\chi\rangle=\langle\tilde{\chi}\rangle$\,\cite{Ahn:2016hbn, Ahn:2017dpf}.
The tension between $\langle\chi\rangle=\langle\tilde{\chi}\rangle$ and $\xi^{\rm FI}_X\neq0$ arises because the FI term cannot be cancelled, unless the VEV of flux in the FI term is below the string scale\,\cite{Achucarro:2006zf, Burgess:2003ic}. The FI term acts as an uplifting potential, raising the Anti-de Sitter minimum to the de Sitter minimum \cite{Burgess:2003ic, Ahn:2023iqa}. To achieve this, the $F$-term must necessarily break SUSY for the $D$-term to act as an uplifting potential.
The PQ scale $\mu_\chi$ can be determined by taking into account both the SUSY-breaking effect, which lifts up the flat direction, and supersymmetric next-leading-order Planck-suppressed terms\,\cite{Nanopoulos:1983sp, Ahn:2016hbn,Ahn:2017dpf}.
The supersymmetric next-to-leading order localized interaction invariant under $A_4\times U(1)_X\times k_I$ is given by
 \begin{eqnarray}
\Delta W_v\simeq\delta(y)\frac{\alpha}{M^2_P}\hat{\chi}_0(\hat{\chi}\hat{\tilde{\chi}})^2\,,
 \label{deW}
 \end{eqnarray}
where $\alpha$ is assumed to be a real-valued constant of order unity.
Given the presence of soft SUSY-breaking terms at the scale relevant to flavor dynamics, the leading order scalar potential for $\chi$ and $\tilde{\chi}$ is
 \begin{eqnarray}
V(\chi,\tilde{\chi})\simeq-\alpha_1m^2_{3/2}|\chi|^2-\alpha_2m^2_{3/2}|\tilde{\chi}|^2+\alpha^2\frac{|\chi|^4|\tilde{\chi}|^4}{M^4_P}\,,
 \label{sof1}
 \end{eqnarray}
where $m_{3/2}$ is the soft SUSY-breaking mass and $\alpha_1, \alpha_2$ are real constants. This results in the PQ breaking scale (equivalently, cutoff scale in the light neutrino operator, see Eqs.(\ref{4d_a2}) and (\ref{MR5}))
 \begin{eqnarray}
\mu_\chi\simeq \big(\frac{c^6_\chi\alpha_1\alpha_2}{16\alpha^4}\big)^{\frac{1}{12}} \big(m_{3/2}M^2_P\big)^{\frac{1}{3}}\,.
 \label{sof2}
 \end{eqnarray} 
The soft SUSY-breaking mass $m_{3/2}$ can be estimated as $m_{3/2}\gtrsim3\times10^4$ TeV (or $m_{3/2}\gtrsim3\times10^{7}$ TeV) for $\langle\chi\rangle\gtrsim3\times10^{14}$ GeV (or $\langle\chi\rangle\gtrsim3\times10^{15}$ GeV) from Eq.(\ref{sof2}), assuming $\alpha_1$ and $\alpha_2$ are of order unity and $\alpha\simeq1$.
 
The model includes the SM gauge singlet scalar fields $\chi$ and $\tilde{\chi}$ charged under $U(1)_X$, which have interactions invariant under $G_{\rm SM}\times U(1)_X\times A_4$  with the transformations Eq.(\ref{tr1}). These interactions result in a chiral symmetry, which is reflected in the form of the kinetic and Yukawa terms, as well as the scalar potential $V_{\rm SUSY}$ in the SUSY limit:
{\begin{eqnarray}
 {\cal L} &\supset& \partial_\mu\chi^\ast\partial^\mu\chi+ \partial_\mu\tilde{\chi}^\ast\partial^\mu\tilde{\chi}+{\cal L}_Y-V_{\rm SUSY}+{\cal L}_{\vartheta}+\overline{\psi}\,i\! \! \not\!\partial \psi+\frac{1}{2}\overline{\nu}\,i\! \! \not\!\partial\nu \,,
\label{lag0}
 \end{eqnarray}}
where $\psi$ denotes Dirac fermions, and $V_{\rm SUSY}$ is replaced by $V_{\rm total}$ when SUSY breaking effects are considered. The above kinetic terms for $\chi(\tilde{\chi})$ are canonically normalized from $\frac{\partial^2 K}{\partial{\cal S}^\ast\partial{\cal S}}\partial_\mu{\cal S}^\ast\partial^\mu{\cal S}$ with K{\"a}hler potential $K\supset|{\cal S}|^2+$higher order terms. Here four component Majorana spinors ($\nu^c=\nu$) are used.
The global $U(1)_X$ PQ symmetry guarantees the absence of bare mass term in the Yukawa Lagrangian ${\cal L}_Y$ in Eq.(\ref{lag0}).
The QCD Lagrangian has a CP-violating term 
{\begin{eqnarray}
{\cal L}_{\vartheta}&=&\vartheta_{\rm QCD}\,\frac{g^{2}_{s}}{32\pi^{2}}\,G^{a\mu\nu}\tilde{G}^{a}_{\mu\nu}
 \end{eqnarray}}
where $g_s$ stands for the gauge coupling constant of $SU(3)_C$, and $G^{a\mu\nu}$ is the color field strength tensor and its dual $\tilde{G}^{a}_{\mu\nu}=\frac{1}{2}\varepsilon_{\mu\nu\rho\sigma}G^{a\mu\nu}$ (here $a$ is an $SU(3)$-adjoint index), coming from the strong interaction. 
Upon acquiring the vacuum expectation value (VEV) $\langle\chi\rangle \neq 0$, the $U(1)_X$ symmetry in this model undergoes spontaneous breaking at a scale $\Lambda_5$, much higher than the EW scale. This breaking is realized by the emergence of a Nambu-Goldstone (NG) mode $A$, which interacts with ordinary quarks and leptons via Yukawa interactions, as described in Eqs.(\ref{fl_in}, \ref{fla_1}, \ref{neut1}). To extract the associated boson resulting from spontaneous breaking of $U(1)_X$, we set the decomposition of complex scalar fields\,\cite{Ahn:2014gva, Ahn:2016hbn, Ahn:2018cau} as follows
 \begin{eqnarray}
\chi=\frac{v_{\chi}}{\sqrt{2}}e^{i\frac{A}{u_{\chi}}}\left(1+\frac{h_{\chi}}{u_{\chi}}\right)\,,\quad\,\tilde{\chi}=\frac{v_{\tilde{\chi}}}{\sqrt{2}}e^{-i\frac{A}{u_{\chi}}}\left(1+\frac{h_{\tilde{\chi}}}{u_{\chi}}\right)\qquad\text{with}~u_{\chi}=\sqrt{v^2_{\chi}+v^2_{\tilde{\chi}}}\,,
  \label{NGboson}
 \end{eqnarray}
in which $A$ is the NG mode and we set $v_\chi=v_{\tilde{\chi}}$ and $h_{\chi}=h_{\tilde{\chi}}$ in the supersymmetric limit. The derivative coupling of NG boson $A$ arises from the kinetic term 
{\begin{eqnarray}
\partial_\mu\chi^\ast\partial^\mu\chi+\partial_\mu\tilde{\chi}^\ast\partial^\mu\tilde{\chi}=\frac{1}{2}(\partial_\mu A)^2\Big(1+\frac{h_\chi}{u_\chi}\Big)^2+\frac{1}{2}(\partial_\mu h_\chi)^2\,.
 \end{eqnarray}}
Performing $u_\chi\rightarrow\infty$, the NG mode $A$, whose interaction is determined by symmetry, is distinguished from the radial mode $h_\chi$, which is invariant under the symmetry $U(1)_X$.

%%%%%%%%%%%%%%%%%%%%%%%%%%%%%%%%%%%%%%%%%%%%%%%%%%%%%%%%%%%%%%%%%%%%%%%%%%
\subsection{4D Yukawa Lagrangian}
\label{4DYL}

By varying the action given in Eq.(\ref{Yu_su}) with respect to the flavored bulk fermions and imposing $(\delta S)_{\rm boundary}=0$ at the boundaries, we derive boundary conditions analogous to those in Ref.\cite{Ahn:2021ndu}. These boundary conditions are adjusted according to Eqs.(\ref{bc_1}, \ref{bc_2}) within the framework of the action described by Eq.(\ref{Yu_su}), resulting in six conditions specifically for $u$-type quarks
\begin{eqnarray}
 &&\Psi_{u3}(x,L)=2\hat{Y}^{(6)}_{\bf 1}Q_3H_u\,,\qquad\qquad\Psi^c_{u3}(x,0)=2\alpha_t\hat{Y}^{(6)}_{\bf 1}t^c{\cal S}\,,\nonumber\\
 &&\Psi_{u2}(x,L)=2\hat{Y}^{(6)}_{\bf 1}Q_2H_u\,,\qquad\qquad\Psi^c_{u2}(x,0)=2\alpha_c\Big(\frac{\cal S}{\Lambda_5}\Big)^{|f_c|}\hat{Y}^{(6)}_{\bf 1}c^c{\cal S}\,,\nonumber\\
 &&\Psi_{u1}(x,L)=2\hat{Y}^{(6)}_{\bf 1}Q_1H_u\,,\qquad\qquad \Psi^c_{u1}(x,0)=2\alpha_u\Big(\frac{\cal S}{\Lambda_5}\Big)^{|f_u|}\hat{Y}^{(6)}_{\bf 1}u^c{\cal S};
 \label{bc_01}
 \end{eqnarray} 
 six for $d$-type quarks
 \begin{eqnarray}
 &&\Psi_{d3}(x,L)=2\hat{Y}^{(6)}_{\bf 1}Q_3H_d\,,\qquad\qquad\Psi^c_{d3}(x,0)=2\alpha_b\Big(\frac{\cal S}{\Lambda_5}\Big)^{|f_b|}(\hat{Y}^{(6)}_{\bf 3}D^c)_{{\bf 1}''}{\cal S}\,,\nonumber\\
 &&\Psi_{d2}(x,L)=2\hat{Y}^{(6)}_{\bf 1}Q_2H_d\,,\qquad\qquad\Psi^c_{d2}(x,0)=2\alpha_s\Big(\frac{\cal S}{\Lambda_5}\Big)^{|f_s|}(\hat{Y}^{(6)}_{\bf 3}D^c)_{{\bf 1}''}{\cal S}\,,\nonumber\\
 &&\Psi_{d1}(x,L)=2\hat{Y}^{(6)}_{\bf 1}Q_1H_d\,,\qquad\qquad\Psi^c_{d1}(x,0)=2\alpha_d\Big(\frac{\cal S}{\Lambda_5}\Big)^{|f_d|}(\hat{Y}^{(6)}_{\bf 3}D^c)_{{\bf 1}''}{\cal S}\,;
 \label{bc_02}
  \end{eqnarray} 
 six for charged-leptons
 \begin{eqnarray}
 &&\Psi_{\tau}(x,L)=2(\hat{Y}^{(2)}_{\bf 3}L)_{{\bf 1}''}H_d\,,\qquad\qquad\Psi^c_{\tau}(x,0)=2\alpha_b\Big(\frac{\cal S}{\Lambda_5}\Big)^{|f_\tau|}\hat{Y}^{(6)}_{\bf 1}\tau^c{\cal S}\,,\nonumber\\
 &&\Psi_{\mu}(x,L)=2(\hat{Y}^{(2)}_{\bf 3}L)_{{\bf 1}'}H_d\,,\qquad\qquad\,\Psi^c_{\mu}(x,0)=2\alpha_s\Big(\frac{\cal S}{\Lambda_5}\Big)^{|f_\mu|}\hat{Y}^{(6)}_{\bf 1}\mu^c{\cal S}\,,\nonumber\\
 &&\Psi_{e}(x,L)=2(\hat{Y}^{(2)}_{\bf 3}L)_{{\bf 1}}H_d\,,\qquad\qquad\,\,\,\Psi^c_{e}(x,0)=2\alpha_d\Big(\frac{\cal S}{\Lambda_5}\Big)^{|f_e|}\hat{Y}^{(6)}_{\bf 1}e^c{\cal S}\,,
\label{bc_03}
\end{eqnarray} 
and one for neutrino
 \begin{eqnarray}
 &&\Psi_{\nu}(x,L)=2\hat{Y}^{(2)}_{\bf 3}LH_u\,.
\label{bc_04}
\end{eqnarray} 
Then the 4D Yukawa interactions can be written as
 \begin{eqnarray}
&-S^{q\ell}_Y=\int d^4x\Big\{\hat{Y}^{(6)}_{\bf 1}\Psi^c_{u3}(x,L)Q_3H_u+\hat{Y}^{(6)}_{\bf 1}\Psi^c_{u2}(x,L)Q_2H_u+\hat{Y}^{(6)}_{\bf 1}\Psi^c_{u1}(x,L)Q_1H_u\nonumber\\
&+\alpha_t\hat{Y}^{(6)}_{\bf 1}t^c\Psi_{u3}(x,0){\cal S}+\alpha_c\hat{Y}^{(6)}_{\bf 1}\Big(\frac{\cal S}{\Lambda_5}\Big)^{|f_c|}c^c\Psi_{u2}(x,0){\cal S}+\alpha_u\hat{Y}^{(6)}_{\bf 1}\Big(\frac{\cal S}{\Lambda_5}\Big)^{|f_u|}u^c\Psi_{u1}(x,0){\cal S}\nonumber\\
&+\hat{Y}^{(6)}_{\bf 1}\Psi^c_{d3}(x,L)Q_3H_d+\hat{Y}^{(6)}_{\bf 1}\Psi^c_{d2}(x,L)Q_2H_d+\hat{Y}^{(6)}_{\bf 1}\Psi^c_{d1}(x,L)Q_1H_d\nonumber\\
&+\alpha_b\Big(\frac{\cal S}{\Lambda_5}\Big)^{|f_b|}(\hat{Y}^{(6)}_{\bf 3}D^c)_{{\bf 1}''}\Psi_{d3}(x,0){\cal S}+\alpha_s\Big(\frac{\cal S}{\Lambda_5}\Big)^{|f_s|}(\hat{Y}^{(6)}_{\bf 3}D^c)_{{\bf 1}'}\Psi_{d2}(x,0){\cal S}\nonumber\\
&+\alpha_d\Big(\frac{\cal S}{\Lambda_5}\Big)^{|f_d|}(\hat{Y}^{(6)}_{\bf 3}D^c)_{{\bf 1}}\Psi_{d1}(x,0){\cal S}\nonumber\\
&+(\hat{Y}^{(2)}_{\bf 3}L)_{{\bf 1}''}\Psi^c_\tau(x,L)H_d+(\hat{Y}^{(2)}_{\bf 3}L)_{{\bf 1}'}\Psi^c_\mu(x,L)H_d+(\hat{Y}^{(2)}_{\bf 3}L)_{{\bf 1}}\Psi^c_e(x,L)H_d\nonumber\\
&+\alpha_\tau\hat{Y}^{(6)}_{\bf 1}\Big(\frac{\cal S}{\Lambda_5}\Big)^{|f_\tau|}\tau^c\Psi_{\tau}(x,0){\cal S}+\alpha_\mu\hat{Y}^{(6)}_{\bf 1}\Big(\frac{\cal S}{\Lambda_5}\Big)^{|f_\mu|}\mu^c\Psi_{\mu}(x,0){\cal S}+\alpha_e\hat{Y}^{(6)}_{\bf 1}\Big(\frac{\cal S}{\Lambda_5}\Big)^{|f_e|}e^c\Psi_{e}(x,0){\cal S}\nonumber\\
&+(\hat{Y}^{(2)}_{\bf 3}\Psi^c_\nu(x,L)L)_{\bf 1}H_u+\frac{1}{2}\hat{y}_\nu\Psi_{\nu}(x,0)\Psi_\nu(x,0)\chi
+{\rm h.c.}\Big\}\,.
\label{4d_0}
\end{eqnarray}

Since we are interested in the energy scale much lower than $1/L$, using the zero modes\,\footnote{For non-zero modes, we impose Dirichlet boundary conditions as in Eq.(\ref{dich}).} in Eq.(\ref{bc_3}) the bulk fermion fields  in Eq.(\ref{KKw1}) at the branes read
 \begin{eqnarray}
 \Psi_{fi}(x,0)=\Psi_{fi}(x,L)\,e^{-2\sigma(L)-M_fL}\,,\qquad \Psi^c_{fi}(x,L)=\Psi^c_{fi}(x,0)\,e^{2\sigma(L)-M_fL}\,.
\label{bfb}
\end{eqnarray} 
Using the above Eq.(\ref{bfb}), substituting Eqs.(\ref{bc_01}, \ref{bc_02}, \ref{bc_03}, \ref{bc_04}) into the above action of Eq.(\ref{4d_0}), and performing the rescaling of dimensionful Yukawa couplings as given in Eq.(\ref{resc1}), we obtain the quark and charged-lepton 4D Yukawa interactions:
\begin{eqnarray}
&&-S^{q\ell}_Y=4\int d^4x\Big\{\alpha_t\hat{Y}^{(6)}_{\bf 1}\hat{Y}^{(6)}_{\bf 1}t^cQ_3H_u+\alpha_c\hat{Y}^{(6)}_{\bf 1}\hat{Y}^{(6)}_{\bf 1}\Big(\frac{\cal S}{\Lambda_5}\Big)^{|f_c|}c^cQ_2H_u\nonumber\\
&&\qquad\qquad\quad+\alpha_u\hat{Y}^{(6)}_{\bf 1}\hat{Y}^{(6)}_{\bf 1}\Big(\frac{\cal S}{\Lambda_5}\Big)^{|f_u|}u^cQ_1H_u
+\alpha_b\Big(\frac{\cal S}{\Lambda_5}\Big)^{|f_b|}\hat{Y}^{(6)}_{\bf 1}(\hat{Y}^{(6)}_{\bf 3}D^c)_{{\bf 1}''}Q_3H_d\nonumber\\
&&\qquad\qquad\quad+\alpha_s\Big(\frac{\cal S}{\Lambda_5}\Big)^{|f_s|}\hat{Y}^{(6)}_{\bf 1}(\hat{Y}^{(6)}_{\bf 3}D^c)_{{\bf 1}'}Q_2H_d+\alpha_d\Big(\frac{\cal S}{\Lambda_5}\Big)^{|f_d|}\hat{Y}^{(6)}_{\bf 1}(\hat{Y}^{(6)}_{\bf 3}D^c)_{{\bf 1}}Q_1H_d\nonumber\\
&&\qquad\qquad\quad+\alpha_\tau\Big(\frac{\cal S}{\Lambda_5}\Big)^{|f_\tau|}\hat{Y}^{(6)}_{\bf 1}(\hat{Y}^{(2)}_{\bf 3}L)_{{\bf 1}''}\tau^cH_d+\alpha_\mu\Big(\frac{\cal S}{\Lambda_5}\Big)^{|f_\mu|}\hat{Y}^{(6)}_{\bf 1}(\hat{Y}^{(2)}_{\bf 3}L)_{{\bf 1}'}\mu^cH_d\nonumber\\
&&\qquad\qquad\quad+\alpha_e\Big(\frac{\cal S}{\Lambda_5}\Big)^{|f_e|}\hat{Y}^{(6)}_{\bf 1}(\hat{Y}^{(2)}_{\bf 3}L)_{{\bf 1}}e^cH_d\Big\}{\cal S}e^{-M_fL}\cosh2\sigma(L)+{\rm h.c.}
\label{4d_a}
\end{eqnarray}
Here, we have used Eq.(\ref{f_M1}) and neglected higher-order operators induced by $\chi\tilde{\chi}$, which can be absorbed by leading-order terms. And the neutrino 4D Yukawa interactions are given by
\begin{eqnarray}
&&-S^{\nu}_Y=\int d^4x\Big\{(\hat{Y}^{(2)}_{\bf 3}\Psi^c_\nu(x,0)L)_{\bf 1}H_u\,e^{2\sigma(L)-M_fL}\nonumber\\
&&\qquad\qquad\quad+\frac{1}{2}\hat{y}_\nu4(\hat{Y}^{(2)}_{\bf 3}LH_u\hat{Y}^{(2)}_{\bf 3}LH_u)\chi\,e^{-4\sigma(L)-2M_fL}\Big\}+{\rm h.c.},
\label{4d_b}
\end{eqnarray} 
where the first operator is a dimension $9/2$ operator and the second operator is a dimension $6$ operator. When ${\cal S}$ acquires VEVs, the $U(1)_X$ symmetry spontaneously broken. After this symmetry breaking, and through the normalization of the Yukawa couplings (see Eq.(\ref{4d_a2})), their operators effectively become dimension $4$ and dimension $5$ operators, respectively. Recalling that in a slice of AdS$_{5}$, the scalars ${\cal S}=\chi\,(\tilde{\chi})$ are confined to the $y=0$ brane, the scalar mass $\mu_\chi$ of Eq.(\ref{vev}) is close to the 5D cutoff scale $\Lambda_5$.

To achieve canonically normalized modular forms (or Yukawa couplings) in the actions Eqs.(\ref{4d_a},\ref{4d_b}), we proceed as follows:
\begin{eqnarray}
\hat{Y}\rightarrow Y\sqrt{\frac{\gamma}{\Lambda_5}}\,,\qquad\qquad \hat{y}\rightarrow y\frac{\gamma_\nu}{\Lambda_5}
\label{nor_1}
\end{eqnarray}
where $\gamma$ and $\gamma_\nu$ are constants\,\footnote{Cf. Ref.\cite{Brignole:1997wnc} computed Yukawa couplings involving chiral matter fields in toroidal compactifications of higher-dimensional super-Yang-Mills theory with magnetic fluxes.}. By setting
\begin{eqnarray}
\gamma\cdot4\frac{\langle{\cal S}\rangle}{\Lambda_5}e^{-M_fL}\cosh2\sigma(L)=1\,,
\label{nor_2}
\end{eqnarray}
we obtain the quark and charged-lepton Yukawa Lagrangian with normalized modular forms:
\begin{eqnarray}
&-{\cal L}^{q\ell}_Y=e^{ig_t\frac{A}{u_\chi}}\alpha_tY^{(6)}_{\bf 1}Y^{(6)}_{\bf 1}t^cQ_3H_u+e^{i\tilde{f}_c\frac{A}{u_\chi}}\alpha_cY^{(6)}_{\bf 1}Y^{(6)}_{\bf 1}\Big(\frac{\langle{\cal S}\rangle}{\Lambda_5}\Big)^{|f_c|}c^cQ_2H_u\nonumber\\
&+e^{i\tilde{f}_u\frac{A}{u_\chi}}\alpha_uY^{(6)}_{\bf 1}Y^{(6)}_{\bf 1}\Big(\frac{\langle{\cal S}\rangle}{\Lambda_5}\Big)^{|f_u|}u^cQ_1H_u\nonumber\\
&+e^{i\tilde{f}_b\frac{A}{u_\chi}}\alpha_b\Big(\frac{\langle{\cal S}\rangle}{\Lambda_5}\Big)^{|f_b|}Y^{(6)}_{\bf 1}(Y^{(6)}_{\bf 3}D^c)_{{\bf 1}''}Q_3H_d+e^{i\tilde{f}_s\frac{A}{u_\chi}}\alpha_s\Big(\frac{\cal S}{\Lambda_5}\Big)^{|f_s|}Y^{(6)}_{\bf 1}(Y^{(6)}_{\bf 3}D^c)_{{\bf 1}'}Q_2H_d\nonumber\\
&+e^{i\tilde{f}_d\frac{A}{u_\chi}}\alpha_d\Big(\frac{\langle{\cal S}\rangle}{\Lambda_5}\Big)^{|f_d|}Y^{(6)}_{\bf 1}(Y^{(6)}_{\bf 3}D^c)_{{\bf 1}}Q_1H_d\nonumber\\
&+e^{i\tilde{f}_\tau\frac{A}{u_\chi}}\alpha_\tau\Big(\frac{\langle{\cal S}\rangle}{\Lambda_5}\Big)^{|f_\tau|}Y^{(6)}_{\bf 1}(Y^{(2)}_{\bf 3}L)_{{\bf 1}''}\tau^cH_d+e^{i\tilde{f}_\mu\frac{A}{u_\chi}}\alpha_\mu\Big(\frac{\langle{\cal S}\rangle}{\Lambda_5}\Big)^{|f_\mu|}Y^{(6)}_{\bf 1}(Y^{(2)}_{\bf 3}L)_{{\bf 1}'}\mu^cH_d\nonumber\\
&+e^{i\tilde{f}_e\frac{A}{u_\chi}}\alpha_e\Big(\frac{\langle{\cal S}\rangle}{\Lambda_5}\Big)^{|f_e|}Y^{(6)}_{\bf 1}(Y^{(2)}_{\bf 3}L)_{{\bf 1}}e^cH_d+{\rm h.c.}.
\label{4d_a1}
\end{eqnarray}
Here, higher-order corrections induced by $\chi \tilde{\chi}$ are neglected, which can be absorbed by leading-order terms (see Eq.(\ref{AFN1})). Additionally,
\begin{eqnarray}
\tilde{f}_\xi=f_\xi+g_\xi\quad\text{with}~|g_\xi|=1~(\xi=u,c,d,s,b,e,\mu,\tau)\,.
\label{4d_f1}
\end{eqnarray}

We are interested in the case where $\sigma(L) = kL \ll 1$, which\footnote{For $\sigma(L) = kL \gg 1$, it is hard to describe the small neutrino mass.} leads to $e^{2\sigma(L)} \approx e^{-2\sigma(L)} \approx 1$. We normalize the modular forms in the neutrino sector (Eq.(\ref{4d_b})) to be consistent with those in the quark and charged-lepton sectors.
Using Eqs.(\ref{nor_1}) and (\ref{nor_2}), for $\sigma(L) = kL \ll 1$, the modular forms in Eq.(\ref{4d_b}) can be canonically normalized:
\begin{eqnarray}
&&\qquad4\hat{y}_\nu\hat{Y}^{(2)}_{\bf 3}\hat{Y}^{(2)}_{\bf 3}\langle\chi\rangle e^{-4\sigma(L)-2M_fL} \rightarrow \frac{y_\nu}{4\langle\chi\rangle}Y^{(2)}_{\bf 3}Y^{(2)}_{\bf 3}\,,\nonumber\\
&&\hat{Y}^{(2)}_{\bf 3}\Psi^c_\nu(x,0)\,e^{2\sigma(L)-M_fL}\rightarrow \Big(\frac{f^{c0}_\nu(0)}{4}\frac{1}{\langle\chi\rangle}\sqrt{\frac{\Lambda_5}{\gamma L}}\Big)Y^{(2)}_{\bf 3}\psi^c_\nu(x)\,,
\end{eqnarray}
where $4\frac{\langle\chi\rangle}{\Lambda_5}e^{-M_fL-2\sigma(L)}\gamma_\nu\approx 1$ and $4\frac{\langle\chi\rangle}{\Lambda_5}\gamma'_\nu\,e^{-2\sigma(L)-M_fL}\approx1$ are used in the first equation, and in the second equation, Eq.(\ref{KKw1}) and $4\frac{\langle\chi\rangle}{\Lambda_5}e^{2\sigma(L)-M_fL}\gamma\approx1$ are used. This approach ensures consistency in normalizing modular forms across different sectors and accurately reflects the small neutrino masses within the context of our extra-dimensional model.
Then, the neutrino Yukawa Lagrangian, given by Eq.(\ref{4d_b}), with the normalized modular forms, is given for zero mode by
\begin{eqnarray}
-{\cal L}^{\nu}_Y\supset\frac{y_\nu e^{i\frac{A}{u_\chi}}}{2}\frac{Y^{(2)}_{\bf 3}LH_uY^{(2)}_{\bf 3}LH_u}{4\langle\chi\rangle}+y_0(Y^{(2)}_{\bf 3}\psi^c_\nu(x)L)_{\bf 1}H_u+{\rm h.c.}\,.
\label{4d_a2}
\end{eqnarray}
Here, the first term is a Majorana neutrino mass term, which is the origin of the light neutrino masses. This is analogous to the Weinberg dimension-5 operator, but with an explicit underlying physical framework, indicating that it originates from new physics at the $U(1)_X$ breaking scale, $\langle\chi\rangle \lesssim \Lambda_5$. Specifically, for the constraint on neutrino masses $m_{\nu}\sim0.05$ eV, this breaking scale is given by
   \begin{eqnarray}
  \langle\chi\rangle\sim10^{15}\,{\rm GeV}\,.
  \label{MR5}
  \end{eqnarray}
The $U(1)_X$ breaking scale $\langle\chi\rangle$ can be identified with the QCD axion decay constant $F_a=2\langle\chi\rangle/N_C$, see Eq.(\ref{qcd_fa}). 
In the second term of Eq.(\ref{4d_a2}), $y_0$ represents the effective Dirac neutrino Yukawa coupling for the zero mode, with $f^{c0}_\nu(0)\equiv f^{0}_{\nu R}(0)$:
 \begin{eqnarray}
y_0=\frac{1}{4}f^{0}_{\nu R}(0)\sqrt{\frac{\Lambda_5L^{-1}}{\gamma \langle\chi\rangle^2}}\simeq\sqrt{\frac{\xi-1}{e^{2\sigma(L)(\xi-1)}-1}}\sqrt{\frac{k}{2\sigma(L)\,e^{M_fL}\langle\chi\rangle}}\,,
\label{4d_a3}
\end{eqnarray}
where $\xi=M_f/k$, and Eqs.(\ref{zerom}) and (\ref{nor_2}) are used. If the bulk fermion masses $M_f$ are close to the fundamental scale $\Lambda_5$, and considering that $\Lambda_5$ must be larger than the curvature scale $k$\,\cite{Randall:1999ee}, the coupling $y_0$ is sufficiently small due to the factor $e^{M_fL}$, particularly with the specific values given in Eq.(\ref{re_M2}).

 When the energies are much larger than the EW scale (in other words, for $E\gtrsim m^\nu_n\gg$ EW scale), all KK modes of the bulk neutrino can be essential.
From the Lagrangian (\ref{4d_0}), and using Eq.(\ref{KKw1}), the neutrino Lagrangian can be expressed as
 \begin{eqnarray}
-{\cal L}^{\nu}_Y&\supset& \frac{e^{\frac{3}{2}\sigma(L)}}{\sqrt{L}}\big(\hat{Y}^{(2)}_{\bf 3}\sum_{n} f^{cn}_\nu(L)\psi^c_\nu(x)L\big)_{\bf 1}H_u\nonumber\\
&+&\frac{1}{2}\hat{y}_\nu\frac{e^{3\sigma(0)}}{L}f^n_\nu(0) f^n_\nu(0)\psi_{\nu}(x)\psi_\nu(x)\chi
+{\rm h.c.}+\psi^c_\nu(x){\cal M}^\nu_n\psi_\nu(x)\,,
\label{4d_nu}
\end{eqnarray}
where ${\cal M}^\nu_n$ is matrix representing the 4D KK mass spectrum, which is proportional to $m^\nu_n$. For the zero modes, the above Lagrangian (\ref{4d_nu}) can be rewritten as in Eq.(\ref{4d_b}).
To derive the KK modes and the 4D KK mass spectrum for bulk neutrinos, we introduce the variable $z=1-\epsilon e^{\sigma(y)}\in[0, 1-\epsilon]$ with $\epsilon=e^{-\sigma(L)}$, and $x_n=m^\nu_n/(k\epsilon)$ for convenience. The first-order equations (\ref{bulk_11}) become 
\begin{eqnarray}
&\big((z-1)\partial_z+\xi-\frac{1}{2}\big)f^n_{\nu R}=(1-z)x_nf^n_{\nu L}\,,\nonumber\\
&\big((z-1)\partial_z-\xi-\frac{1}{2}\big)f^n_{\nu L}=(z-1)x_nf^n_{\nu R}\,.
\label{fo_e}
\end{eqnarray} 
Since the eigenvalues $x_n$ are of order unity, the 4D KK masses are at the curvature scale $k$ by dimensional analysis due to $\epsilon\approx1$.
And rescaling $f^n_{L(R)}(y)\rightarrow\sqrt{\epsilon k L}\,e^{\frac{1}{2}\sigma(y)}f^n_{L(R)}(z)$, the orthonormal condition Eq.(\ref{KKnor}) becomes 
\begin{eqnarray}
\int^{1-\epsilon}_0dz\,f^m_{L(R)}(z)f^n_{L(R)}(z)=\delta_{mn}\,.
\label{KKnor1}
\end{eqnarray} 
Using Eq.(\ref{KKnor1}), the right (left)-handed zero modes can have wave functions
\begin{eqnarray}
f^0_{\nu R}(z)=\sqrt{\frac{2(\xi-1)}{e^{2\sigma(L)(\xi-1)}-1}}(1-z)^{\frac{1}{2}-\xi}\,,\quad f^0_{\nu L}(z)=\sqrt{\frac{2(\xi+1)}{1-e^{2\sigma(L)(\xi+1)}}}(1-z)^{\frac{1}{2}+\xi}\,,
\label{zerom}
\end{eqnarray} 
which do not vanish at the orbifold fixed points. The nonzero KK modes can be obtained by solving the first-order coupled equations of motion for $f^n_{L(R)}$, leading to a pair of decoupled second-order equations:
\begin{eqnarray}
\Big\{(1-z)^2\partial^2_z+(1-z)\partial_z+(1-z)^2x^2_n-\big(\xi^2\pm\xi-\frac{3}{4}\big)\Big\}f^n_{\nu R(L)}=0\,.
\end{eqnarray} 
The solutions of the above differential equations for the case where the eigenvalues $x_n>0$ are Bessel functions:
\begin{eqnarray}
f^n_{\nu L(R)}(z)=N^n_\psi\,(z-1)\Big[J_{\xi\mp\frac{1}{2}}\Big((z-1)\frac{m^\nu_n}{k\epsilon}\Big)+b^n_\psi\,Y_{\xi\mp\frac{1}{2}}\Big((z-1)\frac{m^\nu_n}{k\epsilon}\Big)\Big]\,,
\end{eqnarray} 
where $N^n_\psi$ and $b^n_\psi$ are arbitrary constants. The two functions are not independent, as they are coupled by the first-order differential equations in Eq.(\ref{fo_e}).
Besides the boundary conditions in Eqs.(\ref{bc_1}) and (\ref{bc_2}), we set boundary conditions
\begin{eqnarray}
f^n_{i L}(1-\epsilon)=0\,,\quad f^n_{i R}(0)=0\,\quad(n=1,2,3,...)
\label{dich}
\end{eqnarray} 
which correspond to $f^n_{i L}(y)|_{y=0}=0$ and $f^n_{i R}(y)|_{y=L}=0$ for $n=1,2,3,...$, respectively.  It is evident from Eqs.(\ref{4d_nu}) and (\ref{dich}) that, at the boundaries, the contribution of the action (\ref{4d_0}) for the non-zero modes would be negligible. With the zero mode coupling of Eq.(\ref{4d_a3}) this ensures that any significant mixing between the SM neutrinos $\nu_L$ and the sterile bulk neutrinos $\psi_\nu$ can be negligible. Thus, after EW symmetry breaking, the Dirac neutrino mass terms\,\footnote{For a discussion on generating Dirac neutrino mass without relying on a seesaw mechanism, refer to Ref.\cite{Grossman:1999ra}.} in Eq.(\ref{4d_nu}) do not contribute to the light neutrino mass.
The boundary conditions Eq.(\ref{dich}) at the locations of the 3-branes determine the 4D KK mass $m^n_\nu$. Using the Mehler-Sonine formula for Bessel functions\,\footnote{$J_\nu(x)=\frac{2(x/2)^\nu}{\sqrt{\pi}\Gamma(\nu+\frac{1}{2})}\int^1_0(1-t^2)^{\nu-\frac{1}{2}}\cos(xt)dt$ and $Y_\nu(x)=\frac{2(x/2)^\nu}{\sqrt{\pi}\Gamma(\nu+\frac{1}{2})}\big\{\int^1_0(1-t^2)^{\nu-\frac{1}{2}}\sin(xt)dt-\int^\infty_0 e^{-xt}(1+t^2)^{\nu-\frac{1}{2}}dt\big\}$ where ${\rm Re}[\nu]>-1/2$, and $|{\rm phase}\,\,x|<\pi/2$ for $Y_\nu(x)$.}, it becomes evident that, in the limit $\epsilon\rightarrow1$, only the first-kind Bessel function persists within the wave functions with $b^n_\psi=0$. The constant $N^n_\psi$ is determined by the orthonormality condition Eq.(\ref{KKnor1}). The 4D KK mass spectrum for $n=1,2,3,...$ is given by
\begin{eqnarray}
m^\nu_n\simeq k\,\epsilon\,j_{{\xi},n}\,,
\label{kkmas}
\end{eqnarray} 
where $j_{{\xi},n}={\rm BesselJzero}[{\xi},n]$. 

Below the $U(1)_X$ symmetry breaking scale, the effective interactions of the QCD axion with the weak and hypercharge gauge bosons, as well as with the photon, can be expressed through the chiral rotation of Eq.(\ref{X-tr}) as follows:
\begin{eqnarray}
{\cal L}^{WY}_{A}&=&\frac{A}{f_A}\frac{1}{32\pi^2}\Big\{g^2_W\,N_W\,W^{\mu\nu}\tilde{W}_{\mu\nu}+g^2_Y\,N_Y\,Y^{\mu\nu}\tilde{Y}_{\mu\nu}\Big\}\,,\label{}\\
{\cal L}^{\gamma}_{A}&=&\frac{A}{f_A}\frac{e^2}{32\pi^2}\,E\,F^{\mu\nu}\tilde{F}_{\mu\nu}\,,
\label{ema}
\end{eqnarray}
where $g_{W}$, $g_Y$, and $e$ are the gauge coupling constants for $SU(2)_L$, $U(1)_Y$, and $U(1)_{EM}$, respectively, with the corresponding gauge field strengths $W^{\mu\nu}, Y^{\mu\nu}$, and $F^{\mu\nu}$ and their dual forms $\tilde{W}_{\mu\nu}, \tilde{Y}_{\mu\nu}$, and $\tilde{F}_{\mu\nu}$. Here $N_W\equiv2{\rm Tr}[X_{\psi_f}T^2_{SU(2)}]$ and $N_Y\equiv2{\rm Tr}[X_{\psi_f}(Q_f^Y)^2]$ are the anomaly coefficients for $U(1)_{X}\times[SU(2)_L]^2$ and $U(1)_{X}\times[U(1)_Y]^2$, respectively. The electromagnetic anomaly coefficient $E$ for $U(1)_{X}\times[U(1)_{EM}]^2$ is defined by $E=2\sum_{\psi_f} X_{\psi_f}(Q^{\rm em}_{\psi_f})^2$ where $Q^{\rm em}_{\psi_f}$ is the $U(1)_{\rm EM}$ charge of field $\psi_f$. It can be expressed as 
  \begin{eqnarray}
  E=N_W+N_Y=-2(\tilde{f}_e+\tilde{f}_\mu+\tilde{f}_\tau)-\frac{8}{3}(\tilde{f}_u+\tilde{f}_c+g_t)-\frac{2}{3}(\tilde{f}_d+\tilde{f}_s+\tilde{f}_b)\,.
 \label{eano}
 \end{eqnarray}
The physical quantities of QCD axion, such as its mass $m_a$ and its axion-photon coupling $g_{a\gamma\gamma}$, depend on the ratio of electromagnetic anomaly coefficient $E$ to the color anomaly coefficient $N_C$. The value of $E/N_C$ is determined in terms of the $X$-charges for quarks and leptons, as specified by Eqs.(\ref{eano}) and (\ref{nc1}) or (\ref{gr_ano}),
  \begin{eqnarray}
  \frac{E}{N_C}&=&\frac{2(\tilde{f}_e+\tilde{f}_\mu+\tilde{f}_\tau)+\frac{2}{3}(4\tilde{f}_u+4\tilde{f}_c+4g_t+\tilde{f}_d+\tilde{f}_s+\tilde{f}_b)}{\tilde{f}_u+\tilde{f}_c+g_t+\tilde{f}_d+\tilde{f}_s+\tilde{f}_b}\,\nonumber\\
  &=&\frac{6(\tilde{f}_e+\tilde{f}_\mu+\tilde{f}_\tau)+2(4\tilde{f}_u+4\tilde{f}_c+4g_t+\tilde{f}_d+\tilde{f}_s+\tilde{f}_b)}{-\tilde{f}_e-\tilde{f}_\mu-\tilde{f}_\tau}\,,
 \label{eano}
 \end{eqnarray}
where the first and second equality follow from Eqs.(\ref{nc1}) and (\ref{gr_ano}), respectively, and $\tilde{f}_\xi$ is defined in Eq.(\ref{4d_f1}). For instance, for $|f_b|=5, |f_s|=11, |f_{d}|=14, |f_c|=9$, and $|f_u|=22$ with $|N_C|=1$, which explain the quark data (see Eq.(\ref{para_sp})), equivalently, for $|f_e|=21$, $|f_\mu|=12$, and $|f_\tau|=7$ for the charged-lepton sector (see Eq.(\ref{para_sp_nu})) with those quantum numbers of quarks, there are several possible values of $E/N_C$:
  \begin{eqnarray}
  \frac{E}{N_C}=\frac{44}{3}\,, \frac{56}{3}\,, -\frac{76}{3}\,, -\frac{112}{3}\,, \frac{152}{3}\,, \frac{164}{3}\,, -\frac{196}{3}\,, -\frac{208}{3}\,.
 \label{eano1}
 \end{eqnarray}
These specific values of $E/N_C$ can be tested by near future experiments with the scale of $U(1)_X$ breakdown in Eq.(\ref{MR5}), see Fig.\,\ref{Fig3}.

 %%%%%%%%%%%%%%%%%%%%%%%%%%%%%%%%%%%%%%%%%%%%%%%%%%%%%%%%%%%%%%%%%%%%%%%%%%%
\section{Quark and Lepton interactions with QCD axion}
\label{sec4}
Let us explore how quark and lepton masses and mixings are derived from Yukawa interactions Eqs.(\ref{4d_a1}, \ref{4d_a2}) within a framework based on $A_4\times U(1)_X$ symmetries with modular invariance. 
Non-zero VEVs of scalar fields $\chi(\tilde{\chi})$ spontaneously break  the flavor symmetry\,\footnote{If the $U(1)_{X}$ is broken spontaneously, the massless mode $A$ of the scalar $\chi(\tilde{\chi})$ appear as a phase.} $U(1)_{X}$ at high energies above EW scale.
Then, the effective Yukawa structures in the low-energy limit depend on a small dimensionless parameter $\langle {\cal S}\rangle/\Lambda_5\equiv \Delta_{\cal S}$.
The higher order contributions of superpotentials at the $y=0$ brane are given by
 $\sum^{\infty}_{n=1}\tilde{c}_i \,\Delta^{2n}_\chi\cdot${\it (leading order operators)} where $\tilde{c}_i$ are complex numbers with unit absolute value. These contributions cause a shift in the Yukawa coefficients of the leading order terms in the Lagrangians presented in Eqs.(\ref{4d_a1}, \ref{4d_a2}).
Denoting the effective Yukawa coefficients shifted by higher order contributions as $\alpha_i$, we see that they are constrained as 
\begin{eqnarray}
1-\frac{\Delta^2_\chi}{1-\Delta^2_\chi}\leq|\alpha_i|\leq1+\frac{\Delta^2_\chi}{1-\Delta^2_\chi}\qquad\text{with}~\Delta_\chi\equiv \frac{v_\chi}{\sqrt{2}\,\Lambda_5},
  \label{AFN1}
\end{eqnarray}
where the lower (upper) limit corresponds to the sum of higher order terms. In our framework of $G_F\times${\it extra-dimension}, after the canonical normalization of modular forms, the low-energy effective Yukawa interactions in Eqs.(\ref{4d_a}, \ref{4d_b}, \ref{4d_a1}) take on the standard form. Within this context, the effective Yukawa couplings are expressed as functions of $({\cal S}/\Lambda_5)^{|f_\alpha|}$, effective Yukawa coefficients $\alpha_i$, and modular forms $Y$.
When $H_{u(d)}$ acquire non-zero VEVs, all quarks and leptons obtain masses.
The relevant quark and lepton interactions with their chiral fermions are given by 
 \begin{eqnarray}
  -{\cal L} &\supset&
  \overline{q^{u}_{R}}\,\mathcal{M}_{u}\,q^{u}_{L}+\overline{q^{d}_{R}}\,\mathcal{M}_{d}\,q^{d}_{L}+\frac{g}{\sqrt{2}}W^+_\mu\overline{q^u_L}\gamma^\mu\,q^d_L\nonumber\\
  &+&\overline{\ell_{R}}\,{\cal M}_{\ell}\,\ell_{L} +\frac{1}{2}e^{i\frac{A}{u_\chi}}\,\overline{\nu^c_L}{\cal M}_\nu\,\nu_L +\frac{g}{\sqrt{2}}W^-_\mu\overline{\ell_L}\gamma^\mu\,\nu_L+\text{h.c.}\,,
  \label{AxionLag2}
 \end{eqnarray}
where $g$ is the $SU(2)_L$ coupling constant, $q^{u}=(u,c,t)$, $q^{d}=(d,s,b)$, $\ell=(e, \mu, \tau)$, and $\nu=(\nu_e,\nu_\mu,\nu_\tau)$.  The explicit forms of $\mathcal{M}_{u,d,\ell, \nu}$ will be given later. The above Lagrangian of the fermions, including their kinetic terms of Eq.(\ref{lag0}), should be invariant under $U(1)_X$:
 \begin{eqnarray}
  \psi\rightarrow e^{i(f_{\psi}+g_{\psi})\frac{\gamma_5}{2}\alpha}\psi\,,
 \label{X-tr}
 \end{eqnarray}
where $\psi=\{u,c,t,d,s,b,e,\mu,\tau,\nu\}$ and $\alpha$ is a transformation constant parameter.

%%%%%%%%%%%%%%%%%%%%%%
\subsection{Quark and flavored-QCD axion}
The axion coupling matrices to the up- and down-type quarks, respectively, are diagonalized through bi-unitary transformations: $V^{\psi}_R{\cal M}_{\psi}V^{\psi\dag}_L=\hat{{\cal M}}_\psi$ (diagonal form) with the mass eigenstates $\psi'_R=V^\psi_R\,\psi_R$ and $\psi'_L=V^\psi_L\,\psi_L$. These transformations include, in particular, the chiral transformation of Eq.(\ref{X-tr}), which necessarily makes ${\cal M}_{u,d}$ real and positive. 
This induces a contribution to the QCD vacuum angle in Eq.(\ref{lag0}), given by
 \begin{eqnarray}
  \vartheta_{\rm QCD}\rightarrow\vartheta_{\rm eff}=\vartheta_{\rm QCD}+\arg\{\det({\cal M}_u)\det({\cal M}_d)\}
 \label{}
 \end{eqnarray}
with $-\pi\leq\vartheta_{\rm eff}\leq\pi$. Through a chiral rotation Eq.(\ref{X-tr}), the vanishing QCD anomaly term is obtained as
 \begin{eqnarray}
 {\cal L}_\vartheta=\Big(\vartheta_{\rm eff}+\frac{A}{F_a}\Big)\frac{g^2_s}{32\pi^2}G^{a\mu\nu}\tilde{G}^a_{\mu\nu}\qquad\text{with}~F_a=\frac{u_\chi}{N_C}\,,
 \label{qcd_fa}
 \end{eqnarray}
where $F_a$ is the axion decay constant and $u_\chi$ is defined in Eq.(\ref{NGboson}). At the QCD phase transition, $A$ will get a VEV, $\langle A\rangle=-F_a\vartheta_{\rm eff}$, eliminating the constant $\vartheta_{\rm eff}$ term. The QCD axion then is the excitation of the $A$ field, $a=A-\langle A\rangle$.

By substituting the VEV from Eq.(\ref{vev}) into the Lagrangian (\ref{4d_a1}), the mass matrices ${\cal M}_{u}$ and ${\cal M}_{d}$ for up- and down-type quarks, given in the Lagrangian (\ref{AxionLag2}), are derived as
 \begin{eqnarray}
  &{\cal M}_{u}={\left(\begin{array}{ccc}
 \alpha_u\Delta^{|f_u|}_\chi\,e^{i\tilde{f}_u\frac{A}{u_\chi}} &  0 &  0 \\
 0  &  \alpha_c\Delta^{|f_c|}_\chi\,e^{i\tilde{f}_c\frac{A}{u_\chi}}  &  0   \\
 0  &  0  & \alpha_t\,e^{ig_t\frac{A}{u_\chi}}
 \end{array}\right)}\,Y^{(6)}_{\bf 1}Y^{(6)}_{\bf 1}v_u\,,
 \label{Ch2}\\
 &{\cal M}_{d}=Y^6_1(1+r^3+s^3-3rs)\Big[{\left(\begin{array}{ccc}
 \alpha_d\Delta^{|f_d|}_\chi &  \alpha_s\,r\Delta^{|f_s|}_\chi &  \alpha_b\,s\Delta^{|f_b|}_\chi \\
 \alpha_d\,s\Delta^{|f_d|}_\chi &  \alpha_s\Delta^{|f_s|}_\chi &  \alpha_b\,r\Delta^{|f_b|}_\chi   \\
 \alpha_d\,r\Delta^{|f_d|}_\chi &  \alpha_s\,s\Delta^{|f_s|}_\chi  &  \alpha_b\Delta^{|f_b|}_\chi
 \end{array}\right)}(1+2rs)\nonumber\\
 &\qquad\qquad\qquad\qquad\qquad\qquad\quad+{\left(\begin{array}{ccc}
 \tilde{\alpha}_d\,s\Delta^{|f_d|}_\chi &  \tilde{\alpha}_s\Delta^{|f_s|}_\chi &  \tilde{\alpha}_b\,r\Delta^{|f_b|}_\chi \\
 \tilde{\alpha}_d\,r\Delta^{|f_d|}_\chi &  \tilde{\alpha}_s\,s\Delta^{|f_s|}_\chi &  \tilde{\alpha}_b\Delta^{|f_b|}_\chi   \\
 \tilde{\alpha}_d\Delta^{|f_d|}_\chi &  \tilde{\alpha}_s\,r\Delta^{|f_s|}_\chi  &  \tilde{\alpha}_b\,s\Delta^{|f_b|}_\chi
 \end{array}\right)}(s^2+2r)\Big]\tilde{C}\,v_d\,,  
 \label{Ch1}
 \end{eqnarray}
 with
  \begin{eqnarray}
r\equiv\frac{Y_2}{Y_1}\,,\qquad s\equiv\frac{Y_3}{Y_1}\,.
\label{lepto_09}
\end{eqnarray}
where $v_{d}\equiv\langle H_{d}\rangle=v\cos\beta/\sqrt{2}$, $v_{u}\equiv\langle H_{u}\rangle =v\sin\beta/\sqrt{2}$ with $v\simeq246$ GeV, and 
 \begin{eqnarray}
 \tilde{C}={\rm diag}\big(e^{i\tilde{f}_d\frac{A}{u_{\chi}}} , e^{i\tilde{f}_s\frac{A}{u_{\chi}}},
e^{i\tilde{f}_b\frac{A}{u_{\chi}}}\big)\,.
 \label{c_num}
 \end{eqnarray}
The terms with $\alpha_{d, s, b}$ in Eq.(\ref{Ch1}) arise from the modular form $Y^{(6)}_{{\bf 3},1}$ given in Eq.(\ref{modu6}), whereas the terms with $\tilde{\alpha}_{d, s, b}$ in Eq.(\ref{Ch1}) arise from $Y^{(6)}_{{\bf 3},2}$.

The quark mass matrices ${\cal M}_{u}$ in Eq.(\ref{Ch2}) and ${\cal M}_{d}$ in Eq.(\ref{Ch1}) generate the up- and down-type quark masses:
 \begin{eqnarray}
 \hat{\mathcal{M}}_{u}=V^{u}_{R}\,{\cal M}_{u}\,V^{u\dag}_{L}
 ={\rm diag}(m_{u},m_{c},m_{t})\,,\quad \hat{\mathcal{M}}_{d}=V^{d}_{R}\,{\cal M}_{d}\,V^{d\dag}_{L}
 ={\rm diag}(m_{d},m_{s},m_{b})\,.
 \label{Quark21}
 \end{eqnarray}
Diagonalizing the matrices ${\cal M}_f^\dagger {\cal M}_f$ and ${\cal M}_f {\cal M}_f^\dagger$ ($f=u,d$) determine the mixing matrices $V_L^f$ and $V_R^f$, respectively \cite{Ahn:2011yj}. The left-handed quark mixing matrices $V_L^u$ and $V_L^d$ are components of the CKM matrix $V_{\rm CKM}=V_L^u V_L^{d\dagger}$. The CKM matrix is generated primarily from the down-type quark matrix in Eq.(\ref{Ch1}) due to the diagonal form of the up-type quark mass matrix in Eq.(\ref{Ch2}). The CKM matrix is parameterized using the Wolfenstein parametrization\,\cite{Wolfenstein:1983yz}, and its elements have been determined with high precision\,\cite{Ahn:2011fg}:
 \begin{eqnarray}
 V_{\rm CKM}={\left(\begin{array}{ccc}
 1-\frac{1}{2}\lambda^2 & \lambda & A_d\lambda^3(\rho-i\eta)  \\
-\lambda & 1-\frac{1}{2}\lambda^2 & A_d\lambda^2   \\
 A_d\lambda^3(1-\rho-i\eta) & -A_d\lambda^2  & 1
 \end{array}\right)}+{\cal O}(\lambda^4)
 \label{ckm0}
 \end{eqnarray}
in the Wolfenstein parametrization\,\cite{Wolfenstein:1983yz} and at higher precision\,\cite{Ahn:2011fg}, where $\lambda=0.22475^{+0.00106}_{-0.00018}$, $A_d=0.840^{+0.016}_{-0.043}$, $\bar{\rho}=\rho/(1-\lambda^2/2)=0.158^{+0.036}_{-0.020}$, and $\bar{\eta}=\eta/(1-\lambda^2/2)=0.349^{+0.029}_{-0.025}$ with $3\sigma$ errors\,\cite{ckm}. 
Their corresponding current best-fit values of the CKM mixing angles in the standard parameterization\,\cite{Chau:1984fp} within the $3\sigma$ range\,\cite{ckm} are
 \begin{eqnarray}
  \theta^q_{23}[^\circ]=2.376^{+0.054}_{-0.070}\,,\quad\theta^q_{13}[^\circ]=0.210^{+0.016}_{-0.010}\,,\quad\theta^q_{12}[^\circ]=13.003^{+0.048}_{-0.036}\,,\quad\delta^q_{CP}[^\circ]=65.5^{+3.1}_{-4.9}\,.
 \label{ckmmixing}
 \end{eqnarray} 
The physical structure of the up- and down-type quark Lagrangian should align with the empirical results calculated by the Particle Data Group (PDG) \cite{PDG}:
 \begin{eqnarray}
 m_d&=&4.67^{+0.48}_{-0.17}\,{\rm MeV}\,,\qquad~m_s=93^{+11}_{-5}\,{\rm MeV}\,,\qquad\qquad~~m_b=4.18^{+0.03}_{-0.02}\,{\rm GeV}\,,  \nonumber\\
 m_u&=&2.16^{+0.49}_{-0.29}\,{\rm MeV}\,,\qquad~ m_c=1.27\pm0.02\,{\rm GeV}\,,\qquad~ m_t=173.1\pm0.9\,{\rm GeV}\,,
\label{qumas}
 \end{eqnarray}
where $t$-quark mass is the pole mass, $c$- and $b$-quark masses are the running masses in the $\overline{\rm MS}$ scheme, and the light $u$-, $d$-, $s$-quark masses are the current quark masses in the $\overline{\rm MS}$ scheme at the momentum scale $\mu\approx2$ GeV. Below the scale of spontaneous $SU(2)_L\times U(1)_Y$ gauge symmetry breaking, the running masses of $c$- and $b$-quark receive corrections from QCD and QED loops\,\cite{PDG}. The top quark mass at scales below the pole mass is unphysical since the $t$-quark decouples at its scale, and its mass is determined more directly by experiments\,\cite{PDG}.
 %%%%%%%%%%%%%%%
%   Fig A-2   %
%%%%%%%%%%%%%%%
\begin{figure}[t]
\begin{minipage}[h]{8.0cm}
\includegraphics[width=8.0cm]{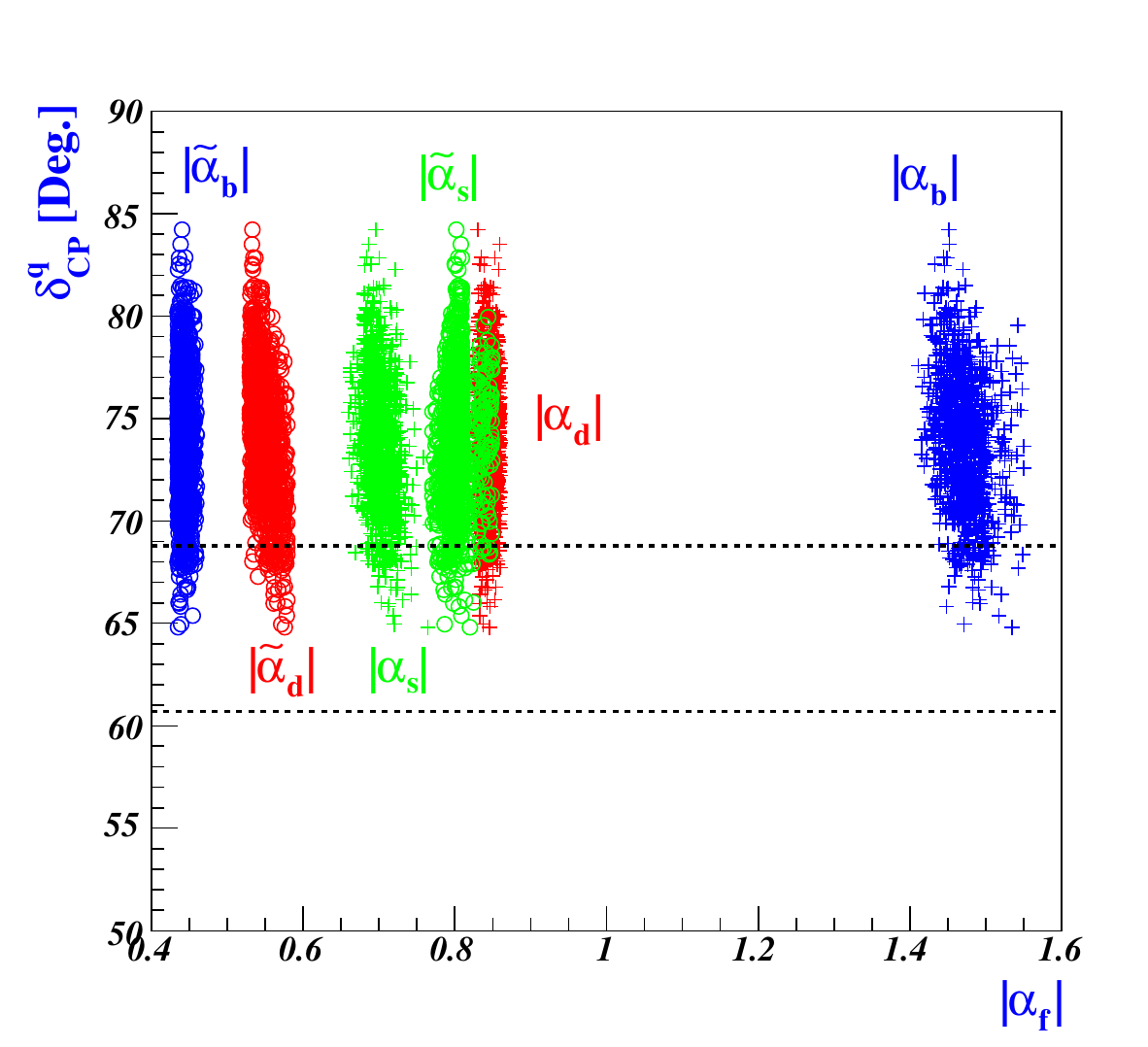}
\end{minipage}
\begin{minipage}[h]{8.0cm}
\includegraphics[width=8.0cm]{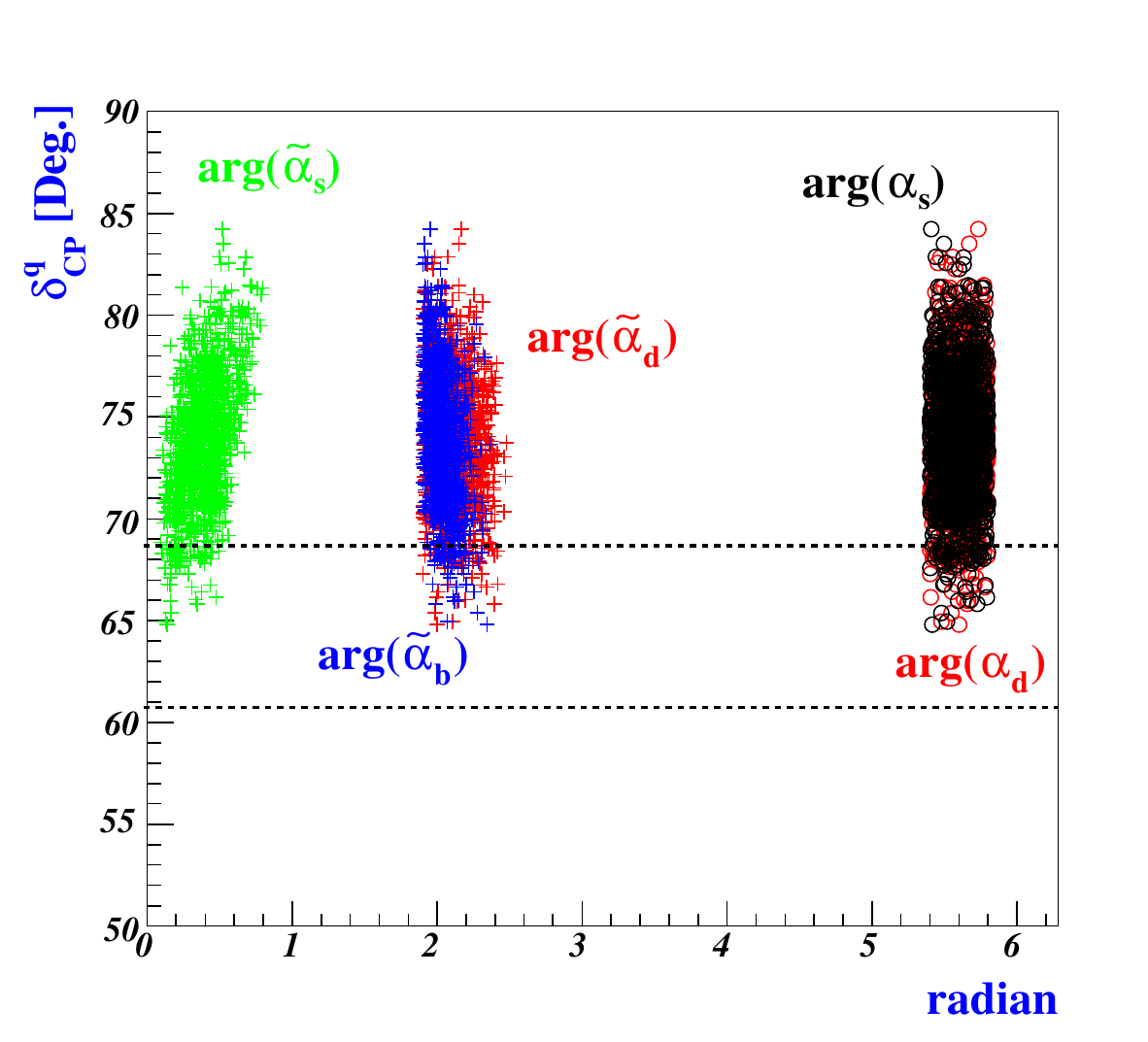}
\end{minipage}\\
\begin{minipage}[h]{8.0cm}
\includegraphics[width=8.0cm]{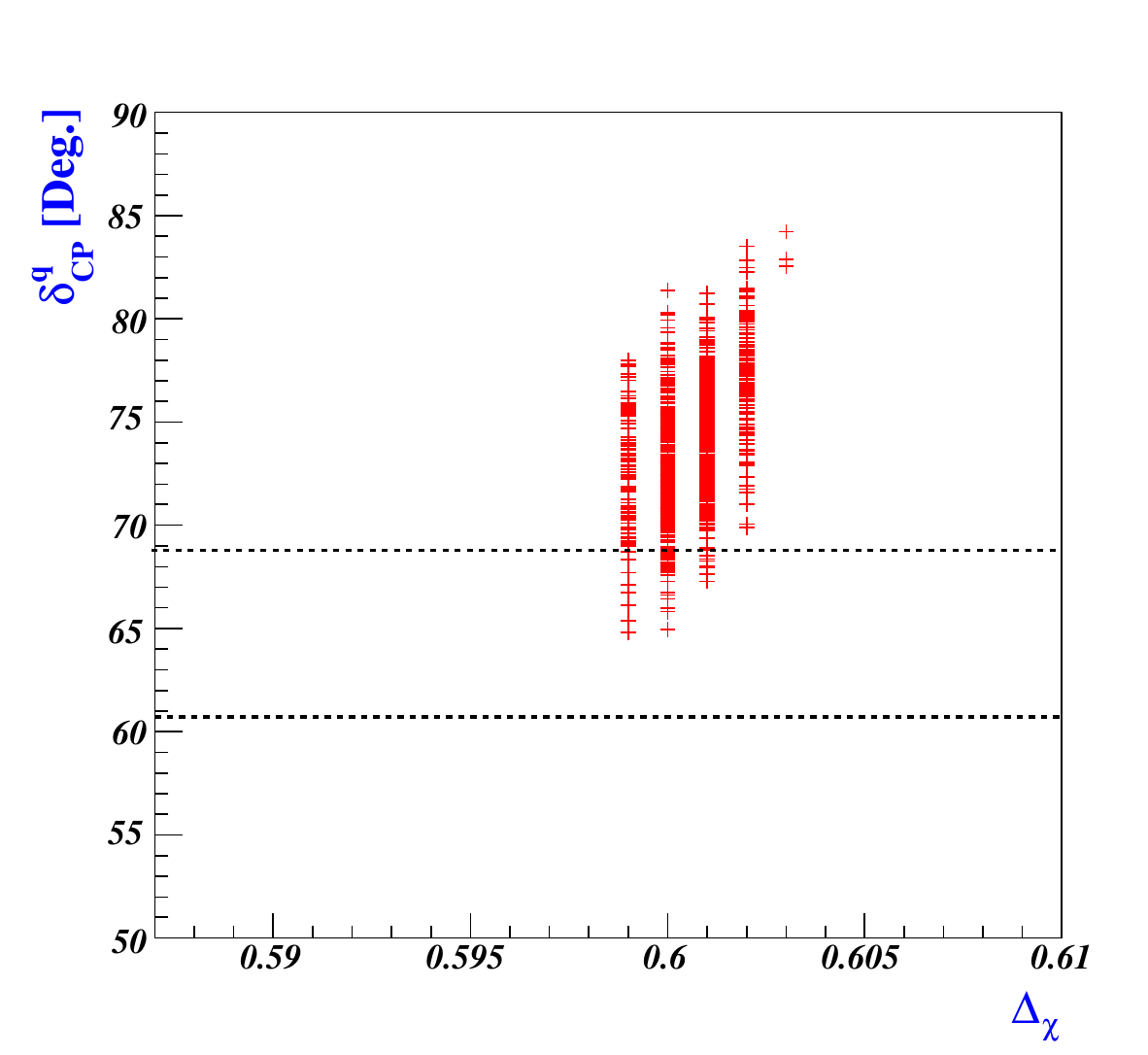}
\end{minipage}
\begin{minipage}[h]{8.0cm}
\includegraphics[width=8.0cm]{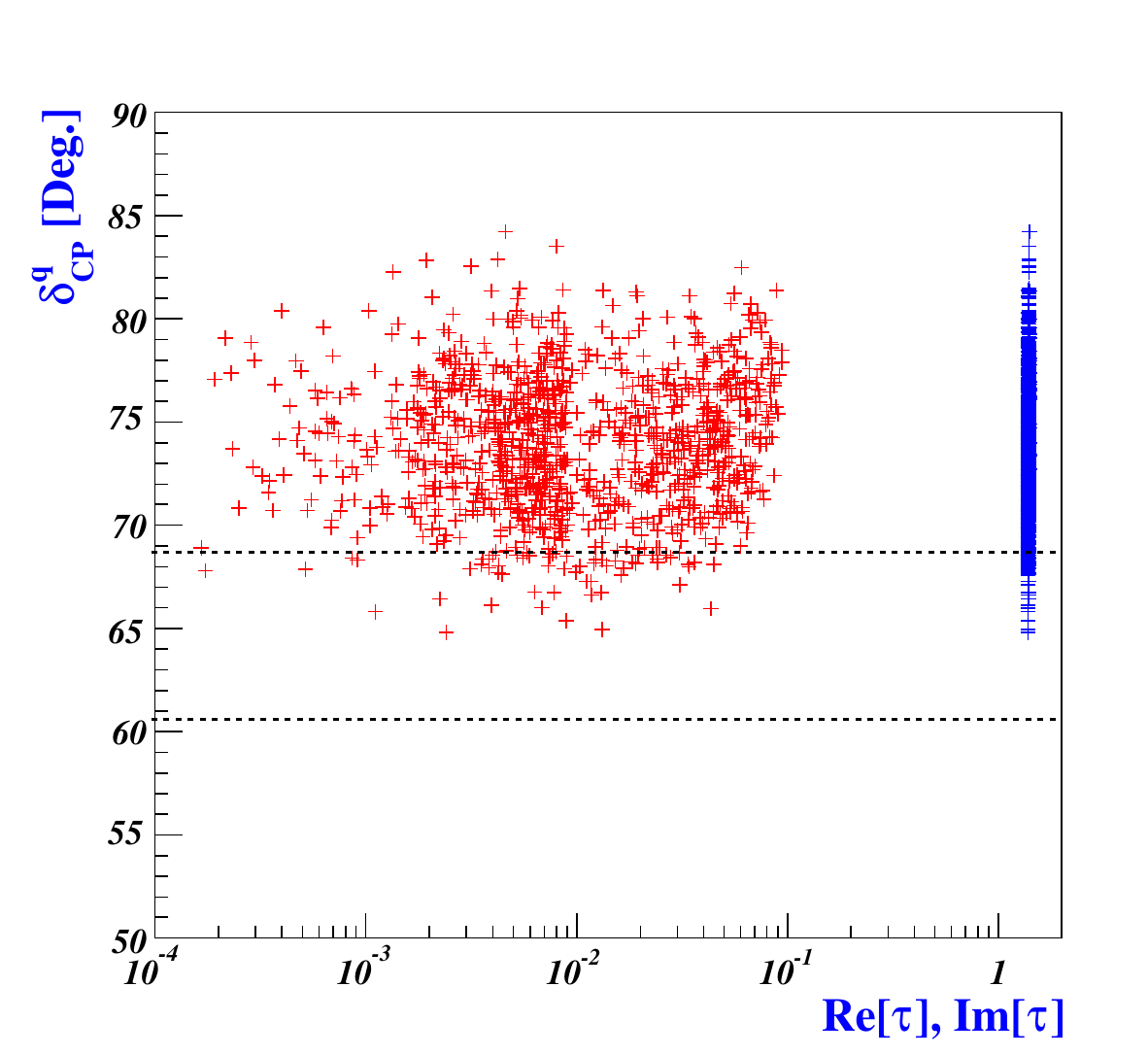}
\end{minipage}
\caption{\label{FigA2}  Model predictions for $\delta^q_{\rm CP}$ are shown, left upper, right upper, and left lower panel, as a function of the parameters that are constrained by other empirical results. The horizontal black-dotted lines indicate the $3\sigma$ experimental bound.}
\end{figure}

{\it Numerical analysis:} To simulate and match experimental results for quarks and leptons, we use linear algebra tools from Ref.\cite{Antusch:2005gp}. We have 10 physical observables in the quark sector: $m_d, m_s, m_b$, $m_u, m_c, m_t$, and $\theta^q_{12}, \theta^q_{23}, \theta^q_{13}, \delta^q_{CP}$. These observables are used to determine 10 effective model parameters among 18 parameters ($|\alpha_d|$, $|\alpha_s|$, $|\alpha_b|$, $|\tilde{\alpha}_d|$, $|\tilde{\alpha}_s|$, $|\tilde{\alpha}_b|$, $\alpha_t$, $\alpha_c$, $\alpha_u$; $\arg(\alpha_d), \arg(\alpha_s), \arg(\tilde{\alpha}_d), \arg(\tilde{\alpha}_s), \arg(\tilde{\alpha}_b)$; $\Delta_\chi$, $\tan\beta$; ${\rm Re[\tau]}, {\rm Im}[\tau]$).
Using highly precise data as constraints for quarks, with the exception of the quark Dirac CP phase, as described in Eqs.(\ref{ckmmixing}), (\ref{qumas}), and (\ref{clepmass}), we scanned all parameter ranges and determined that 
\begin{eqnarray}
&|f_{d}|=14,~|f_{s}|=11,~|f_{b}|=5,~|f_u|=22,~|f_c|=9,\nonumber\\
&\Delta_\chi=[0.599,0.603]\,,\qquad \tan\beta=3\,,\nonumber\\
&\tau=(0.000166\sim0.0940)+(1.3880\sim1.4000)i\,.
\label{para_sp}
\end{eqnarray}
Fig.\ref{FigA2} shows how the quark Dirac CP phase $\delta^q_{CP}$ behaves based on certain constrained parameters. Our model predicts that $\delta^q_{CP}$ falls between $64^\circ$ and $85^\circ$, which aligns well with experimental data. The horizontal black-dotted lines in Fig.\ref{FigA2} represent the $3\sigma$ experimental bound for $\delta^{q}_{CP}$. Notably, the effective Yukawa coefficients satisfying the experimental data fall well within the bound specified in Eq.(\ref{AFN1}), as shown in the top left panel of Fig.\ref{FigA2}. This reflects that these coefficients have a natural size of unity, as stated in Eq.(\ref{AFN1}).
We choose reference values, for example, that satisfy the experimental data :
\begin{eqnarray}
&\Delta_\chi=0.6\,,\qquad\tau=0.00068+1.38302i\,,\qquad\tan\beta=3
\label{delchi}
\end{eqnarray}
which result in effective Yukawa coefficients from Eq.(\ref{AFN1}) satisfying $0.45\lesssim|\alpha_i|\lesssim1.55$.
With the following inputs
 \begin{eqnarray}
 &\arg(\alpha_d)=5.590,~\arg(\alpha_s)=5.749,~\arg(\alpha'_d)=2.183,~\arg(\alpha'_s)=0.437,~\arg(\alpha'_b)=2.134,\nonumber\\
&\alpha_u=1.252,~\alpha_c=0.913,~\alpha_t=1.193,\nonumber\\
&\alpha_d=0.576,~\alpha_s=0.793,~\alpha_b=0.449,~\alpha'_d=0.834,~\alpha'_s=0.717,~\alpha'_b=1.482,
 \label{quarkvalue21}
 \end{eqnarray}
 we obtain the mixing angles and Dirac CP phase $\theta^q_{12}=13.048^{\circ}$, $\theta^q_{23}=2.318^{\circ}$, $\theta^q_{13}=0.205^{\circ}$, $\delta^q_{CP}=65.837^{\circ}$ compatible with the $3\sigma$ Global fit of CKMfitter\,\cite{ckm}, see Eq.(\ref{ckmmixing}); the quark masses $m_d=4.902$ MeV, $m_s=103.046$ MeV, $m_b=4.175$ GeV, $m_u=2.164$ MeV, $m_c=1.271$ GeV, and $m_t=173.1$ GeV compatible with the values in PDG\,\cite{PDG}, see Eq.(\ref{qumas}). 
Here, without loss of generality, the up-type quark masses $m_u$, $m_c$, and $m_t$ are a one-to-one correspondence with $\alpha_u$, $\alpha_c$, and $\alpha_t$, which have been taken real, and we have set $\arg(\alpha_b)=0$.

After diagonalizing the mass matrices in Eqs.(\ref{Ch2}) and (\ref{Ch1}), the flavored-QCD axion to quark interactions at leading order are given by
\begin{eqnarray}
-{\cal L}^{aq}&\simeq&\frac{\partial_\mu a}{2u_{\chi}}\Big\{\tilde{f}_u\,\bar{u}\gamma^\mu\gamma_5 u+\tilde{f}_c\,\bar{c}\gamma^\mu\gamma_5c+g_t\,\bar{t}\gamma^\mu\gamma_5t+\tilde{f}_d\,\bar{d}\gamma^\mu\gamma_5 d+\tilde{f}_s\,\bar{s}\gamma^\mu\gamma_5s+\tilde{f}_b\,\bar{b}\gamma^\mu\gamma_5b\Big\}\nonumber\\
&+&\frac{\partial_\mu a}{2u_{\chi}}\Big\{C_{sd}\,\bar{d}\gamma^\mu\big(1+\gamma_5\big) s+C_{bs}\,\bar{s}\gamma^\mu\big(1+\gamma_5) b+C_{db}\,\bar{b}\gamma^\mu\big(1+\gamma_5\big) d+{\rm h.c.}\Big\}\nonumber\\
&+&m_u\,\bar{u}u+m_c\,\bar{c}c+m_t\,\bar{t}t+m_d\,\bar{d}d+m_s\,\bar{s}s+m_b\,\bar{b}b-\bar{q}i\! \! \not\! D\, q\,,
\label{fl_in}
\end{eqnarray}
with
\begin{eqnarray}
&&C_{sd}=(\tilde{f}_d-\tilde{f}_s)\lambda\Big(1-\frac{\lambda^2}{2}\Big)\,,\nonumber\\
&&C_{bs}=(\tilde{f}_s-\tilde{f}_b)A_d\lambda^2\,,\nonumber\\
&&C_{db}=A_d\lambda^3\Big(\tilde{f}_d(\rho+i\eta)-\tilde{f}_s+\tilde{f}_b(1-\rho-i\eta)\Big)\,,
  \label{}
\end{eqnarray}
where $V^{d\dag}_L= V_{\rm CKM}$ is used by rotating the phases in ${\cal M}_u$ away, which is the result of a direct interaction of the SM gauge singlet scalar field ${\cal S}=\{\chi,\tilde{\chi}\}$ with the SM quarks charged under $U(1)_X$.
The flavored-QCD axion $a$ is produced by flavor-changing neutral Yukawa interactions in Eq.(\ref{fl_in}), which leads to induced rare flavor-changing processes. The strongest bound on the QCD axion decay constant is from the flavor-changing process $K^+\rightarrow\pi^++a$\,\cite{Wilczek:1982rv, Bolton:1988af, Artamonov:2008qb, raredecay}, induced by the flavored-QCD axion $a$.
From Eq.(\ref{fl_in}), the flavored-QCD axion interactions with the flavor-violating coupling to the $s$- and $d$-quark are given by
\begin{eqnarray}
-{\cal L}^{asd}_Y\simeq \frac{i}{2}(\tilde{f}_d-\tilde{f}_s)\frac{a}{N_CF_a}\bar{s}d\,(m_d-m_s)\lambda\Big(1-\frac{\lambda^2}{2}\Big)\,.
  \label{}
\end{eqnarray}
 Then, the decay width of $K^+\rightarrow\pi^++a$ is given by
 \begin{eqnarray}
   \Gamma(K^+\rightarrow\pi^++a)=\frac{m^3_K}{16\pi}\Big(1-\frac{m^2_{\pi}}{m^2_{K}}\Big)^3\Big|\frac{\tilde{f}_d-\tilde{f}_s}{2\,F_{a}N_C}\lambda\Big(1-\frac{\lambda^2}{2}\Big)\Big|^2\,,
  \label{Gkp}
 \end{eqnarray}
 where $m_{K^{\pm}}=493.677\pm0.013$ MeV, $m_{\pi^{\pm}}=139.57061\pm0.00024$ MeV\,\cite{PDG}.
From the present experimental upper bound ${\rm Br}(K^+\rightarrow\pi^+a)<(3-6)\times10^{-11} (1\times 10^{-11})$ for $m_{a}=0-110$ (160-260) MeV at $90\%$ CL with ${\rm Br}(K^+\rightarrow\pi^+\nu\bar{\nu})=(10.6^{+4.0}_{-3.4}|_{\rm stat}\pm0.9_{\rm syst})\times10^{-11}$ at $68\%$CL\,\cite{NA62:2021zjw},
we obtain the lower limit on the QCD axion decay constant,
 \begin{eqnarray}
  F_{a}\frac{2|N_C|}{|\tilde{f}_d-\tilde{f}_s|}\gtrsim(0.86-1.90)\times10^{11}\,{\rm GeV}\,.
 \label{cons_1}
 \end{eqnarray}

The QCD axion mass $m_a$ in terms of the pion mass and pion decay constant reads\,\cite{Ahn:2014gva, Ahn:2016hbn}
 \begin{eqnarray}
 m^{2}_{a}F^{2}_{a}=m^{2}_{\pi^0}f^{2}_{\pi}F(z,w)\,,
\label{axiMass2}
 \end{eqnarray}
where $f_\pi\simeq92.1$ MeV\,\cite{PDG} and
 $F(z,w)=z/(1+z)(1+z+w)$ with $\omega=0.315\,z$.
Here the Weinberg value lies in $z\equiv m^{\overline{\rm MS}}_u(2\,{\rm GeV})/m^{\overline{\rm MS}}_d(2\,{\rm GeV})=0.47^{+0.06}_{-0.07}$\,\cite{PDG}.
After integrating out the heavy $\pi^{0}$ and $\eta$ at low energies, there is an effective low energy Lagrangian with an axion-photon coupling $g_{a\gamma\gamma}$:
${\cal L}_{a\gamma\gamma}= -g_{a\gamma\gamma}\,a\,\vec{E}\cdot\vec{B}$
where $\vec{E}$ and $\vec{B}$ are the electromagnetic field components. The axion-photon coupling is expressed in terms of the QCD axion mass, pion mass, pion decay constant, $z$ and $w$,
 \begin{eqnarray}
 g_{a\gamma\gamma}=\frac{\alpha_{\rm em}}{2\pi}\frac{m_a}{f_{\pi}m_{\pi^0}}\frac{1}{\sqrt{F(z,w)}}\left(\frac{E}{N_C}-\frac{2}{3}\,\frac{4+z+w}{1+z+w}\right)\,.
 \label{gagg}
 \end{eqnarray}
The upper bound on the axion-photon coupling, derived from the recent analysis of the horizontal branch stars in galactic globular clusters\,\cite{Ayala:2014pea}, can be translated to 
 \begin{eqnarray}
 |g_{a\gamma\gamma}|<6.6\times10^{-11}\,{\rm GeV}^{-1}\,(95\%\,{\rm CL})\Leftrightarrow F_a\gtrsim2.525\times10^{7}\Big|\frac{E}{N_C}-1.903\Big|\,{\rm GeV}\,,
 \label{gagg_1}
 \end{eqnarray}
where $z=0.47$ is used.
%%%%%%%%%%%%%%%
%   Fig A-3   %
%%%%%%%%%%%%%%%
\begin{figure}[t]
%%\vspace*{-5.0cm}
\hspace*{-0.5cm}
\begin{minipage}[h]{9.3cm}
\includegraphics[width=9.3cm]{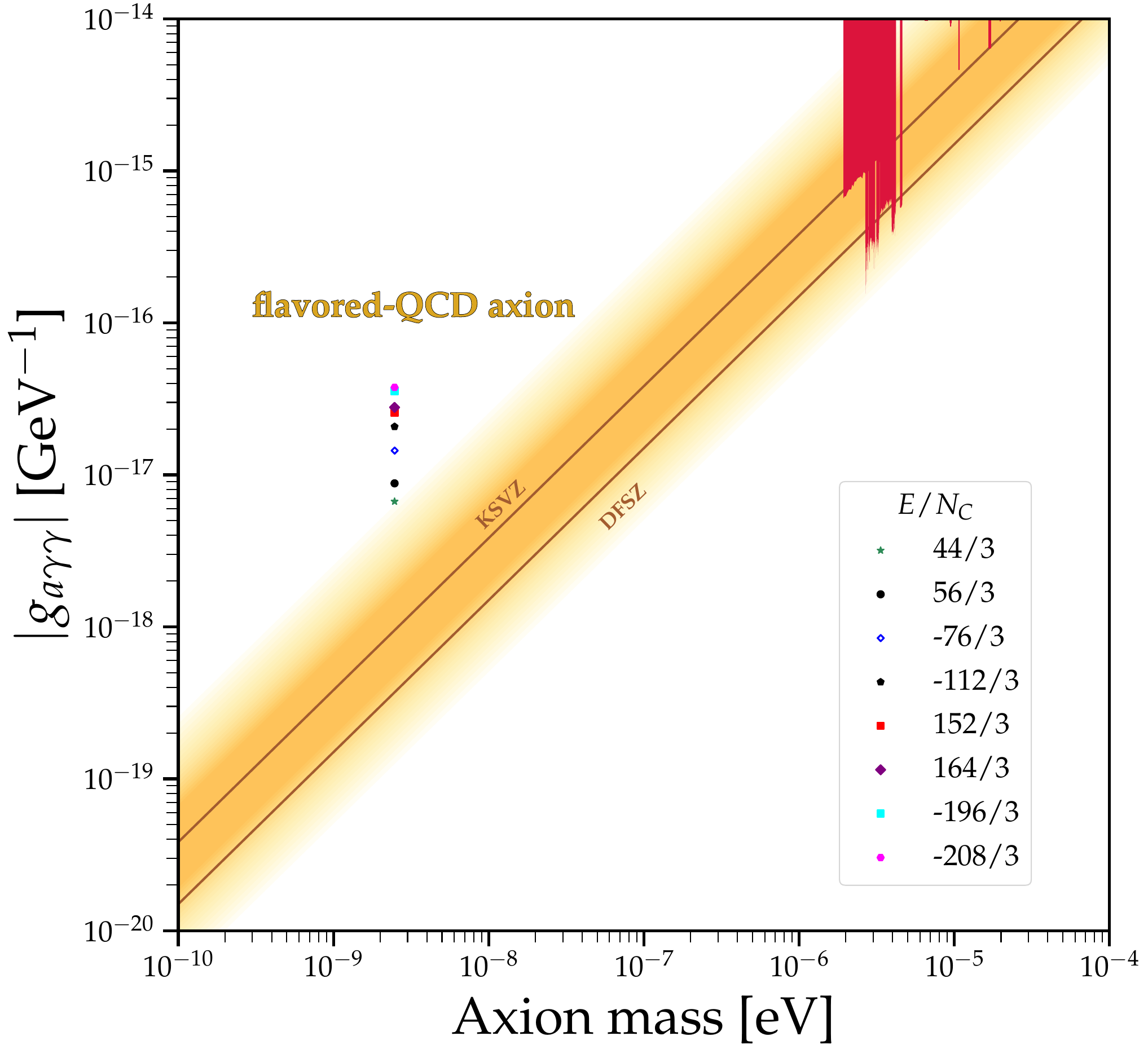}
\end{minipage}
%%\vspace*{-5.5cm}
\caption{\label{Fig3} Plot for axion-photon coupling $|g_{a\gamma\gamma}|$ as a function of the flavored-QCD axion mass $m_{a}$. Orange shaded region and vertical red lines indicate the conventional QCD axion predictions and the exclusion region of various axion search experiments, respectively, see Ref.\cite{PDG}.}
\end{figure}
Fig.\ref{Fig3} illustrates plot of the axion-photon coupling $|g_{a\gamma\gamma}|$ as a function of the flavored-QCD axion mass $m_{a}$. Each plotted point corresponds to values listed in Eq.(\ref{eano1}), using the $U(1)_X$ breaking scale from Eqs.(\ref{MR5}) and (\ref{para_sp_nu1}), and shows $m_a=2.47\times10^{-9}$ eV. This value is consistent with the experimental constraints described in Eqs.(\ref{cons_1}), (\ref{gagg_1}), and (\ref{alee0}).
 
%%%%%%%%%%%%%%%%%%%%%%
\subsection{Charged-Lepton and flavored-QCD axion}
By substituting the VEV from Eq.(\ref{vev}) into the Lagrangian (\ref{4d_a1}), the charged-lepton mass matrix given in the Lagrangian (\ref{AxionLag2}) is derived as
 \begin{eqnarray}
 {\cal M}_{\ell}&=& \tilde{D}{\left(\begin{array}{ccc}
 \alpha_{e}\Delta^{|f_e|}_\chi & \alpha_{e}\Delta^{|f_e|}_\chi s &  \alpha_{e}\Delta^{|f_e|}_\chi r \\
 \alpha_{\mu}\,\Delta^{|f_\mu|}_\chi r &  \alpha_{\mu}\,\Delta^{|f_\mu|}_\chi & \alpha_{\mu}\,\Delta^{|f_\mu|}_\chi s \\
 \alpha_{\tau}\,\Delta^{|f_\tau|}_\chi s & \alpha_{\tau}\,\Delta^{|f_\tau|}_\chi r &  \alpha_{\tau}\,\Delta^{|f_\tau|}_\chi
 \end{array}\right)}Y^{4}_{1}(1+r^3+s^3-3rs)\,v_d\,,
 \label{ChL1}
 \end{eqnarray}
where
\begin{eqnarray}
\tilde{D} &=& {\rm diag} \big(e^{i\tilde{f}_e \frac{A}{u{\chi}}}, e^{i\tilde{f}_\mu \frac{A}{u{\chi}}}, e^{i\tilde{f}_\tau \frac{A}{u{\chi}}} \big)\,.
\label{d_num}
\end{eqnarray}
 In the limit where $\langle\tau\rangle\rightarrow i$, which corresponds to $r,s\rightarrow0$, the above mass matrix can be diagonalized. Recall that the coefficients $\alpha_i$ are complex numbers with an effective absolute value that satisfies Eq.(\ref{AFN1}). 
The charged-lepton mass matrix  ${\cal M}_\ell$ generates the charged-lepton masses:
 \begin{eqnarray}
\hat{\cal M}_\ell=V^\ell_R{\cal M}_{\ell}V^{\ell\dag}_L={\rm diag}(m_e,m_\mu,m_\tau)\,,
\label{cl_ma}
\end{eqnarray}
where $V^\ell_R$ and $V^\ell_L$ are the diagonalization matrices for ${\cal M}_\ell$. Given the hierarchical nature of the charged-lepton masses, we make the reasonable assumption that $|f_{e}|\gg|f_\mu|\gg|f_\tau|$. Under this assumption, $V^\ell_L$ and $V^\ell_R$ can be obtained by diagonalizing the matrices ${\cal M}^\dag_\ell{\cal M}_\ell$ and ${\cal M}_\ell{\cal M}^\dag_\ell$, respectively.
Specifically, the mixing matrix $V^\ell_L$ becomes one of the matrices composing the Pontecorvo-Maki-Nakagawa-Sakata (PMNS) mixing matrix, while $V^\ell_R$ participates in flavor-QCD axion to lepton interactions due to Eq. (\ref{ChL1}).
At leading order, the charged-lepton mixing matrix is given by
 \begin{eqnarray}
  V^\ell_{L}\simeq
  {\left(\begin{array}{ccc}
   1-\frac{1}{2}|r|^2 & |r|e^{i\xi_{3}} & |s|e^{i\xi_{2}} \\
   -|r|e^{-i\xi_3}-|rs| & 1-\frac{1}{2}|r|^2 & |r|e^{i\xi_{1}}  \\
   -|s|e^{-i\xi_2} & -|r|e^{-i\xi_1}-|rs|e^{i(\xi_2-\xi_3)} & 1-\frac{1}{2}|r|^2
   \end{array}\right)}+{\cal O}(|r|^3,|s|^2,|r|^2|s|)\,,
 \label{clm}
 \end{eqnarray}
with the assumption that $1\gg|r|\gg|s|$, where $\xi_1\simeq\xi_3\simeq\arg(r^\ast)$ and $\xi_2\simeq\arg(s^\ast)-\frac{1}{2}\arg(r^\ast)$. An overall phase matrix is rotated away by redefinition of the left-handed charged lepton fields and omitted.
The corresponding charged-lepton masses are approximately given by
 \begin{eqnarray}
 &m_{\ell}\simeq|\alpha_{\ell}|\,|Y^{4}_{1}(1-3rs)|\Delta^{|f_\ell|}_{\chi}\,v_d\,\quad\text{for}\,\,\ell=e,\mu,\tau\,.
 \label{cLep1}
 \end{eqnarray}
These theoretical expressions match the empirical values from the PDG\,\cite{PDG}:
  \begin{eqnarray}
 m_{e}&=&0.511\,{\rm MeV}\,,\qquad m_{\mu}=105.658\,{\rm MeV}\,,\qquad
 m_{\tau}=1776.86\pm0.12\,{\rm MeV}\,.
 \label{clepmass}
 \end{eqnarray}

{\it Numerical analysis}: In the charged-lepton sector, we have three physical parameters: $m_e$, $m_\mu$, and $m_\tau$. These observables are used to determine three effective model parameters out of a total of five: $|\alpha_e|$, $|\alpha_\mu|$, $|\alpha_\tau|$, $\arg(\alpha_e)$, and $\arg(\alpha_\mu)$. Using the numerical results from Eq.(\ref{delchi}) in the quark sector, with the input values
\begin{eqnarray}
&|f_{e}|=21\,,~|f_{\mu}|=12\,,~|f_{\tau}|=7\,,\nonumber\\
&|\alpha_e|=0.533\,,~|\alpha_\mu|=0.966\,,~|\alpha_\tau|=1.193\,,~\arg(\alpha_e)=[0,2\pi]\,,~\arg(\alpha_\mu)=[0,2\pi]\,.
\label{para_sp_nu}
\end{eqnarray}
we obtain the charged-lepton masses, which agree well with the empirical values of Eq.(\ref{clepmass}), and mixing matrix $V^\ell_L$.

Flavored axion typically interacts with charged leptons (electrons, muons, taus) as discussed in\,\cite{Ahn:2014gva, Ahn:2016hbn, Ahn:2018cau, Ahn:2021ndu}. These interactions can occur through processes such as atomic axio-recombination, axio-deexcitation, axio-bremsstrahlung in electron-ion or electron-electron collisions, and Compton scattering\,\cite{Redondo:2013wwa}. The interactions between flavored-QCD axions and charged leptons can be expressed as
\begin{eqnarray}
-{\cal L}^{a\ell}&\simeq&\frac{\partial_\mu a}{2u_{\chi}}\Big\{\tilde{f}_e\,\bar{e}\gamma^\mu\gamma_5 e+\tilde{f}_\mu\,\bar{\mu}\gamma^\mu\gamma_5\mu+\tilde{f}_\tau\,\bar{\tau}\gamma^\mu\gamma_5 \tau\Big\}\nonumber\\
&+&\frac{\partial_\mu a}{2u_{\chi}}\Big\{C_{e\mu}\,\bar{\mu}\gamma^\mu\big(1+\gamma_5\big)e+C_{e\tau}\,\bar{\tau}\gamma^\mu\big(1+\gamma_5\big) e+C_{\mu\tau}\,\bar{\tau}\gamma^\mu\big(1+\gamma_5\big) \mu+{\rm h.c.}\Big\}\nonumber\\
&+&m_e\,\bar{e}e+m_\mu\,\bar{\mu}\mu+m_\tau\,\bar{\tau}\tau-\bar{\ell}i\! \! \not\! D\, \ell\,,
  \label{fla_1}
\end{eqnarray}
where 
\begin{eqnarray}
 &&C_{e\mu}=(\tilde{f}_e-\tilde{f}_\mu)e^{i\omega_1}\big|\frac{\alpha_e q_\ell}{\alpha_\mu p_\ell}\big|\Delta^{|f_e|-|f_\mu|}_\chi\,,\nonumber\\
 &&C_{e\tau}=(\tilde{f}_e-\tilde{f}_\tau)e^{i\omega_2}\big|\frac{\alpha_e q^\ast_\ell}{\alpha_\tau p_\ell}\big|\Delta^{|f_e|-|f_\tau|}_\chi\,,\nonumber\\
 &&C_{\mu\tau}=(\tilde{f}_\mu-\tilde{f}_\tau)e^{i\omega_3}\big|\frac{\alpha_\mu q_\ell}{\alpha_\tau p_\ell}\big|\Delta^{|f_\mu|-|f_\tau|}_\chi\,,
\end{eqnarray}
and the phases $\omega_1=\arg(\alpha_e\alpha^\ast_\mu q_\ell)$, $\omega_2=\arg(\alpha_e\alpha^\ast_\tau q^\ast_\ell)$, and $\omega_3=\arg(\alpha_\mu\alpha^\ast_\tau q_\ell)$ arise from $V^\ell_R$ with $p_\ell=1+|r|^2+|s|^2$ and $q_\ell=r^\ast+s+rs^\ast$ (see Eq.(\ref{vRl})).
Similar to rare neutral flavor-changing decays in particle physics, the interaction of the flavored QCD axion $a$ with leptons enables the search for the QCD axion in astroparticle physics through its effects on stellar evolution. The strongest constraint among the decay widths for the process $\ell_i \rightarrow \ell_j + a$ for $m_{\ell} \gg m_a$ comes from the branching ratio proportional to $1/F^2_a$, {\it i.e.}, ${\rm Br}(\mu \rightarrow e + a) < {\cal O}(10^{-6})$, which translates into an upper bound on the flavored QCD axion decay constant of $F_a \gtrsim {\cal O}(10^9)\,{\rm GeV}$\,\cite{Feng:1997tn}. This constraint is much weaker than the one in Eq.(\ref{cons_1}) from the quark sector.

%%%%%%%%%%%%%%%
%   Fig A-4   %
%%%%%%%%%%%%%%%
\begin{figure}[t]
%%\vspace*{-5.0cm}
\hspace*{-0.5cm}
\begin{minipage}[h]{9.3cm}
\includegraphics[width=9.3cm]{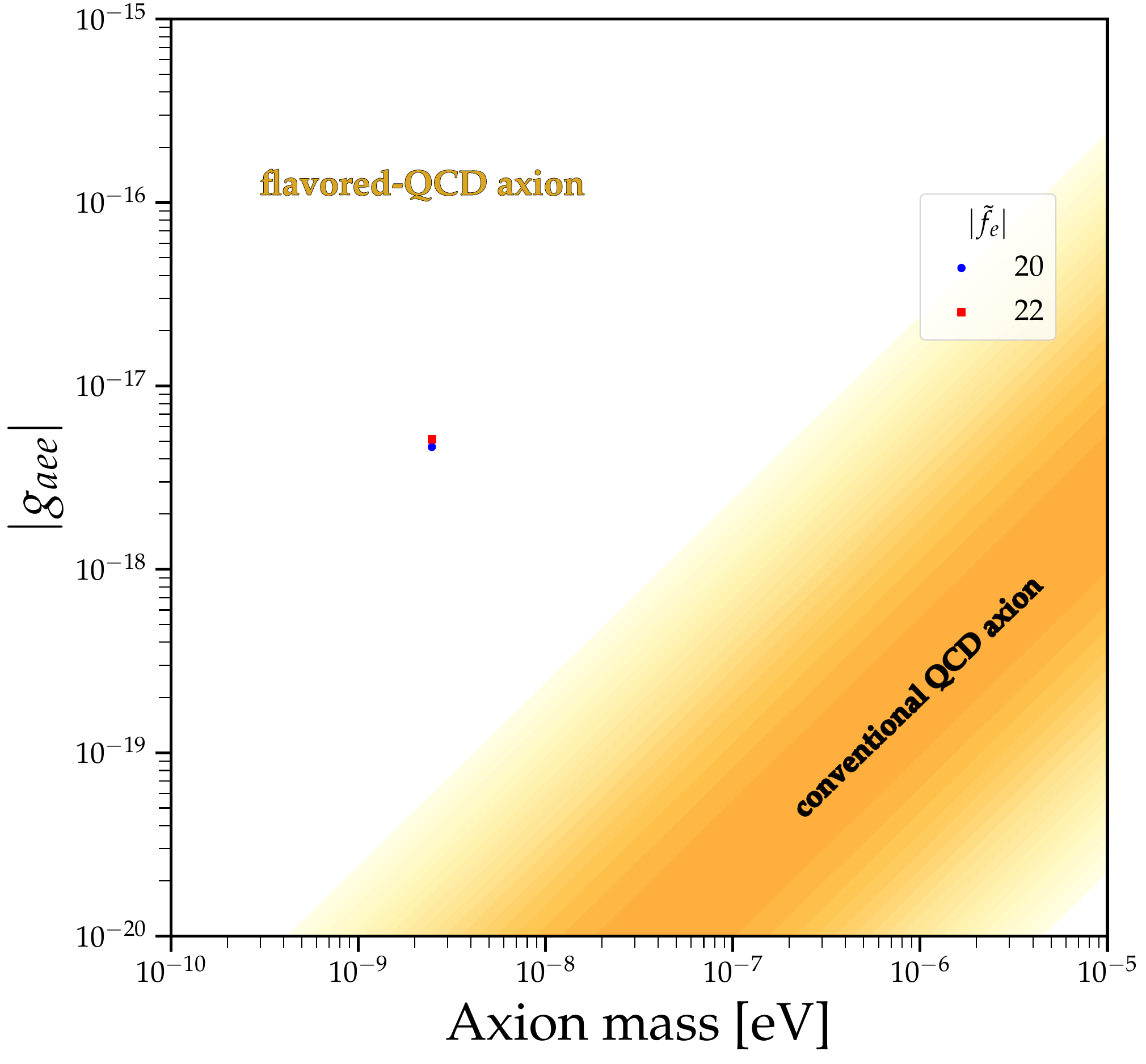}
\end{minipage}
%%\vspace*{-5.5cm}
\caption{\label{Fig4} Plot for axion-electron coupling $|g_{aee}|$ as a function of the flavored-QCD axion mass $m_{a}$. Orange shaded region indicates the conventional QCD axion predictions, see Ref.\cite{PDG}.}
\end{figure}
The flavored-QCD axion coupling to electrons is given by
\begin{eqnarray}
g_{aee}&=&|\tilde{f}_e|\,\frac{m_e}{u_\chi}\,,
\label{gaTe}
\end{eqnarray}
where $|\tilde{f}_e|=20$ or $22$ considering Eqs.(\ref{para_sp_nu}) and (\ref{4d_f1}). Fig.\,\ref{Fig4} shows axion-electron coupling $|g_{aee}|\sim5\times10^{-18}$ as a function of the flavored-QCD axion mass $m_{a}$ for $|\tilde{f}_e|=20$ (blue spot) or $22$ (red rectangle), considering the $U(1)_X$ breakdown scale in Eqs.(\ref{MR5}) and (\ref{para_sp_nu1}).
Stars in the red giant branch (RGB) of color-magnitude diagrams in globular clusters provide strict constraints on axion-electron couplings, which leads to a lower bound on the axion decay constant. This constraint is expressed as\,\cite{Viaux:2013lha}
\begin{eqnarray}
 |g_{aee}|<4.3\times10^{-13}\quad(95\%\,{\rm CL})~\Leftrightarrow N_CF_a\gtrsim1.19|\tilde{f}_e|\times10^9\,{\rm GeV}\,.
   \label{alee0}
\end{eqnarray}
Bremsstrahlung off electrons ($e+Ze\rightarrow Ze+e+a$) in white dwarfs (WDs) is an effective process for detecting axions, as the Primakoff and Compton processes are suppressed due to the large plasma frequency. Comparing the theoretical and observed WD luminosity functions (WDLFs) provides a way to place limits on\,\footnote{Note that Refs.\,\cite{WD01,Bertolami:2014wua} have pointed out features in some WDLFs\,\cite{DeGennaro:2007yw, Rowell:2011wp} that could imply axion-electron couplings in the range $7.2\times10^{-14}\lesssim|g_{aee}|\lesssim2.2\times10^{-13}$.}  on $|g_{aee}|$\,\cite{Raffelt:1985nj}. Recent analyses of WDLFs, using detailed WD cooling treatment and new data on the WDLF of the Galactic disk, suggest electron couplings $|g_{aee}|\lesssim2.8\times10^{-13}$\,\cite{Bertolami:2014wua}. However, these results come with large theoretical and observational uncertainties.

%%%%%%%%%%%%%%%%%%%%%%
\subsection{Neutrino and flavored-QCD axion}
In our framework, the spontaneous $U(1)_X$ breaking generates the light neutrino operator with the cutoff set by the $U(1)_X$ breaking scale, as shown in Eq.(\ref{4d_a2}), through the normalization of modular forms in Eq.(\ref{4d_b}). After EW symmetry breaking, the light neutrino operator with the cutoff scale $\langle\chi\rangle$ in Eq.(\ref{4d_a2}) generates the neutrino masses. The light neutrino mass matrix ${\cal M}_\nu$ given in the Lagrangian\,(\ref{AxionLag2})
is derived from Eq.(\ref{4d_a2}) as
\begin{eqnarray}
 {\cal M}_{\nu}&=&Y^2_1\,\small\Big[ \gamma_1(1+2rs){\left(\begin{array}{ccc}
1 & 0 &  0 \\
0 & 0 & 1  \\
0 & 1  & 0
 \end{array}\right)}+\gamma_2(r^2+2s){\left(\begin{array}{ccc}
0 & 1 &  0 \\
1 & 0 & 0  \\
0 & 0  & 1
 \end{array}\right)}+\gamma_3(s^2+2r){\left(\begin{array}{ccc}
0 & 0 &  1 \\
0 & 1 & 0  \\
1 & 0  & 0
 \end{array}\right)}\nonumber\\
 &&+\gamma_4{\left(\begin{array}{ccc}
1 & s &  r \\
s & s^2 & rs  \\
r & rs  & r^2
 \end{array}\right)}+\gamma_5{\left(\begin{array}{ccc}
rs & \frac{1}{2}(r^2+s) &  \frac{1}{2}(s^2+r) \\
\frac{1}{2}(r^2+s) & r & \frac{1}{2}(1+rs)  \\
\frac{1}{2}(s^2+r) & \frac{1}{2}(1+rs)  & s
 \end{array}\right)}\nonumber\\
 &&+\small\frac{\gamma_6}{3}{\left(\begin{array}{ccc}
4+2rs & -s-2r^2 & -r-2s^2  \\
-s-2r^2 & s^2-4r & 1+5rs \\
-r-2s^2 & 1+5rs  & r^2-4s
 \end{array}\right)}\Big]\frac{v^2_u}{\langle\chi\rangle}=U^{\ast}_{\nu}{\rm diag}(m_{\nu_1}, m_{\nu_2}, m_{\nu_3})U^{\dag}_{\nu}\,,
 \label{neut2}
  \end{eqnarray}
where $U_{\nu}$ is the rotation matrix diagonalizing ${\cal M}_{\nu}$ and $m_{\nu_i}$ ($i=1,2,3$) are the light neutrino masses. 
The matrix terms in Eq.(\ref{neut2}) are derived by taking $A_4$ singlet and triplet combinations $\gamma_1(Y^{(2)}_{\bf 3}Y^{(2)}_{\bf 3})_{\bf 1}(LL)_{\bf 1}+\gamma_2(Y^{(2)}_{\bf 3}Y^{(2)}_{\bf 3})_{{\bf 1}'}(LL)_{{\bf 1}''}+\gamma_3(Y^{(2)}_{\bf 3}Y^{(2)}_{\bf 3})_{{\bf 1}''}(LL)_{{\bf 1}'}+\gamma_4(Y^{(2)}_{\bf 3}L)_{{\bf 1}}(Y^{(2)}_{\bf 3}L)_{{\bf 1}}+\gamma_5(Y^{(2)}_{\bf 3}L)_{{\bf 1}'}(Y^{(2)}_{\bf 3}L)_{{\bf 1}''}+\gamma_6[(Y^{(2)}_{\bf 3}L)_{\bf 3s}(Y^{(2)}_{\bf 3}L)_{\bf 3s}]_{\bf 1}$.
Eq.(\ref{neut2}) has  6 unknown complex parameters, $\gamma_i$, where one complex parameter contributes as an overall factor. Other variables such as $r, s, Y_1$  are determined from the analysis for the quark and charged-lepton sectors. 
Once $\langle\chi\rangle$ is determined by neutrino experimental bounds, for instance $\langle\chi\rangle\simeq1.1\times10^{15}$ GeV for $m_{\nu_3}\sim0.05$ eV, the 5D Planck mass $\Lambda_5$ is fixed as $\simeq1.8\times10^{15}$ GeV for $\langle\chi\rangle/\Lambda_5\simeq0.6$ by Eq.(\ref{AFN1}) through quark and charged lepton flavor physics. And the soft SUSY-breaking mass $m_{3/2}$ in Eq.(\ref{sof2}) can be estimated, for instance, $(2.6\sim8.2)\times10^8$ GeV for $\langle\chi\rangle\simeq10^{15}$ GeV.

From Eq.(\ref{AxionLag2}) with Eqs.(\ref{clm}) and (\ref{neut2}), the PMNS mixing matrix becomes
  \begin{eqnarray}
 U_{\rm PMNS}=V^\ell_LU^\dag_\nu .
   \label{pmns0}
 \end{eqnarray}
 The matrix $U_{\rm PMNS}$ is expressed in terms of three mixing angles, $\theta_{12}, \theta_{13}, \theta_{23}$, and a Dirac type \cp ~violaitng phase $\delta_{CP}$ and two additional \cp~ violating phases $\varphi_{1,2}$ if light neutrinos are Majorana particle as\,\cite{PDG}
 \begin{eqnarray}
  U_{\rm PMNS}=
  {\left(\begin{array}{ccc}
   c_{13}c_{12} & c_{13}s_{12} & s_{13}e^{-i\delta_{CP}} \\
   -c_{23}s_{12}-s_{23}c_{12}s_{13}e^{i\delta_{CP}} & c_{23}c_{12}-s_{23}s_{12}s_{13}e^{i\delta_{CP}} & s_{23}c_{13}  \\
   s_{23}s_{12}-c_{23}c_{12}s_{13}e^{i\delta_{CP}} & -s_{23}c_{12}-c_{23}s_{12}s_{13}e^{i\delta_{CP}} & c_{23}c_{13}
   \end{array}\right)}Q_{\nu}\,,
 \label{PMNS1}
 \end{eqnarray}
where $s_{ij}\equiv \sin\theta_{ij}$, $c_{ij}\equiv \cos\theta_{ij}$ and $Q_{\nu}={\rm Diag}(e^{-i\varphi_1/2}, e^{-i\varphi_2/2},1)$.

The observed hierarchy $|\Delta m^{2}_{\rm Atm}|= |m^{2}_{\nu_3}-(m^{2}_{\nu_1}+m^{2}_{\nu_2})/2|\gg\Delta m^{2}_{\rm Sol}\equiv m^{2}_{\nu_2}-m^{2}_{\nu_1}>0$ and the requirement of a Mikheyev-Smirnov-Wolfenstein resonance\,\cite{Wolfenstein:1977ue} for solar neutrinos lead to two possible neutrino mass spectra: normal mass ordering (NO) $m^2_{\nu_1}<m^2_{\nu_2}<m^2_{\nu_3}$ and inverted mass ordering (IO) $m^2_{\nu_3}<m^2_{\nu_1}<m^2_{\nu_2}$. Nine physical observables can be derived from Eqs.(\ref{PMNS1}) and (\ref{neut2}): $\theta_{23}$, $\theta_{13}$, $\theta_{12}$, $\delta_{CP}$, $\varphi_1$, $\varphi_2$, $m_{\nu_1}$, $m_{\nu_2}$, and $m_{\nu_3}$. 
\begin{table}[h]
%\begin{widetext}
%\begin{center}
\caption{\label{exp_nu} The global fit of three-flavor oscillation parameters at the best-fit and $3\sigma$ level with Super-Kamiokande atmospheric data\,\cite{Esteban:2020cvm}. NO = normal neutrino mass ordering; IO = inverted mass ordering. And $\Delta m^{2}_{\rm Sol}\equiv m^{2}_{\nu_2}-m^{2}_{\nu_1}$, $\Delta m^{2}_{\rm Atm}\equiv m^{2}_{\nu_3}-m^{2}_{\nu_1}$ for NO, and  $\Delta m^{2}_{\rm Atm}\equiv m^{2}_{\nu_2}-m^{2}_{\nu_3}$ for IO.}
\begin{ruledtabular}
\begin{tabular}{cccccccccccc} &$\theta_{13}[^{\circ}]$&$\delta_{CP}[^{\circ}]$&$\theta_{12}[^{\circ}]$&$\theta_{23}[^{\circ}]$&$\Delta m^{2}_{\rm Sol}[10^{-5}{\rm eV}^{2}]$&$\Delta m^{2}_{\rm Atm}[10^{-3}{\rm eV}^{2}]$\\
\hline
$\begin{array}{ll}
\hbox{NO}\\
\hbox{IO}
\end{array}$&$\begin{array}{ll}
8.58^{+0.33}_{-0.35} \\
8.57^{+0.37}_{-0.34}
\end{array}$&$\begin{array}{ll}
232^{+118}_{-88} \\
276^{+68}_{-82}
\end{array}$&$33.41^{+2.33}_{-2.10}$&$\begin{array}{ll}
42.2^{+8.8}_{-2.5} \\
49.0^{+2.5}_{-9.1}
\end{array}$&$7.41^{+0.62}_{-0.59}$
 &$\begin{array}{ll}
2.507^{+0.083}_{-0.080} \\
2.486^{+0.084}_{-0.080}
\end{array}$ \\
\end{tabular}
\end{ruledtabular}
%\end{center}
%\end{widetext}
\end{table}
Recent global fits\,\cite{Esteban:2018azc, deSalas:2017kay, Capozzi:2018ubv} of neutrino oscillations have enabled a more precise determination of the mixing angles and mass squared differences, with large uncertainties remaining for $\theta_{23}$ and $\delta_{CP}$ at 3$\sigma$. The most recent analysis\,\cite{Esteban:2020cvm} lists global fit values and $3\sigma$ intervals for these parameters in Table-\ref{exp_nu}.
Furthermore, recent constraints on the rate of $0\nu\beta\beta$ decay have added to these findings. Specifically, the most tight upper bounds for the effective Majorana mass (${\cal M}_{\nu})_{ee}$, which is the modulus of the $ee$-entry of the effective neutrino mass matrix, are given by
  \begin{eqnarray}
 ({\cal M}_{\nu})_{ee}< 0.036-0.156\,{\rm eV}\,~ (^{136}\text{Xe-based experiment\,\cite{KamLAND-Zen:2022tow}})
 \label{nubb}
 \end{eqnarray}
at $90\%$ confidence level.
$0\nu\beta\beta$ decay is a low-energy probe of lepton number violation and its measurement could provide the strongest evidence for lepton number violation at high energy. Its discovery would suggest the Majorana nature of neutrinos.
 
Transforming the neutrino fields by chiral rotations of Eq.(\ref{X-tr}) under $U(1)_X$ we obtain the flavored-QCD axion interactions to neutrinos
 \begin{eqnarray}
  -{\cal L}^{a\nu} &\simeq&\frac{1}{2}\overline{\nu^c_L}{\cal M}_\nu\,\nu_L +{\rm h.c.}-\frac{1}{2}\overline{\nu}\,i\! \! \not\!\partial\nu 
    +\frac{\partial a}{2u_{\chi}}\big(\frac{1}{2}\overline{\nu^c}\gamma_\mu\gamma_{5}\nu\big)\,.
  \label{neut1}
 \end{eqnarray}
Given that the light neutrino mass is less than 0.1 eV, the coupling between the flavored-QCD axion and light neutrinos is stringently constrained by Eqs.(\ref{MR5}) and (\ref{cons_1}), which significantly suppresses the interaction. Therefore, we will not take it into consideration. Ref.\cite{Kharusi:2021jez} provides the latest experimental constraints on Majoron-neutrino coupling, which are below the range of $(0.4-0.9)\times10^{-5}$.
 %%%%%%%%%%%%%%%
%   Fig A-5   %
%%%%%%%%%%%%%%%
\begin{figure}[t]
%\vspace*{-5.0cm}
%\hspace*{-1cm}
\begin{minipage}[t]{8.0cm}
\includegraphics[width=8.0cm]{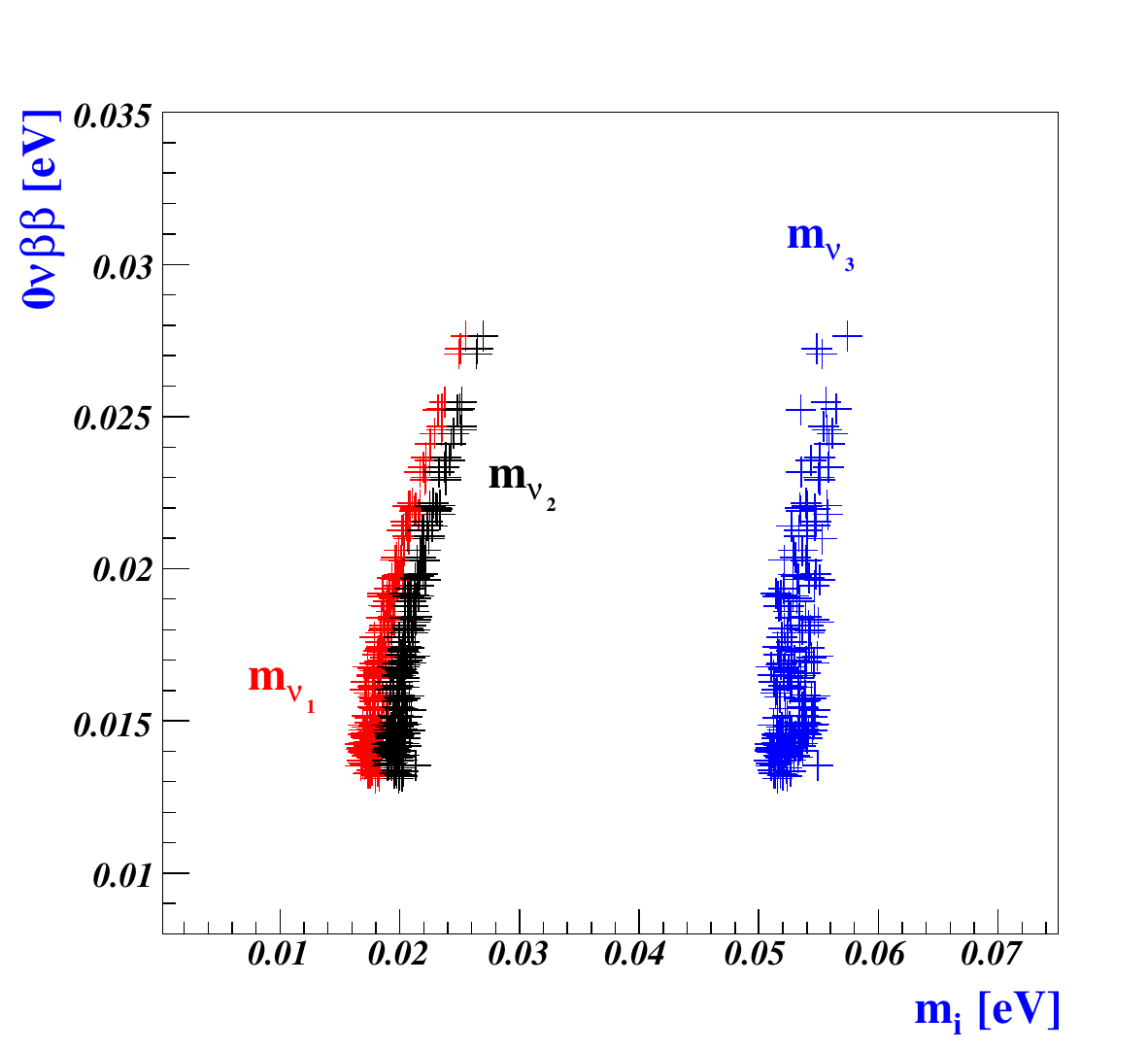}
\end{minipage}
%\hspace*{1.0cm}
\begin{minipage}[t]{8.0cm}
\includegraphics[width=8.0cm]{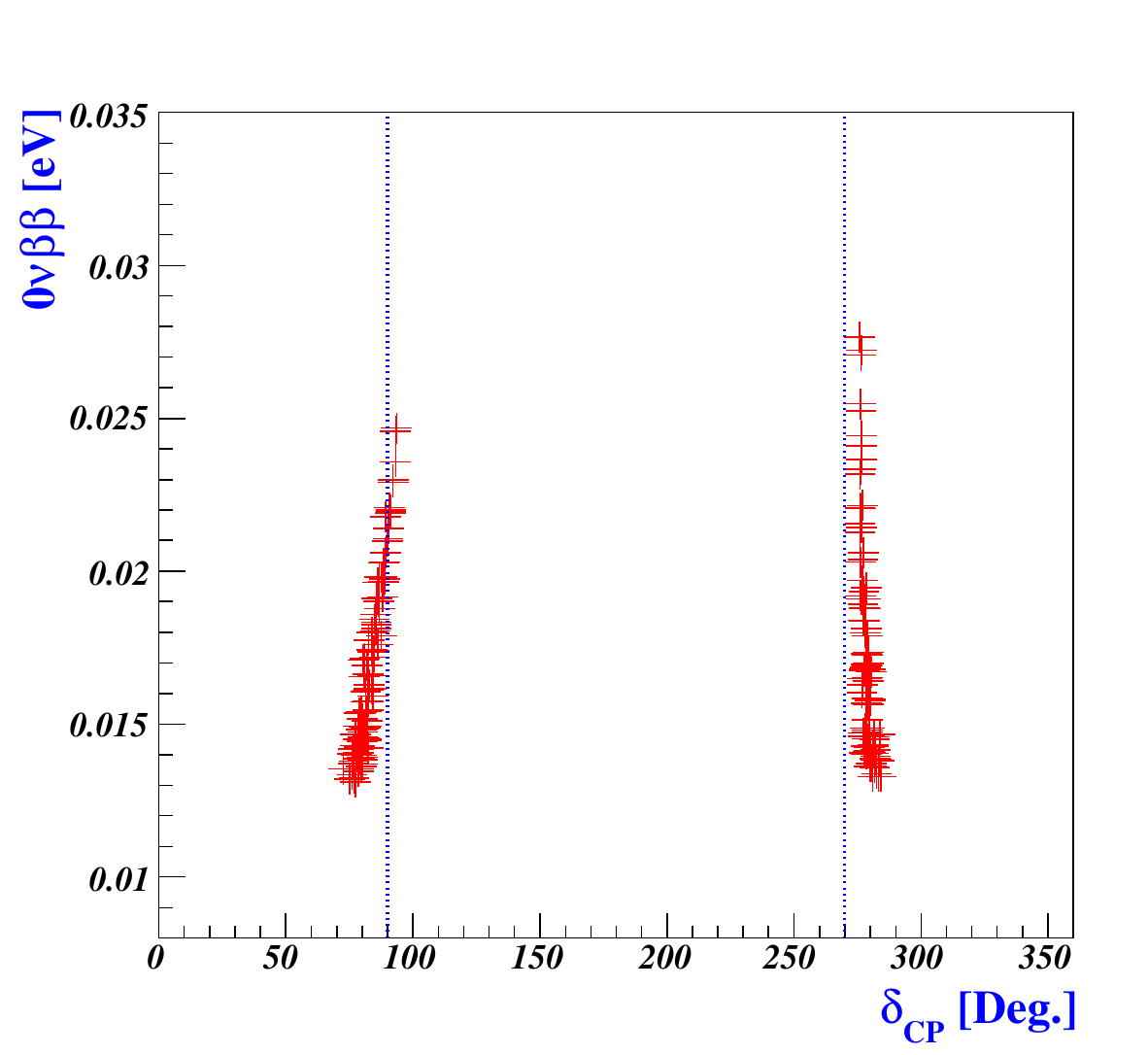}
\end{minipage}\\
\begin{minipage}[t]{8.0cm}
\includegraphics[width=8.0cm]{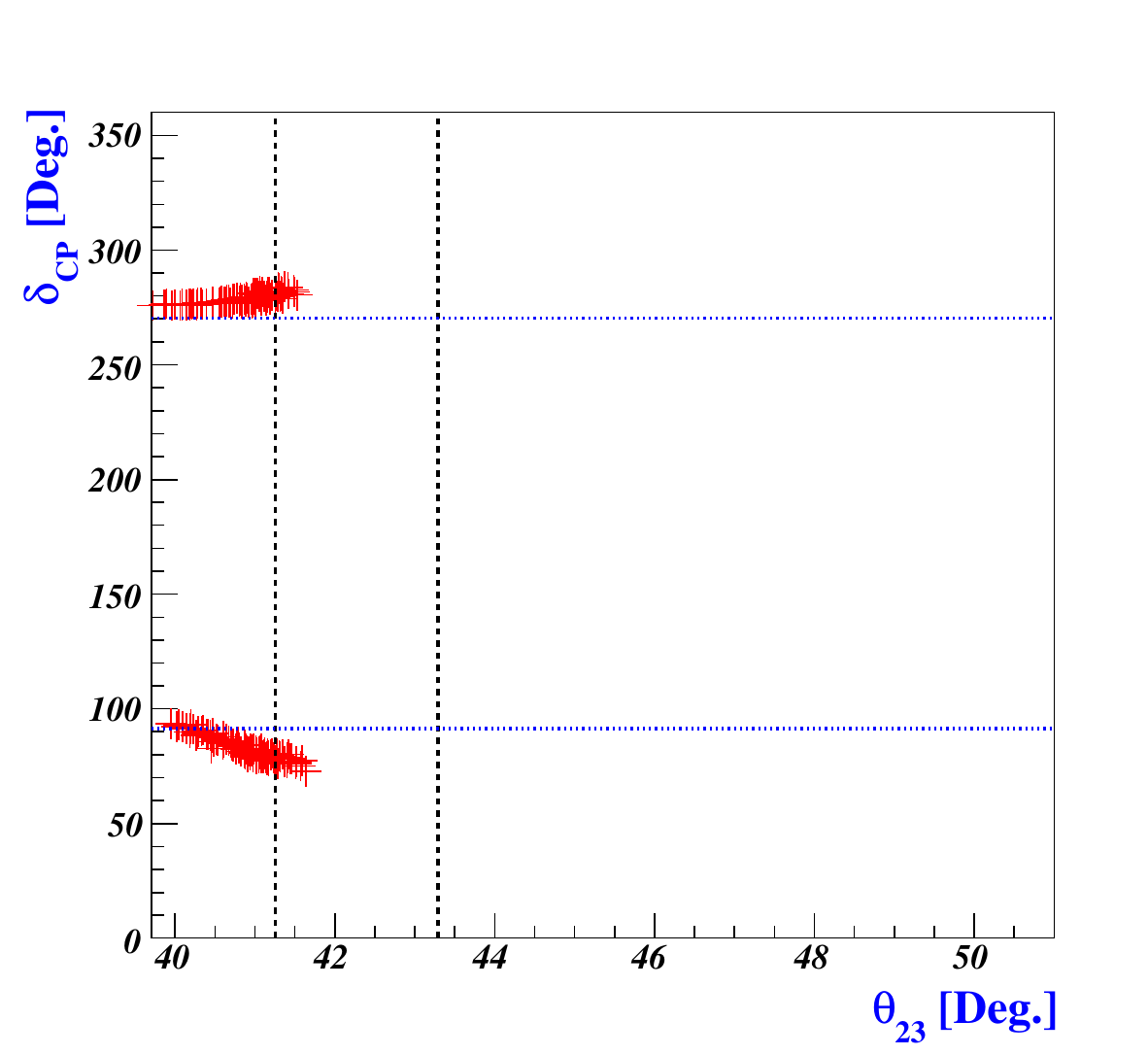}
\end{minipage}
%\hspace*{1.0cm}
\begin{minipage}[t]{8.0cm}
\includegraphics[width=8.0cm]{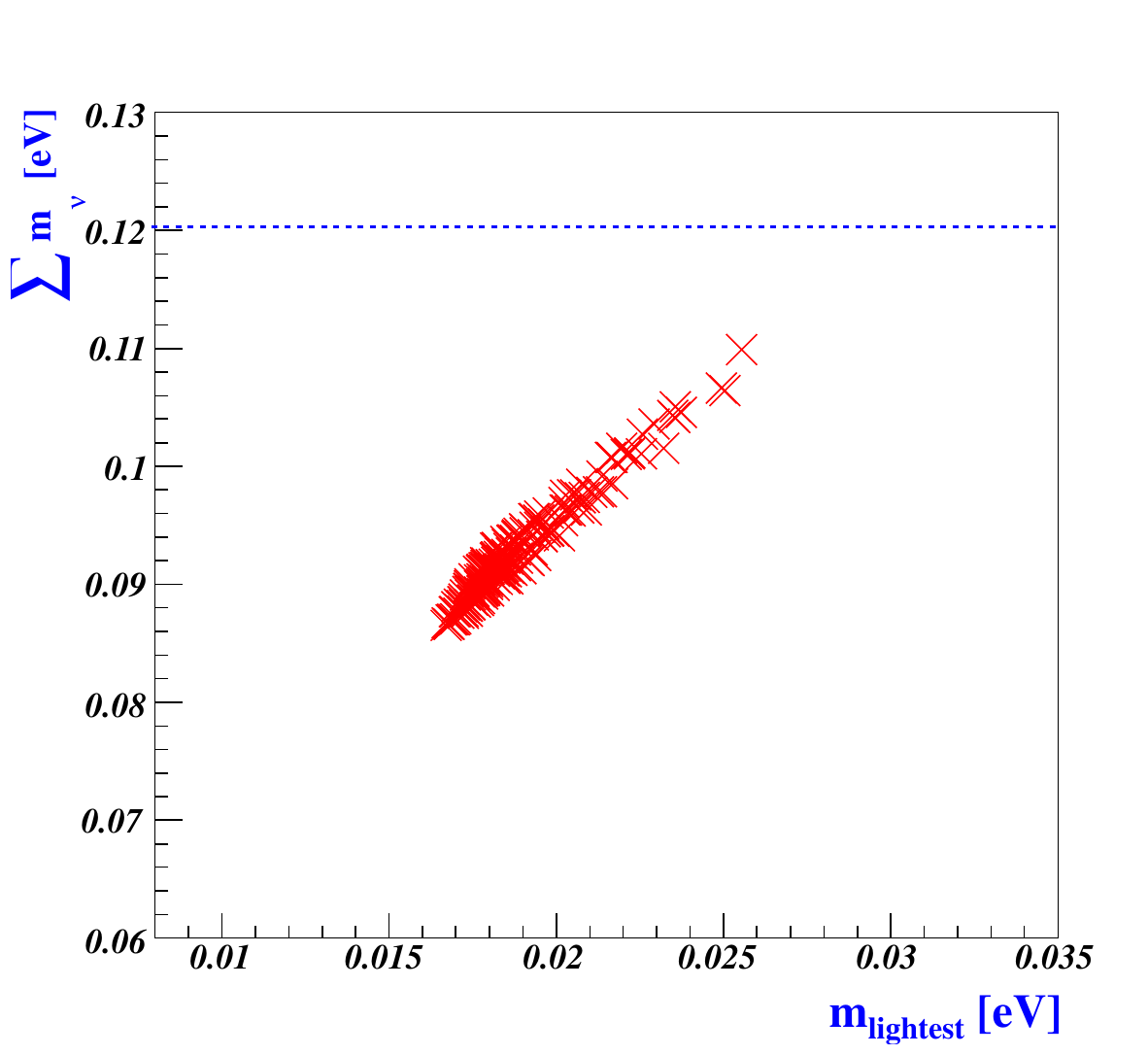}
\end{minipage}
%\vspace*{-5.5cm}
\caption{\label{Fig5} Plots for $0\nu\beta\beta$-decay rate as a function of the neutrino masses $m_{\nu_i}$ (upper left panel), $0\nu\beta\beta$-decay rate as a function of leptonic Dirac CP phase $\delta_{CP}$ (upper right panel), leptonic Dirac CP phase $\delta_{CP}$ as a function of atmospheric mixing angle $\theta_{23}$ (lower left panel), and the sum of neutrino mass as a function of the lightest neutrino mass (lower right panel). Vertical dashed lines (lower left panel) represent the $1\sigma$ bounds for $\theta_{23}$, in Table\,\ref{exp_nu}.}
\end{figure}

{\it Numerical analysis} : Using the linear algebra tools from Ref.\cite{Antusch:2005gp}, we analyze the neutrino sector, where we consider nine physical parameters: $m_{\nu_1}$, $m_{\nu_2}$, $m_{\nu_3}$, $\theta_{13}$, $\theta_{12}$, $\theta_{23}$, $\delta_{CP}$, $\varphi_1$, and $\varphi_2$. These observables are used to determine nine effective model parameters out of a total of thirteen: $\langle\chi\rangle$, $|\gamma_1|$, $|\gamma_2|$, $|\gamma_3|$, $|\gamma_4|$, $|\gamma_5|$, $|\gamma_6|$, $\arg(\gamma_1)$, $\arg(\gamma_2)$, $\arg(\gamma_3)$, $\arg(\gamma_4)$, $\arg(\gamma_5)$, and $\arg(\gamma_6)$.
The neutrino mass generator in Eq.(\ref{4d_a2}) operates at the $U(1)_X$ symmetry breaking scale, while its implications are measured by experiments below the EW scale. Here, we will ignore quantum corrections to neutrino masses and mixing angles since our numerical results show a normal hierarchical neutrino mass spectrum, as shown in Fig.\,\ref{Fig5}. Neutrino oscillation experiments currently aim to make precise measurements of the Dirac CP-violating phase $\delta_{CP}$ and atmospheric mixing angle $\theta_{23}$. To explore the parameter spaces, we scan the precision constraints $\{\theta_{13}, \theta_{23}, \theta_{12}, \Delta m^2_{\rm Sol}, \Delta m^2_{\rm Atm}\}$ at $3\sigma$ from Table\,\ref{exp_nu}. Using the references values of Eqs.(\ref{delchi}) and (\ref{para_sp_nu}) in the quark and charged-lepton sectors, we determine the input parameter spaces 
\begin{eqnarray}
&\langle\chi\rangle=1.1\times10^{15}\,{\rm GeV}\nonumber\\
&|\gamma_1|=[0.75, 1.30]\,,~|\gamma_2|=[0.43, 0.80]\,,~|\gamma_3|=[0.80,1.30]\,,\nonumber\\
&|\gamma_4|=[0.43, 0.75]\,,~|\gamma_5|=[0.50, 1.00]\,,~|\gamma_6|=[0.60, 1,30]\,,\nonumber\\
&\arg(\gamma_1)=[1.70, 2.95]\,,~\arg(\gamma_2)=[0, 1.50]\,,~\arg(\gamma_3)=[1.80, 2.42]\,,\nonumber\\
&\arg(\gamma_4)=[4.90, 6.00]\,,~\arg(\gamma_5)=[1.91, 1.80]\,,~\arg(\gamma_6)=[4.70, 5.30]\,.
\label{para_sp_nu1}
\end{eqnarray}
We find numerically that only the normal mass hierarchy is allowed.
In Fig.\,\ref{Fig5}, the upper panel shows the $0\nu\beta\beta$-decay rate as functions of the neutrino masses (left) and the Dirac CP phase (right). The lower panel shows the Dirac CP phase as a function of $\theta_{23}$ (left) and the sum of neutrino masses as a function of the lightest neutrino mass (right). The upper bound on the sum of the three active neutrino masses can be summarized as $\sum m_{\nu}=m_{\nu_1}+m_{\nu_2}+m_{\nu_3}<0.120$ eV at $95\%$ CL for TT, TE, EE+lowE+lensing+BAO\,\cite{Planck:2018vyg}. According to the $3\sigma$ allowed regions outlined in Table\,\ref{exp_nu}, our model most favorably aligns with $\delta_{CP} \sim 280^\circ$ and $\theta_{23} \sim 41.5^\circ$. Ongoing experiments like DUNE\,\cite{DUNE:2018tke} and proposed next-generation experiments such as Hyper-K\,\cite{Hyper-Kamiokande:2018ofw} are anticipated to significantly reduce uncertainties in the values of $\theta_{23}$ and $\delta_{CP}$, providing a rigorous test of our model.

 %%%%%%%%%%%%%%%%%%%%%%%%%%%%%%%%%%%%%%%%%%%%%%%%%%%%%%%%%%%%%%%%%%%%%%%%%%%
\section{Conclusion}
We have proposed a unified Standard Model (SM) framework featuring the SM fields on two 3-branes within an extra-dimensional setup. This model incorporates $G_F = U(1)_X \times \Gamma_N$ symmetry, with a modulus and a scalar field responsible for symmetry breaking. 
In a slice of AdS$_5$, bulk fermions propagate in a 5-dimensional space. These fermions, singlets under $SU(2)$ with hypercharge $Y_f$ and masses $M^f_i$, interact with ordinary matter fields confined to the branes at $y=0$ and $y=L$. The bulk fermions can exchange information, such as the breaking of flavor symmetry $G_F$ and the quantum numbers of SM fields, between the two branes. To conserve charge under $G_{\rm SM} \times G_F$, two types of $SU(2)$ singlet bulk fermions are introduced: bulk fermions and their mirror counterparts. 

Upon compactification to four dimensions, the Yukawa couplings, initially expressed as modular forms, are normalized to conform to the canonical 4D theory, with the Yukawa coefficients being complex numbers of unit absolute value. We have demonstrated that this framework provides a natural explanation for the mass and mixing hierarchies of quarks and leptons, addresses the strong CP problem, and inherently satisfies the domain-wall number condition $N_{\rm DW} = 1$ due to the presence of an additional scalar field ${\cal S}$ charged under $U(1)_X$ in the $y=0$ brane operators. Additionally, the Higgs mass parameter, if radiatively generated at the electroweak scale, remains invariant under the rescaling of dimensionful parameters, potentially offering a natural solution to the hierarchy problem.

In the absence of right-handed neutrinos in the SM, no corresponding right-handed neutrino exists at the $y=0$ brane. To satisfy the $U(1)_X$ mixed gravitational anomaly-free condition, electrically neutral mirror bulk fermions must couple to the normal neutrino field on the 3-brane, while electrically neutral bulk fermion can couple to itself with the scalar field ${\cal S}$, facilitating a mechanism for generating light neutrino masses. In AdS$_5$, the scalar mass $\mu_{\cal S}$ is near the 5D cutoff scale, as expected for a scalar field. This scalar is responsible for the spontaneous breaking of the $U(1)_X$ flavor symmetry at a high energy scale $\Lambda_5$, resulting in a flavored-QCD axion that interacts with ordinary quarks and leptons via Yukawa interactions. 
We have shown that the scale of $U(1)_X$ breaking, which can be interpreted as the flavored-QCD axion decay constant, is determined by experimental bounds on neutrinos. This leads to $\langle\chi\rangle \sim 10^{15}$ GeV, which in turn implies a QCD axion mass of $m_a \simeq 2.5 \times 10^{-9}$ eV.

We explored numerical values of physical parameters that align with the highly precise experimental data on the masses of quarks and charged leptons, as well as the quark mixing angles, with the exception of the quark Dirac CP phase. Our model predicts that the value of $\delta^q_{CP}$ falls within the range of $64^\circ$ to $85^\circ$, which is consistent with current experimental observations. 
Using precise neutrino oscillation data as constraints, we found that only the normal mass hierarchy is numerically viable within our model. We investigated how the $0\nu\beta\beta$-decay rate and the Dirac CP phase could be determined in the neutrino sector. According to the $3\sigma$ allowed regions outlined in Ref.\cite{Esteban:2020cvm}, our model most favorably aligns with $\delta_{CP} \sim 280^\circ$ and $\theta_{23} \sim 41.5^\circ$. Ongoing experiments like DUNE\,\cite{DUNE:2018tke} and proposed next-generation experiments such as Hyper-K\,\cite{Hyper-Kamiokande:2018ofw} are anticipated to significantly reduce uncertainties in the values of $\theta_{23}$ and $\delta_{CP}$, providing a rigorous test of our model.

%%%%%%%%%%%%%%%%%%%%%%%%%%%%%%%%%%%%%%%%%%%%%%%%%%%%%%%%%%%%%%%%%%%%%%%%%%%%%%%%%%%%%%%%%%%%%%%%%%%%%%%%%%%
\acknowledgments{}
We are grateful to YaDong Yang and Eung Jin Chun for useful comments and discussions. This work is supported by  NSFC 12135006.
%\newpage
\appendix
%%%%%%%%%%%%%%%%%%%%%%%%%%%%%%%%%%%%%%%%%%%%%%%%%%%%%%%%%%%%%%%%%%%%%%%%%%%
\section{The group $A_4$ and the modular forms}
\label{A4_i}
We shortly review the modular symmetry. The infinite groups $\Gamma(N)$, called principal congruence subgroups of level $N=1,2,3,...$, are defined by
{\begin{eqnarray}
 \Gamma(N)=\Big\{\begin{pmatrix} a & b  \\ c & d  \end{pmatrix}\in SL(2, Z), \begin{pmatrix} a & b  \\ c & d  \end{pmatrix}=\begin{pmatrix} 1 & 0  \\ 0 & 1  \end{pmatrix}\quad(\text{mod}~N)\Big\}\,,
\end{eqnarray}}
which are normal subgroups of homogeneous modular group $\Gamma\equiv\Gamma(1)\simeq SL(2,Z)$, where $SL(2, Z)$ is the group of $2\times2$ matrices with integer entries and determinant equal to one. The finite modular groups are defined by the quotient $\Gamma_N\equiv\bar{\Gamma}/\bar{\Gamma}(N)$. The groups $\Gamma_N$ are isomorphic to the permutation groups $S_3$, $A_4$, $S_4$, and $A_5$ for $N = 2, 3, 4, 5$, respectively\,\cite{deAdelhartToorop:2011re}.

For our purpose, we take into account $\Gamma(3)$ modular symmetry, which gives the modular forms of level 3. The group $\Gamma_3$ is isomorphic to $A_4$ which is the symmetry group of the tetrahedron and the finite groups of the even permutation of four objects having four irreducible representations. The group $A_4$ has two generators, denoted $S$ and $T$, satisfying the relations $S^2=T^3=(ST)^3={\bf 1}$. Its irreducible representations are three singlets ${\bf 1}, {\bf 1}'$, and ${\bf 1}''$ and one triplet ${\bf 3}$ with the multiplication rules ${\mathbf3}\otimes{\mathbf3}={\mathbf3}_{s}\oplus{\mathbf3}_{a}\oplus{\mathbf1}\oplus{\mathbf1}'\oplus{\mathbf1}''$ and ${\mathbf1}'\oplus{\mathbf1}'={\mathbf1}''$, where the subscripts $s$ and $a$ denote symmetric and antisymmetric combinations respectively. Let $(a_1, a_2, a_3)$ and $(b_1, b_2, b_3)$ denote the basis vectors for two ${\mathbf3}$'s. Then we have 
{\begin{eqnarray}
 (a\otimes b)_{{\mathbf 3}_s}&=&\frac{1}{\sqrt{3}}\big(2a_1b_1-a_2b_3-a_3b_2, ~2a_3b_3-a_1b_2-a_2b_1, ~2a_2b_2-a_3b_1-a_1b_3\big)\,,\nonumber\\
 (a\otimes b)_{{\mathbf 3}_a}&=&\big(a_2b_3-a_3b_2, ~a_1b_2-a_2b_1, ~a_3b_1-a_1b_3\big)\,,\nonumber\\
 (a\otimes b)_{\mathbf 1}&=& a_1b_1+a_2b_3+a_3b_2\,,\nonumber\\
 (a\otimes b)_{{\mathbf 1}'}&=& a_3b_3+a_1b_2+a_2b_1\,,\nonumber\\
  (a\otimes b)_{{\mathbf 1}''}&=& a_2b_2+a_3b_1+a_1b_3\,.
  \label{A4x}
\end{eqnarray}} 
The modular forms $f(\tau)$ of level $3$ and weight $k$, such as Eq.(\ref{mof}), are holomorphic functions of the complex variable $\tau$ with well-defined transformation properties 
{\begin{eqnarray}
 f(\gamma\tau)=(c\tau+d)^{k}f(\tau)\quad \gamma=\begin{pmatrix} a & b  \\ c & d  \end{pmatrix}\in\Gamma_3
 \end{eqnarray}}
with an integer $k\geq0$, under the group $\Gamma_3$. The three linearly independent weight 2 and level-3 modular forms are given by\,\cite{Feruglio:2017spp}
{\begin{eqnarray}
&&Y_1(\tau)=\frac{i}{2\pi}\Big[\frac{\eta'(\frac{\tau}{3})}{\eta(\frac{\tau}{3})}+\frac{\eta'(\frac{\tau+1}{3})}{\eta(\frac{\tau+1}{3})}+\frac{\eta'(\frac{\tau+2}{3})}{\eta(\frac{\tau+2}{3})}-\frac{27\eta'(3\tau)}{\eta(3\tau)}\Big]\,,\nonumber\\
&&Y_2(\tau)=\frac{-i}{\pi}\Big[\frac{\eta'(\frac{\tau}{3})}{\eta(\frac{\tau}{3})}+\omega^2\frac{\eta'(\frac{\tau+1}{3})}{\eta(\frac{\tau+1}{3})}+\omega\frac{\eta'(\frac{\tau+2}{3})}{\eta(\frac{\tau+2}{3})}\Big]\,,\nonumber\\
&&Y_3(\tau)=\frac{-i}{\pi}\Big[\frac{\eta'(\frac{\tau}{3})}{\eta(\frac{\tau}{3})}+\omega\frac{\eta'(\frac{\tau+1}{3})}{\eta(\frac{\tau+1}{3})}+\omega^2\frac{\eta'(\frac{\tau+2}{3})}{\eta(\frac{\tau+2}{3})}\Big]\,,
\end{eqnarray}}
where $\omega=-1/2+i\sqrt{3}/2$ and $\eta(\tau)$ is the Dedekind eta-function defined by
{\begin{eqnarray}
 \eta(\tau)=q^{1/24}\prod^{\infty}_{n=1}(1-q^n)\quad\text{with}~q\equiv e^{i2\pi\tau}~\text{and}~{\rm Im}(\tau)>0\,.
\end{eqnarray}}
The Dedekind eta-function satisfies
{\begin{eqnarray}
 \eta(-1/\tau)=\sqrt{-i\tau}\,\eta(\tau)\,,\qquad\eta(\tau+1)=e^{i\pi/12}\,\eta(\tau)\,.
\end{eqnarray}}
The three linear independent modular functions transform as a triplet of $A_4$, {\it i.e.} $Y^{(2)}_{\bf 3}=(Y_1, Y_2, Y_3)$. The $q$-expansion of $Y_i(\tau)$ reads
{\begin{eqnarray}
 &&Y_1(\tau)=1+12q+36q^2+12q^3+...\nonumber\\
 &&Y_2(\tau)=-6q^{1/3}(17q+8q^2+...)\nonumber\\
 &&Y_3(\tau)=-18q^{2/3}(1+2q+5q^2+...)\,.
\label{mf1}
\end{eqnarray}}
$Y^{(2)}_{\bf 3}$ is constrained by the relation,
{\begin{eqnarray}
 (Y^{(2)}_{\bf 3}Y^{(2)}_{\bf 3})_{{\bf 1}''}=Y^2_2+2Y_1Y_3=0\,.
\end{eqnarray}}

%%%%%%%%%%%%%%%%%%%%%%%%%%%%%%%%%%%%%%%%%%%%%%%%%%%%%%%%%%%%%%%%%%%%%%%%%%%
\section{Modular invariance of the superpotential}
\label{A4_2}
Ref.\cite{Ahn:2023iqa} shows that the weight of modular forms is corrected by the K{\"a}hler transformation; see the details in the paper. 
The action we consider is required to be invariant under the modular transformation of the modulus $\tau$,
{\begin{eqnarray}
 \tau\rightarrow\gamma\tau=\frac{a\tau+b}{c\tau+d}\,,\quad(a, b, c, d\in{Z},~~ ad-bc=1)\,.
  \label{mt1}
\end{eqnarray}}
and the K{\"a}hler transformation
 {\begin{eqnarray}
&K(\Phi, \bar{\Phi}e^{2A})\rightarrow K(\Phi, \bar{\Phi}e^{2A})+\big(g(\tau)+g(\bar{\tau})\big)M^2_P\,,\nonumber\\
&W(\Phi)\rightarrow W(\Phi)e^{-g(\tau)}
\label{tr1}
\end{eqnarray}}
with $g(\tau)=h\ln(c\tau+d)$.
The modular invariance $W(\Phi)$ under the modular group $\Gamma_N$ ($N\geq2$) provides a strong restriction on the flavor structure\,\cite{Feruglio:2017spp}. The superpotential $W(\Phi)$ can be expanded in power series of the multiplets $\varphi$ which are separated into brane sectors $\varphi_{(I)}$
 {\begin{eqnarray}
W(\Phi)=\sum_n Y_{I_1...I_n}(\tau)\varphi_{(I_1)}\cdot\cdot\cdot\varphi_{(I_n)}\,,
\label{ms1}
\end{eqnarray}}
where the functions $Y_{I_1...I_n}(\tau)$ are generically $\tau$-dependent in type IIA intersecting D-brane models\,\cite{string_book, Cremades:2004wa}.
To ensure that the superpotential $W(\Phi)$ has modular invariance under Eq.(\ref{tr1}),  two conditions must be satisfied:
 (i) the matter superfields $\varphi_{I_i}$ of the brane sector $I_i$ should transform
 {\begin{eqnarray}
\varphi_{(I_i)}\rightarrow (c\tau+d)^{-k_{I_i}}\rho_{(I_i)}(\gamma)\varphi_{(I_i)}
\label{mt3}
\end{eqnarray}}
in a representation $\rho_{(I_i)}(\gamma)$ of the modular group $\Gamma_N$, where $-k_{I_i}$ is the modular weight of sector $I_i$, and (ii) the functions $Y_{I_1...I_n}(\tau)$ should be modular forms of weight $k_Y(n)$ transforming in the representation $\rho(\gamma)$ of $\Gamma_N$,
{\begin{eqnarray}
Y_{I_1...I_n}(\gamma\tau)=(c\tau+d)^{k_Y(n)}\rho(\gamma)Y_{I_1...I_n}(\tau)\,,
\label{mof}
\end{eqnarray}}
with the requirements
{\begin{eqnarray}
&&k_Y(n)-h=k_{I_1}+...+k_{I_n}\,,\nonumber\\
&&\rho(\gamma)\otimes\rho_{(I_1)}\otimes\cdot\cdot\cdot\otimes\rho_{(I_n)}\ni{\bf 1}\,,
\label{req}
\end{eqnarray}}
where ${\bf 1}$ is an invariant singlet.

 %%%%%%%%%%%%%%%%%%%%%%%%%%%%%%%%%%%%%%%%%%%%%%%%%%%%%%%%%%%%%%%%%%%%%%%%%%%
\section{}
The mixing matrix $V^\ell_R$ diagonalizing ${\cal M}_\ell{\cal M}^\dag_{\ell}$ is given, to a good approximation for $1\gg|r|\gg|s|$, by
\begin{eqnarray}
  V^\ell_{R}&\simeq&
  {\left(\begin{array}{ccc}
   e^{-i(\alpha_2+\alpha_3)} & -\big|\frac{\alpha_e}{\alpha_\mu}\frac{q_\ell}{p^\ast_\ell}\big|\Delta_\chi^{|f_e|-|f_\mu|}\,e^{-i(\alpha_2+\alpha_3)} & -\big|\frac{\alpha_e}{\alpha_\tau}\frac{q_\ell}{p^\ast_\ell}\big|\Delta_\chi^{|f_e|-|f_\tau|}\,e^{-i\alpha_2} \\
   \big|\frac{\alpha_e}{\alpha_\mu}\frac{q_\ell}{p^\ast_\ell}\big|\Delta_\chi^{|f_e|-|f_\mu|}\,e^{\alpha_3-\alpha_1} & e^{i(\alpha_3-\alpha_1)} & -\big|\frac{\alpha_\mu}{\alpha_\tau}\frac{q_\ell}{p_\ell}\big|\Delta_\chi^{|f_\mu|-|f_\tau|}\,e^{i(\alpha_2-\alpha_1)}  \\
   \big|\frac{\alpha_e}{\alpha_\tau}\frac{q_\ell}{p^\ast_\ell}\big|\Delta_\chi^{|f_e|-|f_\tau|}\,e^{i(\alpha_1+\alpha_2-\alpha_3)}& \big|\frac{\alpha_\mu}{\alpha_\tau}\frac{q_\ell}{p_\ell}\big|\Delta_\chi^{|f_\mu|-|f_\tau|}\,e^{i(\alpha_1+\alpha_3)} & e^{i(\alpha_1+\alpha_2)}
   \end{array}\right)}\nonumber\\
   &+&{\cal O}\Big(\big|\frac{\alpha_e}{\alpha_\mu}\frac{q_\ell}{p^\ast_\ell}\big|^2\Delta_\chi^{2(|f_e|-|f_\mu|)}, \big|\frac{\alpha_e}{\alpha_\tau}\frac{q_\ell}{p^\ast_\ell}\big|^2\Delta_\chi^{2(|f_e|-|f_\tau|)}, \big|\frac{\alpha_\mu}{\alpha_\tau}\frac{q_\ell}{p_\ell}\big|^2\Delta_\chi^{2(|f_\mu|-|f_\tau|)}\Big)\,,
 \label{vRl}
 \end{eqnarray}
where $\alpha_1=\frac{1}{2}\arg(\alpha_\mu\alpha^\ast_\tau \tilde{q}_\ell)$, $\alpha_2\simeq\arg(\alpha_e\alpha^\ast_\mu \tilde{q}_\ell)-\arg(\alpha_e\alpha^\ast_\tau \tilde{q}^\ast_\ell)+\arg(\alpha_\mu\alpha^\ast_\tau \tilde{q}_\ell)$, $\alpha_3\simeq\frac{1}{2}\arg(\alpha_e\alpha^\ast_\mu \tilde{q}_\ell)+\frac{\alpha_1-\alpha_2}{2}$
with $\tilde{q}_\ell =q_\ell\,Y^4_1(1+r^3+s^3-3rs)$.
\newpage
%%%%%%%%%%%%%%%%%%%%%%%%%%%%%%%%%%%%%%%%%%%%%%%%%%%%%%%%%%%%%%%%%%%%%%%%%%%%%%%%%%%%%%

\end{document}